\documentclass[10pt,twocolumn,twoside]{IEEEtran}
\usepackage{amsmath,graphicx}
\usepackage{epstopdf}
\usepackage{multirow}
\usepackage{balance}
\usepackage{algorithm}
\usepackage{cite}
\usepackage[caption=false,font=footnotesize]{subfig}
\usepackage{stfloats}
\usepackage{amssymb}
\usepackage{algorithmicx}
\usepackage{algpseudocode}
\usepackage{amsfonts}
\usepackage{marvosym}
\graphicspath{{figure/}}

\usepackage{xcolor}

\def\END{\color{black}}

\begin{document}
	\title{Joint DoA-Range Estimation Using Space-Frequency Virtual Difference Coarray}
	
	\author{Zihuan~Mao,~\IEEEmembership{Student Member,~IEEE},
		Shengheng~Liu,~\IEEEmembership{Senior Member,~IEEE},\\
		Yimin~D.\ Zhang,~\IEEEmembership{Fellow,~IEEE},
		Leixin~Han,~\IEEEmembership{Student Member,~IEEE},
		Yongming~Huang,~\IEEEmembership{Senior Member,~IEEE}
		
		\vspace{-1em}		
		\thanks{Z.~Mao, S.~Liu, L.~Han, and Y.~Huang were supported in part by the National Natural Science Foundation of China under Grant Nos.\ 62001103 and U1936201 and the Basic Research Program of Jiangsu Province under Grant No.\ BK20190338. Part of this work was presented at the 2021 IEEE Radar Conference \cite{Cao2021Doubly}. (Corresponding authors: S.~Liu and Y.~Huang.)}
		
		\thanks{Z.~Mao, S.~Liu, L.~Han, and Y.~Huang are with the School of Information Science and Engineering, Southeast University, Nanjing 210096, China, and the Purple Mountain Laboratories, Nanjing 211111, China (email: \{s.liu; huangym\}@seu.edu.cn).}
		
		\thanks{Y.~D.\ Zhang is with the Department of Electrical and Computer Engineering, College of Engineering, Temple University, Philadelphia, PA 19122 USA  (email: ydzhang@temple.edu).}
		
		\thanks{This paper has supplementary downloadable material available at http://ieeexplore.ieee.org, provided by the author. The material includes a PDF file, which is referred to as Supplement File in the manuscript, containing additional proofs and numerical simulations. This material is 4 pages in size.}	
	}
	
	\markboth{IEEE Transactions on Signal Processing,~Vol.~XX, No.~XX, XXX~2021}%
	{MAO \MakeLowercase{\textit{et al.}}: Joint DoA-Range Estimation Using Space-Frequency Virtual Difference Coarray}
	
	\maketitle	
	\begin{abstract}		
		In this paper, we address the problem of joint direction-of-arrival (DoA) and range estimation using frequency diverse coprime array (FDCA). By incorporating the coprime array structure and coprime frequency offsets, a two-dimensional space-frequency virtual difference coarray corresponding to uniform array and uniform frequency offset is considered to increase the number of degrees-of-freedom (DoFs). However, the reconstruction of the doubly-Toeplitz covariance matrix is computationally prohibitive. To solve this problem, we propose an interpolation algorithm based on decoupled atomic norm minimization (DANM), which converts the coarray signal to a simple matrix form. On this basis, a relaxation-based optimization problem is formulated to achieve joint DoA-range estimation with enhanced DoFs. The reconstructed coarray signal enables application of existing subspace-based spectral estimation methods. The proposed DANM problem is further reformulated as an equivalent rank-minimization problem which is solved by cyclic rank minimization. This approach avoids the approximation errors introduced in nuclear norm-based approach, thereby achieving superior root-mean-square error which is closer to the Cram\'er-Rao bound. The effectiveness of the proposed method is confirmed by theoretical analyses and numerical simulations.
	\end{abstract}
	
	\vspace{-1em}	
	
	\section{Introduction}
	\IEEEPARstart {L}OCALIZATION is a fundamental problem in radar, sonar, wireless communications, and navigation. It plays an increasingly important role in various emerging applications such as autonomous driving\cite{Patole2017Automotive}, 5G and next-generation communications\cite{Garcia2017Direct}, and Internet of Things (IoT)\cite{Wan2018Context}. In order to localize targets in both direction-of-arrival (DoA) and range, beam steering of phased arrays should be achieved across a certain signal bandwidth. Recently, a special type of phased array with frequency diversity, namely, frequency diverse array (FDA), has attracted great attention \cite{Wang2013Phased, Liao2019Frequency}. The concept of FDA was first proposed by Antonik et al. \cite{Antonik2006Range, Antonik2007Frequency}. By exploiting a small frequency increment across the array elements, FDA achieves a focused beam to localize targets in both angle and range dimensions \cite{Wang2014Transmit, Xu2015Joint, Lan2020Trans, Xiong17}. On the other hand, an analogous technique called {SpotFi}\cite{Kotaru2015SpotFi} also aroused much attention in the wireless communication community. {SpotFi} is a localization system using channel state information acquired from commercial off-the-shelf Wi-Fi devices, which use orthogonal frequency-division multiplexing (OFDM) signals to estimate the DoA and range of targets relative to the access point. Essentially, the ranging capability of both FDA and {SpotFi} originates from the utilization of frequency diverse signals, which produce a range-related phase difference between subcarriers. The combination of spatial sampling and frequency diversity yields a space-frequency diverse system and, thereby, facilitates joint DoA-range estimation.
	
	Future localization technology is envisioned to accommodate a massive number of devices and support fine-grain positioning. This imposes much higher requirements in terms of the number of spatial degrees-of-freedom (DoFs) and signal bandwidth. The prototype FDA configuration \cite{Antonik2006Range,Antonik2007Frequency}, which consists of $N$ antenna elements with each element using a subcarrier of a fixed frequency offset, is able to resolve up to $N^2-1$ targets. As the number of DoFs, which determines the maximum number of detectable targets, is fundamentally limited by the number of sensors and the order of frequency diversity, increasing the number of DoFs and range resolution with a small number of sensors and low spectrum occupancy is a problem of great interest \cite{Zheng2020Padded}. In the literature, several types of sparse arrays have been designed to provide a larger array aperture and increase the number of DoFs for a given number of physical sensors\cite{Zhang2013Sparsity,Qin2015Genralized,Liu2016Super,Guo2018DOA,Zheng2018DoA}. Extending this concept to FDAs rendered the development of coprime array with coprime frequency offsets \cite{Qin2017Frequency,Ni2021Range} that exploits the virtual difference coarray concept \cite{Vaidyanathan2011Theory} to circumvent the limitation of physical sampling. In this paper, we refer to the array configurations as the frequency diverse coprime array (FDCA), and the corresponding coarray processing as its space-frequency virtual difference coarray.
	
	Under the space-frequency difference equivalence of an FDCA, the derived virtual coarray has multiple missing elements, commonly referred to as `holes', which lead to the model mismatch\cite{BouDaher2015Multi} for decoherence. To fully utilize the array aperture and the DoF potentials of a space-frequency difference coarray, the complex multi-task Bayesian compressive sensing (CMT-BCS) framework is employed  in \cite{Qin2017Frequency} to enhance the robustness to dictionary coherence.  However, this method has two major issues: (a) The maximum number of DoFs is still limited by the number of non-negative unique lags \cite{Qin2015Genralized}, and (b) Pre-defined spatial grids are required. As will be demonstrated later, in addition to its high computationally complexity, CMT-BCS is grid-based and suffers from performance loss when there is basis mismatch.
	
	Solutions to avoiding `holes' and providing extra DoFs in a coarray so as to enable gridless DOA estimation can be largely divided into two classes. The first one extends the maximum contiguous segment by proper non-uniform configuration design\cite{Zheng2020Padded,misc,Cohen2020Sparse}, at the expense of imposing extra requirement on hardware complexity. The other approach is to fill in the holes based on Toeplitz covariance matrix reconstruction, such as those using the nuclear norm minimization (NNM) algorithm \cite{Liu2016Coprime,Qiao2017Unified}, the maximum entropy algorithm \cite{Qiao2017Unified}, and the atomic-norm minimization (ANM) algorithm \cite{Zhou2018Direction,Zheng2020DoA}. These algorithms recover missing array data for a general class of non-uniform array configurations. To ensure that these algorithms work properly, the array manifold must satisfy the Vandermonde property, which is an intrinsic property of the uniform linear array (ULA). However, as the FDCA receive signals are coupled in direction and range, such prerequisite is not met. More specifically, the covariance matrix of an FDCA is obtained from the Kronecker product of two analogous Toeplitz matrices, which are respectively constructed by spatial diversity of sensor position and frequency diversity of different carrier frequency. As such, the resultant covariance matrix is a doubly-Toeplitz matrix instead of a Toeplitz one.
	
	Motivated by the above facts, in this paper, we propose a low-complexity interpolation framework for gridless joint DoA-range estimation using the space-frequency difference coarray of FDCA. First, we derive the space-frequency difference coarray from the physical FDCA receive signal. By respectively applying two-dimensional (2D) spatial smoothing technique (SST)\cite{Kotaru2015SpotFi} and 2D MUltiple SIgnal Classification (MUSIC) algorithm\cite{Schmidt1986Multiple} on the difference coarray signal, a DoF-enhanced joint DoA-range estimation framework is established. Then, we propose a DANM-based interpolation method to fully utilize the information in the presence of `holes' as well as to reduce the computational complexity of doubly-Toeplitz reconstruction. To avoid the approximation loss introduced in the convex relaxation by DANM, we convert the original ANM problem into a 2D rank-minimization problem. By further incorporating the cyclic rank minimization approach\cite{Liu2021Rank} into the rank reformulation framework\cite{Daniel2019Low}, we reconstruct the coarray signal more accurately with lower parameter estimation errors. The enhanced estimation performance in terms of the higher number of DoFs, improved estimation accuracy, and lower computational complexity is verified by the numerical results. The main technical contributions of this work are summarized as follows:
	\begin{itemize}
		\item We design a 2D SST method as the foundation to develop a generic coarray-based gridless joint DoA-range estimation method. A baseline method for DoF-enhanced estimation is formulated using the consecutive part of the space-frequency virtual difference coarray.
		\item We decouple the underlying doubly-Toeplitz reconstruction problem, which is computationally costly, and formulate the DANM approach based on decoupled atomic norm minimization. DANM can fully utilize the space-frequency difference coarray signals with `holes' to increase the number of DoFs.
		\item To avoid the approximation loss of the DANM induced by relaxation, we reformulate a dual-variable rank minimization problem, which is exactly equivalent to the original atomic $l_0$-minimization and thus yields more accurate parameter estimation. We solve the problem
		via cyclic minimization and further design an alternating direction method of multipliers (ADMM)-based solver, which has closed-form solutions and is computationally efficient.
		\item We provide comprehensive theoretical analyses of the proposed joint estimation framework. Based on the interpolated coarray manifold, we first explore the identifiabilty conditions of the space-frequency coarray, and then prove the convergence of the proposed CRM method. Furthermore, to analytically evaluate the performance of the interpolation, the respective bounds of the reconstruction performance using DANM and CRM are derived.
	\end{itemize}
	
	The rest of this paper is organized as follows. In Section \ref{sec:SM}, we first describe the array configuration and the signal model under investigation. Then, we reconstruct the consecutive coarray signal into a spatial-smoothed form as a baseline for DoF-enhanced joint estimation. In Section \ref{sec:DANM}, we reduce the complexity of the underlying doubly-Toeplitz reconstruction problem by decoupling the coarray signal and the atomic norm is defined. We then interpolate the `holes' using DANM and achieve DoF-enhanced joint DoA-range estimation. In Section \ref{sec:rank}, we propose a loss-free cyclic rank-minimization (CRM)-based method to reconstruct the signal matrix and further design an ADMM-based highly efficient closed-form solver. In Section \ref{sec:Performance}, we analyze the identifiabilty of the coarray-based estimation framework, the convergence of the CRM method, and the theoretical reconstruction performance of the interpolation methods. We evaluate and analyze the performance of the proposed scheme via extensive numerical simulations in Section \ref{sec:simu}. The conclusion of this paper is drawn in Section \ref{sec:conclusion}.
	
	\begin{figure*}[!hb]
		\centering
		\subfloat[]{\includegraphics[width=0.5\linewidth]{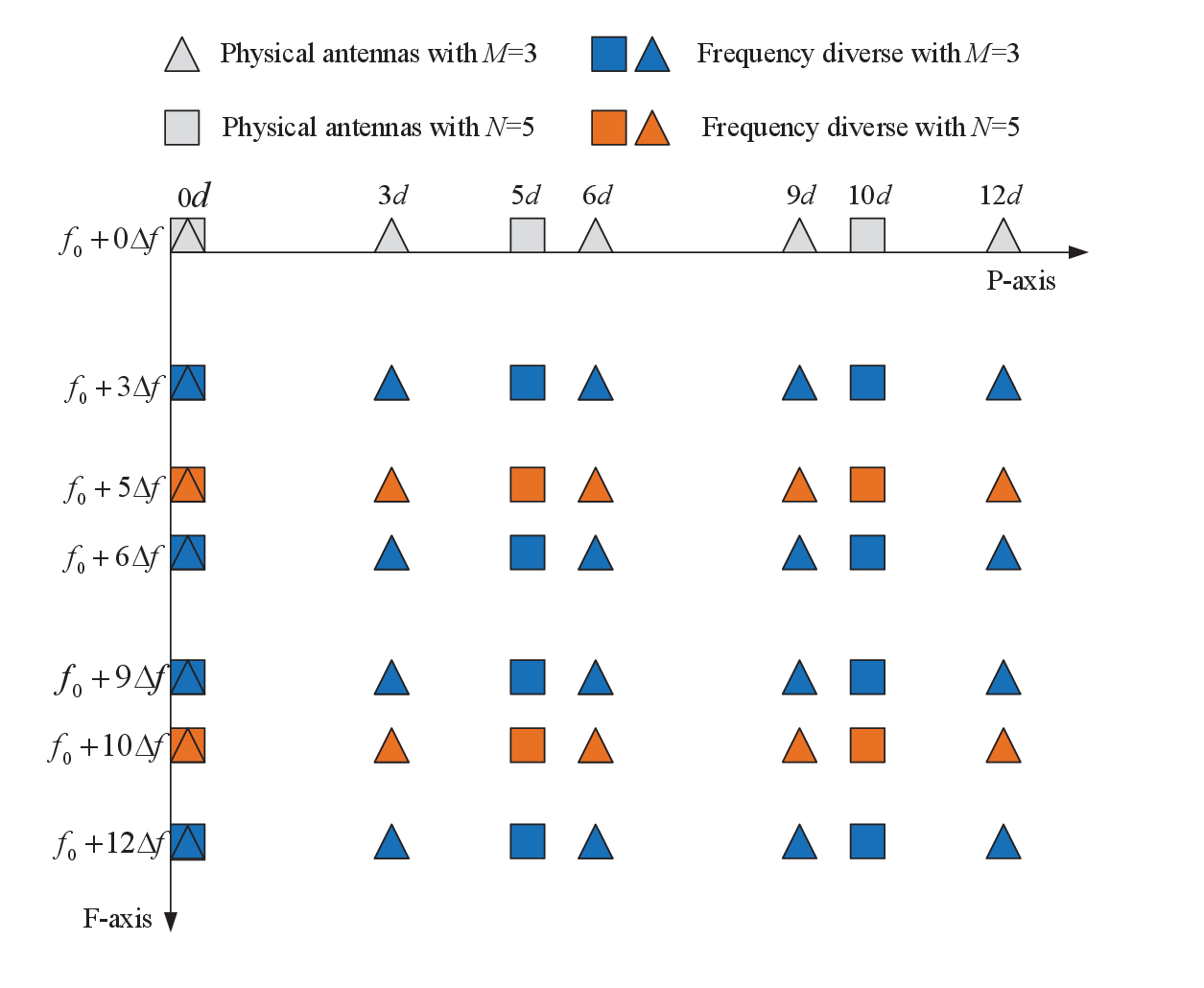}}
		\hfill
		\vspace{-0.5em}	
		\subfloat[]{\includegraphics[width=0.43\linewidth]{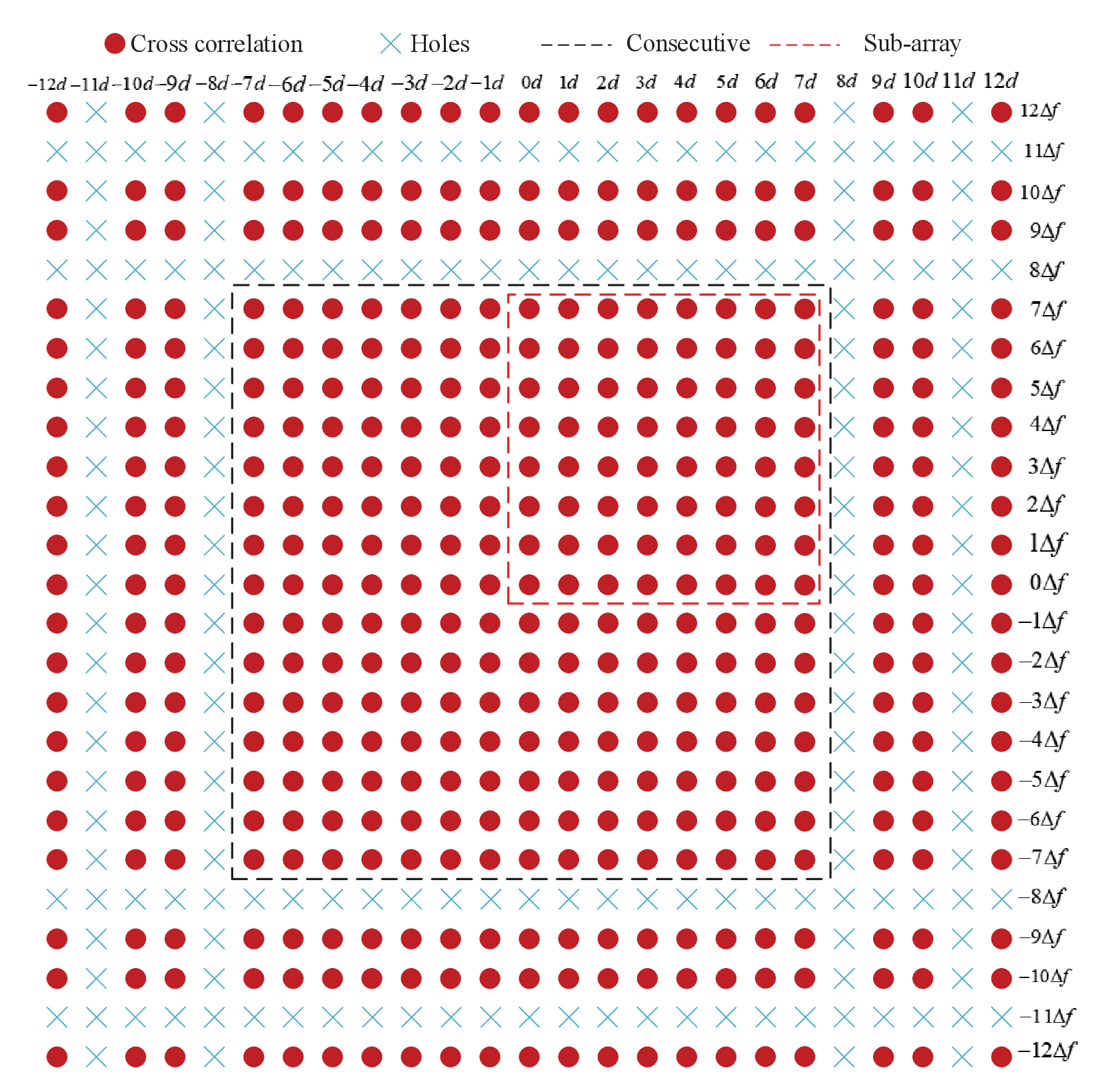}}
		\caption{Illustrative example of an FDCA with $M\!=\!3$ and $N\!=\!5$. (a) Physical array. (b) Space-frequency virtual difference coarray.}\label{FDCA}
		\vspace{-1em}
	\end{figure*}
	Notations: Lower (upper)-case bold characters are used to denote vectors (matrices), and the vectors are by default in column orientation.  Both $A_{i,j}$ and $(\mathbf{A})_{i,j}$ represent the $(i, j)$-th entry of matrix $\mathbf{A}$, while $[\mathbf{A}]_{i,j}$ denotes the $(i,j)$-th submatrix, i.e., $\mathbf{A}$ with the $i$-th row and the $j$-th column deleted. Blackboard-bold characters denote standard sets of numbers and, in particular, $\mathbb{C}$ denotes the set of complex numbers. The operator ${\rm {vec}}(\cdot)$ vectorizes a certain matrix and
	${\rm{matrix}}(\mathbf{a})$ reshapes a vector $\mathbf{a}\in{\mathbb{C}}^{L^2}$ to a square \END matrix  $\mathbf{A}\in{\mathbb{C}}^{L\times L}$. The superscripts $(\cdot)^{\rm T}$, $(\cdot)^*$, and $(\cdot)^{\rm H}$ represent the transpose, complex conjugate, and conjugate transpose operators, respectively. ${\rm {tr}}(\cdot)$ returns the trace of a matrix, and ${\rm{rank}}(\cdot)$ returns the rank of a matrix. ${\rm{diag}}\{\cdot\}$ represents a diagonal matrix that uses the entries of a vector as its diagonal entries. ${\bf{I}}$ denotes an identity matrix, and $\mathbf{i} = {\rm {vec}}(\mathbf{I})$. $\mathbf{T}(\mathbf{x})$ denotes a Hermitian and Toeplitz matrix with $\mathbf{x}$ as the first column. Symbols $\circ$, $\odot$ and $\otimes$ stand for Hadamard, Khatri-Rao and Kronecker products, respectively. The notation $\| \cdot \|_{\mathcal{F}}$ represents the Frobenius norm, and ${\rm {E}}[\cdot]$ denotes the expected value of a discrete random variable. The curled symbol $\succeq$ denotes linear matrix inequality. $\mathcal{N}(\mu,\sigma^2)$ denotes a Gaussian distribution with mean $\mu$ and variance $\sigma^2$. $\jmath = \sqrt{-1}$ is the imaginary unit.

\section{Preliminary}\label{sec:SM}

\subsection{FDCA Signal Model}\label{physical}

In this section, the signal model of FDCA under investigation is provided. The receive array in an FDCA transceiver system consists of two uniform sparse subarrays. One subarray is equipped with $M$ sensors spaced by $N$ units, and the other has $N$ sensors spaced by $M$ units, where $M$ and $N$ are a pair of coprime integers with $M < N$, and the unit inter-element spacing $d$ typically takes the value of half wavelength to avoid spatial ambiguity. In the following, we consider a simple signal model in which the carrier offsets of the transmit frequency diverse signal and the sensor spacing share the same coprime pattern. As we will show in the sequel, this setting produces a clear symmetric doubly-Toeplitz structure. Note that the derivation can be easily extended to a general case where the sparse pattern for the sensor positions and that for the carrier frequencies are different. Overall, the space-frequency structure of the FDCA can be described as the following prototype coprime integers set
		\begin{equation}
		\mathbb{S}=\mathbb{S}_1\cup\mathbb{S}_2
		\end{equation}
	with
	\begin{subequations}\label{coprimeset}
		\vspace{-0.5em}
		\begin{equation}
		\mathbb{S}_1=\{Nm,0\le m\le M-1\},
		\end{equation}
		\begin{equation}
		\mathbb{S}_2=\{Mn,0\le n\le N-1\}.
		\end{equation}
	\end{subequations}
	As such, the coprime integer set yields the set of all sensor positions $\mathbb{P}=\mathbb{S}d$ and the set of all carrier frequencies $\mathbb{F}=f_0+\mathbb{S}\Delta f$, where $f_0$ and $\Delta f$ are respectively the reference frequency and the unit frequency offset. As an example, the FDCA model corresponding to $M=3$ and $N=5$ is illustrated in Fig.~\ref{FDCA}(a). Without loss of generality, we limit our discussion to far-field targets with only azimuth DoAs. Assume $K$ uncorrelated targets at positions $(\theta_k,r_k), k=1,\cdots,K$, where $\theta_k$ and $r_k$ respectively represent the DoA and range of the $k$-th target. The $q$-th frequency component reflected by the $k$-th target and received by the $i$-th receive sensor is expressed as
	\begin{align}
	x_{i,q}(t)&=\sum\nolimits_{k=1}^{K}s_k(t){\rm{e}}^{\jmath\frac{4\pi f_q r_k}{\rm{c}}}{\rm{e}}^{-\jmath\frac{2\pi d_i}{\lambda_q}\sin\theta_k}+n_{i,q}(t)\nonumber \\
	&\approx\sum\nolimits_{k=1}^{K}s_k(t){\rm{e}}^{\jmath\frac{4\pi f_q r_k}{\rm{c}}}{\rm{e}}^{-\jmath\frac{2\pi d_i}{\lambda}\sin\theta_k}+n_{i,q}(t),
	\end{align}
	where $s_k(t)$ is the reflection complex envelope of the $k$-th target for $t=1, \cdots, T$, and $f_q=f_0+s_q\Delta f$ is the $q$-th carrier frequency, where $s_q$ is the $q$-th element of coprime integer set $\mathbb{S}$ and $\rm{c}$ is the speed of light. Likewise, $d_i=s_id$ is the position of the $i$-th receive sensor, where $s_i$ is the $i$-th element of $\mathbb{S}$. Finally, $n_{i,q}(t)$ denotes the additive white Gaussian noise (AWGN). Stacking $x_{i,q}(t)$ for all $i,q=1,\cdots,M+N-1$, the receive signal vector of the FDCA is obtained as
	\begin{align}\label{signalmodel} \mathbf{x}(t)\!=\!\sum\nolimits_{k=1}^{K}s_k(t)\mathbf{h}_{\rm{p,f}}(\theta_k,r_k)\!+\!\mathbf{n}(t)\!=\!\mathbf{H}_{\rm{p,f}}\mathbf{s}(t)\!+\!\mathbf{n}(t),
	\end{align}
	where $\mathbf{h}_{\rm{p},\rm{f}}(\theta_{k},r_{k})=\mathbf{h}_{\rm{p}}(\theta_k)\otimes\mathbf{h}_{\rm{f}}(r_k)$ represents the steering vector associated with the position $(\theta_k,r_k)$, with $\mathbf{h}_{\rm{p}}(\theta_k)$ and $\mathbf{h}_{\rm{f}}(r_k)$ denoting the steering vectors corresponding to DoA $\theta_k$ and range $r_k$, respectively. In addition, the following vectors and matrix are used in (\ref{signalmodel}):
	\begin{subequations}
		\begin{equation}
		\mathbf{s}(t)=\left[s_1(t),\cdots,s_K(t)\right]^{\rm{T}},
		\end{equation}
		\vspace{-1.5em}
		\begin{equation}
		\mathbf{H}_{\rm{p,f}}=\left[\mathbf{h}_{\rm{p,f}}(\theta_1,r_1),\cdots,\mathbf{h}_{\rm{p,f}}(\theta_K,r_K)\right],
		\end{equation}
		\vspace{-1.5em}
		\begin{equation}
		\mathbf{h}_{\rm{p}}(\theta_k)=\left[1,\cdots,\exp({-\jmath{2\pi d_{M+N-1}\sin\theta_k}/{\lambda}})\right]^{\rm{T}},
		\end{equation}
		\vspace{-1.5em}
		\begin{equation}
		\mathbf{h}_{\rm{f}}(r_k)=\left[1,\cdots,\exp({\jmath{4\pi f_{M+N-1}r_k}/{\rm{c}}})\right]^{\rm{T}}.
		\end{equation}
	\end{subequations}
	The covariance matrix of the FDCA signal vector $\mathbf{x}(t)$ can be written as
	\begin{align}\label{covariance}
	\mathbf{R}_{\mathbf{x}}&={\rm E}\left[\mathbf{x}(t)\mathbf{x}^{\rm{H}}(t)\right]=\mathbf{H}_{\rm{p,f}}\mathbf{R}_{\mathbf{s}}\mathbf{H}_{\rm{p,f}}^{\rm{H}}+\sigma_{\rm{n}}^2\mathbf{I}\nonumber \\
	&=(\mathbf{H}_{\rm{p}}\odot\mathbf{H}_{\rm{f}})\mathbf{R}_{\mathbf{s}}(\mathbf{H}_{\rm{p}}\odot\mathbf{H}_{\rm{f}})^{\rm{H}}+\sigma_{\rm{n}}^2\mathbf{I},
	\end{align}
	where $\mathbf{R}_{\mathbf{s}}$ represents the covariance matrix of the target reflection signals $\mathbf{s}(t)$ and $\sigma_{\rm{n}}^2$ denotes the noise power. When the target signals are uncorrelated, $\mathbf{R}_\mathbf{s}={\rm{diag}}\left\{\sigma_1^2,\cdots,\sigma_K^2\right\}$ with $\sigma_k^2$ representing the reflection power of the $k$-th target. Since $\mathbf{R}_\mathbf{x}$ is unavailable in practice, it is substituted by its asymptotically unbiased estimate from $T$ available snapshots,
	\begin{align}\label{covariancematrix}	\hat{\mathbf{R}}_{\mathbf{x}}=\frac{1}{T}\sum\nolimits_{t=1}^{T}\mathbf{x}(t)\mathbf{x}^{\rm{H}}(t).
	\end{align}
	\subsection{Space-Frequency Difference Coarray}\label{sec:virtualcoarray}	
	
	In this subsection, we first derive the virtual domain signal corresponding to the space-frequency difference coarray of the FDCA based on the estimated covariance matrix $\hat{\mathbf{R}}_{\mathbf{x}}$. Vectorizing $\hat{\mathbf{R}}_{\mathbf{x}}$ by stacking the columns of the matrix, the signal in the virtual domain can be written as
	\begin{align}\label{vec}
	\breve{\mathbf{x}}_{\rm{v}}={\rm{vec}}(\hat{\mathbf{R}}_{\mathbf{x}})=\mathbf{A}_{\rm{v}}\mathbf{p}+\sigma_{\rm{n}}^2\mathbf{i},
	\end{align}
	where $\mathbf{A}_{\rm{v}}=\mathbf{H}_{\rm{p,f}}\otimes\mathbf{H}_{\rm{p,f}}^{*}$
	and $\mathbf{p}=[\sigma_1^2,\cdots,\sigma_K^2]^{\rm T}$. Owing to the cross-correlation between different sensors, each element of $\breve{\mathbf{x}}_{\rm{v}}$ corresponds to a sensor in the virtual space-frequency domain. The positions and frequency offsets of virtual sensors are determined by their respective difference sets  $\mathbb{P}$ and $\mathbb{F}$, i.e.,

		\begin{equation}
		\mathbb{P}_{\text{diff}}=\mathbb{S}_{\text{diff}}\cdot d,
\quad 		\mathbb{F}_{\text{diff}}=\mathbb{S}_{\text{diff}}\cdot \Delta f,
		\end{equation}
	where $\mathbb{S}_{\text{diff}}$ is the difference set of the two coprime-integer sets, expressed as
	\begin{align}
	\mathbb{S}_{\text{diff}}=\left\{l\mid l=\pm(Nm-Mn)\vert m=0,1,\cdots,M-1,\right.\nonumber\\
	n=0,1,\cdots,N-1\left.\right\}.
	\end{align}
	The signal output of the virtual sensor at position $l_1d$ with frequency offset $l_2\Delta f$ is computed as the cross-correlation of two physical sensors spacing $l_1d$ with frequency difference $l_2\Delta f$. Thus, the virtual signal vector can be obtained by selecting the element of $\breve{\mathbf{x}}_{\rm{v}}$ as
	\vspace{-0.5em}
	\begin{align}
	\check{\mathbf{x}}_{\rm{v}}=\check{\mathbf{A}}_{\rm{v}}\mathbf{p}+\sigma_{\rm{n}}^2\check{\mathbf{i}}.
	\end{align}
	The entries of submatrix $\check{\mathbf{A}}_{\rm{v}}$ and subvector $\check{\mathbf{i}}$ are respectively selected from ${\mathbf{A}}_{\rm{v}}$ and $\mathbf{i}$ by removing redundant elements.
	
	For the FDCA illustrated in Fig.~\ref{FDCA}(a), the yielding space-frequency difference coarray is shown in Fig.~\ref{FDCA}(b). The dimension of the signal vector rapidly increases from 49 to 441 in the coarray derivation, and the number of non-negative lags increases from 49 to 121 as well. The missing sensors at $\{-11,-8,8,11\}$ correspond to the missing elements in the difference set of coprime integers, which are commonly referred to as `holes' in the literature. If we can fill the `holes' shown in Fig.~\ref{FDCA}(b), as we will discuss in the sequel, the number of non-negative lags will be further increased to 169.
	
\begin{figure}[!hb]
		\centering
		\includegraphics[width=1.05\linewidth]{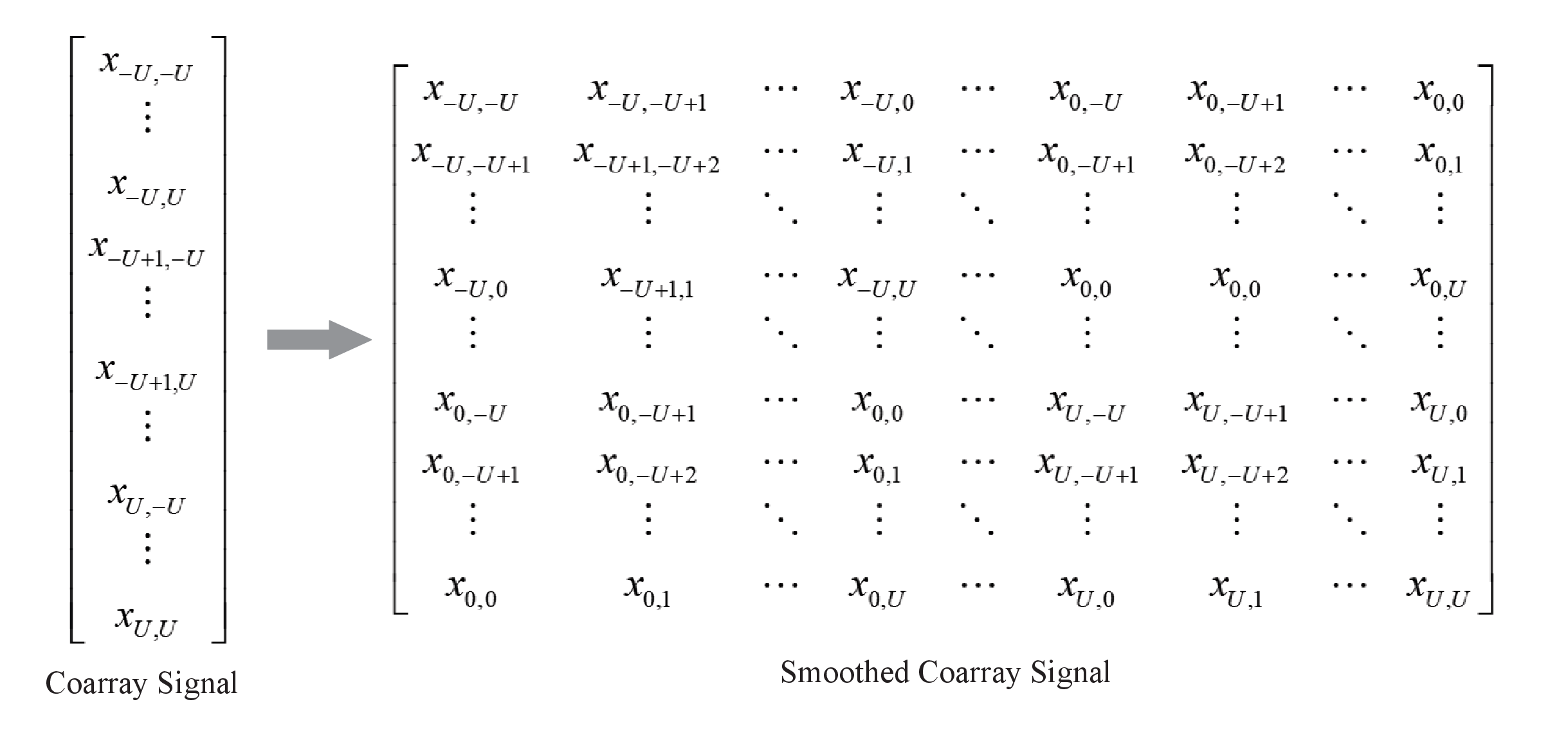}
\vspace{-2em}
		\caption{Construction of smoothed coarray signal from input coarray signal. }
		\label{SST_figure}\vspace{-1em}
	\end{figure}

	Note that, the virtual signal vector $\check{\mathbf{x}}_{\rm{v}}$ is obtained from the covariance matrix which represents the second-order statistics of the signal and, thus, is equivalent to a single-snapshot signal which suffers from the problem of coherence. That is, any uncorrelated signal in the physical FDCA domain becomes coherent in the difference coarray. As a result, the covariance matrix of the virtual signal is rank-deficient and decoherence techniques such as the SST~\cite{Kotaru2015SpotFi} must be applied. The SST divides a uniform linear array into multiple overlapping subarrays to solve the rank-deficient problem. For difference coarray containing `holes', only the consecutive part located within the black dashed square of Fig.~\ref{FDCA}(b) is extracted. The size of subarrays is set as the red dashed square in Fig.\;\ref{FDCA}(b), and a full-rank measurement can be constructed via SST as illustrated in Fig.\;\ref{SST_figure}, where  $x_{j,l}$ is the virtual signal with frequency offset $j\Delta f$ at position $ld$, and $U$ is the maximum index of the consecutive part. 2D MUSIC algorithm can be directly applied to the smoothed coarray signal to estimate the range and DoA of all targets. We denote the smoothed coarray signal in Fig.~\ref{SST_figure} as $\mathbf{R}_{\rm{ss}}$, which is a spatial smoothed covariance matrix of the receive signal. Denote $\mathbf{U}_{\rm{N}}$ as the noise subspace formed by eigenvectors that correspond to the smallest eigenvalues of $\mathbf{R}_{\rm{ss}}.$ Then, the 2D MUSIC algorithm is expressed as
	\begin{equation}\label{2DMUSIC}
	(\hat{\theta},\hat{r})=\arg\mathop{\max}_{\theta,r}\frac{1}{\tilde{\mathbf{a}}^{\rm{H}}_{\rm{p,f}}(\theta,r)\mathbf{U}_{\rm{N}}\mathbf{U}^{\rm{H}}_{\rm{N}}{\tilde{\mathbf{a}}}_{\rm{p,f}}(\theta,r)},
	\end{equation}
	where
	\begin{subequations}
		\begin{equation}
		{\tilde{\mathbf{a}}}_{\rm{p,f}}(\theta,r)={\tilde{\mathbf{a}}}_{\rm{p}}(\theta)\otimes{\tilde{\mathbf{a}}}_{\rm{f}}(r),
		\end{equation}
		\begin{equation}
		{\tilde{\mathbf{a}}}_{\rm{p}}(\theta)=\left[1,\cdots,\exp({-\jmath\pi U\sin\theta})\right]^{\rm{T}},
		\end{equation}
		\begin{equation}
		{\tilde{\mathbf{a}}}_{\rm{f}}(r)=\left[1,\cdots,\exp({{\jmath 4\pi U\Delta f r}/{\rm{c}}})\right]^{\rm{T}}.
		\end{equation}
	\end{subequations}

	By applying the coarray SST, the number of DoFs is effectively increased. For the FDCA example illustrated in Fig.\;\ref{FDCA}, the number of DoFs increases from 48 to 63 in theory. However, such method underutilizes the array aperture and frequency bandwidth as it discards the non-consecutive part, which inevitably results in performance loss.
	
	\section{DoF-enhanced joint DoA-Range Estimation Using DANM}
	\label{sec:DANM}
	
	In this section, we develop a DoF-enhanced joint DoA-range estimation method based on DANM. In order to make full use of the information contained in the non-consecutive coarray, the proposed method first creates an interpolated signal vector. To reduce the computational complexity of doubly-Toeplitz covariance matrix reconstruction,  the coarray signal is converted into a matrix form to decouple the array position and frequency offsets. Based on the atomic norm-based formulation of the signal matrix, a relaxation-based reconstruction method is proposed. Once the augmented signal matrix is recovered, the SST and a subspace-based spectral estimation method are used to jointly estimate the range and DoA of the targets.
	\vspace{-1em}
	\subsection{Initialization of Coarray Signal Matrix}
	The interpolated coarray signal $\tilde{\mathbf{x}}_{\rm{v}}$ can be initialized as
	\begin{align}\label{initialization}
	\left[\tilde{\mathbf{x}}_{\rm{v}}\right]_{j,l}=
	\begin{cases}
	\quad \left[\check{\mathbf{x}}_{\rm{v}}\right]_{j,l},&\quad j,l\in\mathbb{S}_{\rm{diff}},\\
	\quad 0,&\quad j\;\text{or}\; l\in\mathbb{S}_{\rm{I}}-\mathbb{S}_{\rm{diff}},
	\end{cases}
	\end{align}
	where $\mathbb{S}_{\rm{I}}$ denotes the full integer set between $-M(N-1)$ and $M(N-1)$, and $[\check{\mathbf{x}}_{\rm{v}}]_{j,l}$ represents the virtual sensor with frequency offset $j\Delta f$ at position $ld$. Denote $\mathbf{b}$ as a binary vector with one indicating derived statistics in $\tilde{\mathbf{x}}_{\rm{v}}$ and zero indicating interpolated results. Then, we obtain
	$\tilde{\mathbf{x}}_{\rm{v}}\circ\mathbf{b}=\check{\mathbf{x}}_{\rm{v}}$.
	
	Interpolating the `holes' mathematically comes down to a Hermitian and Toeplitz covariance matrix  reconstruction problem in single-parameter estimation (e.g., DoA estimation in an array), which can be solved by semi-definite programming (SDP) approaches such as NNM and ANM. In the space-frequency coarray of an FDCA, on the other hand, the signal covariance matrix is a high-dimensional doubly Toeplitz matrix which is generated by the Kronecker product of two Toeplitz matrices. Consider a noiseless case, the covariance matrix of coarray signal vector  $\mathbf{x}_{\rm{v}}=\mathbf{A}_{\rm{p,f}}\mathbf{p}=(\mathbf{A}_{\rm{p}}\odot\mathbf{A}_{\rm{f}})\mathbf{p}$ after applying SST is expressed as
	\begin{align}\label{Doubly-Toeplitz}	\mathbf{R}_{\rm{v}}&=\mathbf{A}^{\rm{H}}_{\rm{p,f}}\mathbf{R}_{\rm{s}}\mathbf{A}_{\rm{p,f}}=(\mathbf{A}_{\rm{p}}\odot\mathbf{A}_{\rm{f}})^{\rm{H}}\mathbf{R}_{\rm{s}}(\mathbf{A}_{\rm{p}}\odot\mathbf{A}_{\rm{f}}), 
	\end{align}
	where the array manifolds and steering vectors are defined as
	\begin{subequations}
		\begin{equation}
		\mathbf{A}_{\rm{P}}=\left[\mathbf{a}_{\rm{p}}(\theta_1),\cdots,\mathbf{a}_{\rm{p}}(\theta_K)\right]\in\mathbb{C}^{M(N-1)\times K},
		\end{equation}
		\begin{equation}
		\mathbf{A}_{\rm{f}}=\left[\mathbf{a}_{\rm{f}}(r_1),\cdots,\mathbf{a}_{\rm{r}}(r_K)\right]\in\mathbb{C}^{M(N-1)\times K},
		\end{equation}
		\begin{equation}
		\mathbf{a}_{\rm{p}}(\theta_k)=\left[1,\cdots,\exp\left(-\jmath\pi M(N-1)\sin\theta_k\right)\right]^{\rm{T}},
		\end{equation}
		\begin{equation}
		\mathbf{a}_{\rm{f}}\left(r_k\right)=\left[1,\cdots,\exp\left({{\jmath4\pi M(N-1)\Delta f r_k}/{\rm{c}}}\right)\right]^{\rm{T}}.
		\end{equation}
	\end{subequations}
	According to \eqref{Doubly-Toeplitz}, the dimension of $\mathbf{R}_{\rm{v}}\in\mathbb{C}^{M^2(N-1)^2\times M^2(N-1)^2}$ exponentially increases with $M$ and $N$. Therefore, the computational complexity of reconstructing $\mathbf{R}_{\rm{v}}$ is intractable for large-scale antenna arrays\cite{Tian2017Low}.
	
	Note that the coarray signal is obtained from the second-order covariances and is equivalent to a single-snapshot signal as mentioned in Section \ref{sec:SM}. Hence, the augmented virtual coarray signal can also be expressed in a matrix form, i.e.,
	\begin{align}\label{signalmatrix} \mathbf{X}_{\rm{v}}\!=\!{\rm{matrix}}(\!\mathbf{x}_{\rm{v}}\!)\!=\!\sum\nolimits_{k=1}^{K}p_k\mathbf{a}_{\rm{p}}(\theta_k)\mathbf{a}^{\rm{H}}_{\rm{f}}(r_k)
	\!=\!\mathbf{A}_{\rm{p}}\mathbf{P}\mathbf{A}^{\rm{H}}_{\rm{f}},
	\end{align}
	where $\mathbf{P}={\rm{diag}}(\mathbf{p})$.
\vspace{-1em}
\subsection{Atomic Norm of Coarray Signal Matrix}\label{sec:DANM-analysis}
	In view of the definition of the atomic norm of $\mathbf{X}_{\rm{v}}$, an atom for representing $\mathbf{X}_{\rm{v}}$ can be expressed as
	\begin{equation}
	\mathbf{G}(\theta,r)={\mathbf{a}}_{\rm{p}}(\theta)\mathbf{a}^{\rm{H}}_{\rm{f}}(r),
	\end{equation}
	where ${\mathbf{a}}_{\rm{p}}(\theta)=[\exp\left({ \jmath\pi L\sin\theta}\right),\cdots,\exp({-\jmath\pi L\sin\theta})]^{\rm{T}}$, ${\mathbf{a}}_{\rm{f}}(r)=[\exp({{-\jmath 4\pi L\Delta f r}/{\rm{c}}}),\cdots,\exp({{\jmath 4\pi L\Delta f r}/{\rm{c}}})]^{\rm{T}}$ and $L=M(N-1)$. Thus, the corresponding atom set is
	\begin{align} \mathbb{A}=\left\{\mathbf{G}\left(\theta,r\right)\big|\theta\in\left[-90^{\circ},90^{\circ}\right],r\in\left[0,\frac{\rm{c}}{2\Delta f}\right]\right\}.
	\end{align}
	The atomic $l_0$-norm is the smallest number of atoms needed to represent the coarray signal matrix $\mathbf{X}_{\rm{v}}$, defined as
	\begin{align}\label{atomicl0norm} \vert\vert\mathbf{X}_{\rm{v}}\vert\vert_{\mathbb{A},0}\!=\!\mathop{\inf}_{K}\left\{\mathbf{X}_{\rm{v}}=\sum\nolimits_{k=1}^Kp_k\mathbf{G}(\theta_k,r_k),p_k\ge 0\right\}.
	\end{align}
	Reconstructing $\mathbf{X}_{\rm{v}}$ is equivalent to finding the combination of the smallest number of $K$
	atoms. Till this point, the reconstruction problem is loss-free. However, such $l_0$-minimization problem is NP-hard \cite{Zhou2018Direction}. One straightforward and widely accepted alternative is to introduce convex relaxation, rendering the following $l_1$-minimization problem:
	\begin{align}	\vert\vert\mathbf{X}_{\rm{v}}\vert\vert_{\mathbb{A}}\!=\!\mathop{\inf}\left\{\sum\nolimits_{k=1}^Kp_k\big|\mathbf{X}_{\rm{v}}=\!\!\sum\nolimits_{k=1}^Kp_k\mathbf{G}(\theta_k,r_k),p_k\ge0\right\}.
	\end{align}
	Utilizing the initialized signal in \eqref{initialization} as the reference of reconstruction, the optimization problem is formulated as
	\begin{align}\label{ANM} \hat{\mathbf{X}}_{\rm{v}}=\arg&\mathop{\min}_{\mathbf{X}_{\rm{v}}}\|\mathbf{X}_{\rm{v}}\|_{\mathbb{A}} \nonumber \\
	\text{subject to} \quad& \|\mathbf{X}_{\rm{v}}\circ\mathbf{B}-\tilde{\mathbf{X}}_{\rm{v}}\|^2_{\mathcal{F}}\le\eta,
	\end{align}
	where $\tilde{\mathbf{X}}_{\rm{v}}={\rm{matirx}}(\tilde{\mathbf{x}}_{\rm{v}})$, $\mathbf{B}={\rm{matrix}}(\mathbf{b})$, and $\eta$ represents an upper bound of the noise effect. The decoupled ANM problem in \eqref{ANM} is equivalent to a trace-minimization SDP problem as \cite{Yang2016Vandermonde}
	\begin{align} \label{DANM1}
	&\min \limits_{\mathbf{z}_{\rm{p}},\mathbf{z}_{\rm{f}},\mathbf{X}_{\rm{v}}}  \frac{1}{2L}\left({\rm{tr}}\left[\mathbf{T}(\mathbf{z}_{\rm{p}})\right]+{\rm{tr}}\left[\mathbf{T}(\mathbf{z}_{\rm{f}})\right]\right)+\mu\|\mathbf{X}_{\rm{v}}\circ\mathbf{B}-\tilde{\mathbf{X}}_{\rm{v}}\|^2_{\mathcal{F}}\nonumber \\
	&\qquad \qquad \quad \text{subject to} \quad
	\begin{bmatrix}
	{\bf{T({\mathbf{z}}_{\rm{p}}}}_{\rm}) & {\mathbf{X}_{\rm{v}}}    \\
	{\mathbf{X}}_{\rm{v}}^{\rm H} & {\bf{T({\bf{z}}_{\rm{f}}}})
	\end{bmatrix}
	\succeq 0,
	\end{align}
	where $\mu$ is a regularization coefficient. Readers are referred to \cite{Yang2016Vandermonde} for more details on the problem conversion, where a generic framework for multilevel Toeplitz matrix decomposition is introduced. The optimization problem \eqref{DANM1} is convex and can be solved efficiently by using available SDP solvers (e.g., CVX). This work concerns more on how to fully utilize the partially observed 2D coarray signal in joint parameter estimation based on low-complexity and effective interpolation. After obtaining the augmented signal matrix $\mathbf{X}_{\rm{v}}$, the `holes' in the space-frequency difference coarray are effectively interpolated. By observing the optimization in \eqref{DANM1}, the following remarks are in order.
	
	$\textit{Remark 1}$:
	Note that the Hermitian and Toeplitz matrices $\mathbf{T}(\mathbf{z}_{\rm{p}})$ and $\mathbf{T}(\mathbf{z}_{\rm{f}})$ are decoupled from a Kronecker product in \eqref{DANM1}, and three matrices $\mathbf{T}(\mathbf{z}_{\rm{p}})$,  $\mathbf{T}(\mathbf{z}_{\rm{f}})$, $\mathbf{X}_{\rm{v}}$ have the same dimension ${(2L+1)\times(2L+1)}$. Compared to the doubly-Toeplitz covariance matrix $\mathbf{R}_{\rm{v}}\in \mathbb{C}^{L^2\times L^2}$, the number of the optimization variables decreases from $\mathcal{O}(L^4)$ to  $\mathcal{O}(L^2)$. Different from the decoupling operation which directly adopts a single-snapshot signal collected by physical sensors, $\mathbf{x}_{\rm{v}}$ is obtained by averaging the covariance matrices of multiple snapshots. The yielded coarray signal fully utilizes all accessible snapshots and is yet given in a single-snapshot form. Hence, the decoupling in \eqref{DANM1} guarantees the number of DoFs and estimation performances with significantly reduced computational complexity.

	$\textit{Remark 2}$:
	The full observation of the space-frequency difference coarray is obtained in $\mathbf{X}_{\rm{v}}$. To jointly estimate the DoA and range of targets, a method similar to that described in Section~\ref{sec:virtualcoarray} can be applied to overcome the rank deficiency, which increases the number of DoFs to $L^2+2L$. If we instead apply cascaded 2D SST and MUSIC to the consecutive part (hereafter referred to as SST for short), the achievable number of DoFs is only $U^2+2U$; while that of the CMT-BCS \cite{Qin2017Frequency} with all $D$ non-negative unique lags is $D^2+2D$. For instance, when $M=3$ and $N=5$, SST only detects at most $63$ targets, while the CMT-BCS can detect at most $121$ targets. In comparison, the DANM-based interpolation method can detect at most $168$ targets. However, the maximum number of DoFs is not always attainable since the interpolation is governed by the partial observation, as described in \eqref{DANM1}. Detailed consideration of the identifiability will be given in Section \ref{sec:Performance}-A.

	$\textit{Remark 3}$: Compared to the SST which drops the non-consecutive part, the DANM exploits all information in the partially observed difference coarray. As a larger aperture and a higher bandwidth of FDCA are utilized in the DANM than the SST, the estimation performance can be improved by exploiting signal matrix interpolation.
	
	$\textit{Remark 4}$: If we consider a multiple measurement vector (MMV) form of the virtual signal\cite{Zhou2018Direction}, the covariance-based formulation and similar DANM methods \cite{Lu2020Efficient} can also be developed. To further reduce the computational complexity, the methodology of compressed sensing can be combined with DANM \cite{Wang2020Efficient}.
	
	The joint DoA-range estimation method using DANM is summarized in {\bf Algorithm \ref{alg1}}.

	\section{Cyclic Rank-Minimization Based Reconstruction Method}\label{sec:rank}
	
	In the previous section, the original atomic $l_0$-norm minimization problem is solved by relaxing it to an $l_1$-norm minimization problem. However, the relaxation-based solution inevitably leads to performance loss. To avoid such performance loss, we propose a cyclic low-rank algorithm to recover the two decoupled Toeplitz matrices.
	
	\subsection{Reformulation of Multi-Convex Optimization Problem}\label{sec:rankreformulation}
	
	We first prove the equivalence between the atomic $l_0$-norm and a dual-variable rank-minimization problem. On this basis, we recast the underlying rank-minimization problem as an equivalent multi-convex form, which is solved using cyclic iterations. To formulate the CRM problem, the following proposition is first derived.

	$\textit{Proposition 1}$:
	Let $\mathbf{T}(\mathbf{z}_{\rm{p}})$ and $\mathbf{T}(\mathbf{z}_{\rm{f}})\in\mathbb{C}^{2L+1\times2L+1}$ be two Hermitian and Toeplitz matrices. Then, the original atomic $l_0$-norm minimization problem
	\eqref{atomicl0norm} is equivalent to
	\begin{align} \label{rankminimization1}
	&\min \limits_{\mathbf{z}_{\rm{p}},\mathbf{z}_{\rm{f}},\mathbf{X}_{\rm{v}}}  \frac{1}{2}\left({\rm{rank}}\left[\mathbf{T}(\mathbf{z}_{\rm{p}})\right]+{\rm{rank}}\left[\mathbf{T}(\mathbf{z}_{\rm{f}})\right]\right) \nonumber \\
	&\text{subject to} \quad
	\begin{bmatrix}
	{\bf{T({\mathbf{z}}_{\rm{p}}}}_{\rm}) & {\mathbf{X}_{\rm{v}}}    \\
	{\mathbf{X}}_{\rm{v}}^{\rm H} & {\bf{T({\bf{z}}_{\rm{f}}}})
	\end{bmatrix}
	\succeq 0.
	\end{align}
	\begin{IEEEproof}
		Denote by $r^{\rm{p}}_{\text{opt}}$ and $r^{\rm{f}}_{\text{opt}}$ the minimum ranks of $\mathbf{T}(\mathbf{z}_{\rm{p}})$ and $\mathbf{T}(\mathbf{z}_{\rm{f}})$ obtained from \eqref{atomicl0norm}, respectively. We first show that $(1/2)({\rm{rank}}[\mathbf{T}(\mathbf{z}_{\rm{p}})]+{\rm{rank}}[\mathbf{T}(\mathbf{z}_{\rm{f}})])\le\|\mathbf{X}_{\rm{v}}\|_{\mathbb{A},0}$. Let $\|\mathbf{X}_{\rm{v}}\|_{\mathbb{A},0}=m$ and assume that $\mathbf{X}_{\rm{v}}=\sum\nolimits_{k=1}^mp_k\mathbf{a}_{\rm{p}}(\theta_k)\mathbf{a}^{\rm{H}}_{\rm{f}}(r_k)$ with $p_k>0$ achieves $\|\mathbf{X}_{\rm{v}}\|_{\mathbb{A},0}$. We further assume that  $\mathbf{T}(\mathbf{z}_{\rm{p}})=\sum\nolimits_{k=1}^mp_k\mathbf{a}_{\rm{p}}(\theta_k)\mathbf{a}^{\rm{H}}_{\rm{p}}(\theta_k)$ and $\mathbf{T}(\mathbf{z}_{\rm{f}})=\sum\nolimits_{k=1}^mp_k\mathbf{a}_{\rm{f}}(r_k)\mathbf{a}^{\rm{H}}_{\rm{f}}(r_k)$. Then, the matrix in the constraint of \eqref{atomicl0norm} can be expressed as
		\begin{align}\label{subproof1}
		\mathbf{C}&\triangleq
		\begin{bmatrix}
		{\bf{T({\bf{z}}_{\rm{p}}}}) & {\mathbf{X}_{\rm{v}}}    \\
		{\mathbf{X}_{\rm{v}}^{\rm H}} & {\bf{T({\bf{z}}_{\rm{f}}})}
		\end{bmatrix}
		\nonumber \\
		&=\begin{bmatrix}
		\sum\nolimits_{k=1}^mp_k\mathbf{a}_{\rm{p}}(\theta_k)\mathbf{a}^{\rm{H}}_{\rm{p}}(\theta_k) &   \sum\nolimits_{k=1}^mp_k\mathbf{a}_{\rm{p}}(\theta_k)\mathbf{a}^{\rm{H}}_{\rm{f}}(r_k)  \\
		\sum\nolimits_{k=1}^mp_k\mathbf{a}^{\rm{H}}_{\rm{p}}(\theta_k)\mathbf{a}_{\rm{f}}(r_k) & \sum\nolimits_{k=1}^mp_k\mathbf{a}_{\rm{f}}(r_k)\mathbf{a}^{\rm{H}}_{\rm{f}}(r_k)
		\end{bmatrix}
		\nonumber \\
		&=\sum\nolimits_{k=1}^{m}p_k
		\begin{bmatrix}
		\mathbf{a}_{\rm{p}}(\theta_k)   \\
		\mathbf{a}_{\rm{f}}(r_k)
		\end{bmatrix}
		\begin{bmatrix}
		\mathbf{a}_{\rm{p}}^{\rm{H}}(\theta_k)&\mathbf{a}_{\rm{f}}^{\rm{H}}(r_k)
		\end{bmatrix}
		\succeq 0.
		\end{align}
		This implies that $\mathbf{T}(\mathbf{z}_{\rm{p}})$ and $\mathbf{T}(\mathbf{z}_{\rm{f}})$ are two submatrices of positive semi-definite (PSD) matrix $\mathbf{C}$ which can be written as an $m$-fold factorization. As such, neither $r^{\rm{p}}_{\text{opt}}$ nor $r^{\rm{f}}_{\text{opt}}$ is greater than $m$. Thus, $(1/2)(r^{\rm{p}}_{\text{opt}}+r^{\rm{f}}_{\text{opt}})\le m=\|\mathbf{X}_{\rm{v}}\|_{\mathbb{A},0}$.
		
		On the other hand, suppose that the optimal solutions of \eqref{atomicl0norm} are $\mathbf{z}^{\rm{p}}_{\text{opt}}$ and $\mathbf{z}^{\rm{f}}_{\text{opt}}$. If $\mathbf{T}(\mathbf{z}^{\rm{p}}_{\text{opt}})$ and $\mathbf{T}(\mathbf{z}^{\rm{f}}_{\text{opt}})$ respectively follow the Vandermonde decomposition $\mathbf{T}(\mathbf{z}^{\rm{p}}_{\text{opt}})=\mathbf{D}_{\rm{p}}\mathbf{E}_{\rm{p}}\mathbf{D}^{\rm{H}}_{\rm{p}}$ and $\mathbf{T}(\mathbf{z}^{\rm{f}}_{\text{opt}})=\mathbf{D}_{\rm{f}}\mathbf{E}_{\rm{f}}\mathbf{D}^{\rm{H}}_{\rm{f}}$, the PSD-ness of matrix $\mathbf{C}$ implies that $\mathbf{X}_{\rm{v}}$ is in the column space of $\mathbf{D}_{\rm{p}}$ and in the row space of $\mathbf{D}_{\rm{f}}$. This in turn implies that $\mathbf{X}_{\rm{v}}$ can be expressed as a combination of at most $r^{\rm{p}}_{\text{opt}}$ or $r^{\rm{f}}_{\text{opt}}$ atoms, i.e., both $r^{\rm{p}}_{\text{opt}}$ and $r^{\rm{f}}_{\text{opt}}$ are greater than or equal to $m$. It can be further written as $1/2(r^{\rm{p}}_{\text{opt}}+r^{\rm{f}}_{\text{opt}})\ge m=\|\mathbf{X}_{\rm{v}}\|_{\mathbb{A},0}$. The proof is finished.
	\end{IEEEproof}
	\medskip

	\begin{algorithm}[!ht]
		{	
			\caption{ DoF-enhanced joint DoA-range estimation method based on DANM.}
			\label{alg1}
			\textbf{Input:} Received signal $\mathbf{x}(t)^{T}_{t=1}.$\\
			\textbf{Output:} DoAs $\theta_k$ and ranges $r_k,k=1,\cdots,K$.\\
			\textbf{Initialize:} Define $\mu$.
			\begin{algorithmic}[1]
				\item Derive the covariance matrix $\hat{\mathbf{R}}_{\mathbf{x}}$ using \eqref{covariancematrix};
				\item Obtain the equivalent coarray signal vector $\breve{\mathbf{x}}_{\rm{v}}$ using \eqref{vec};
				\item Initialize the interpolated coarray signal vector $\tilde{\mathbf{x}}_{\rm{v}}$ using \eqref{initialization};
				\item Reshape $\tilde{\mathbf{x}}_{\rm{v}}$ as coarray signal matrix $\tilde{\mathbf{X}}_{\rm{v}}={\rm{matrix}}(\tilde{\mathbf{x}}_{\rm{v}})$;
				\item Define binary matrix $\mathbf{B}$ to distinguish the elements in the interpolated coarray;
				\item Solve the SDP problem to obtain the full observation $\mathbf{X_{\rm{v}}}$ according to \eqref{DANM1};
				\item Perform SST and 2D MUSIC on  $\mathbf{X_{\rm{v}}}$ in order to estimate $\theta_{k}$ and $r_k$ according to Fig.\;\ref{SST_figure} and \eqref{2DMUSIC}.
		\end{algorithmic}}
	\end{algorithm}
	
	The nuclear norm is commonly adopted to solve the rank-optimization problem, which is essentially a variation of $l_1$-norm relaxation and, thus, also introduces approximation loss. Inspired by the rank function proposed in \cite{Daniel2019Low}, we adopt an exactly-equivalent reformulation of \eqref{rankminimization1} to solve the optimization problem without approximation. Let $\gamma>0$ be a positive constant, $\mathbf{W}\succeq 0$, and define function ${\rm{f}}[\mathbf{W},\mathbf{T}(\mathbf{z}),\gamma]$ as
	\begin{align}\label{rankformulation}
	{\text{f}}[\mathbf{W},\mathbf{T}(\mathbf{z}),\gamma]={\gamma}^{-2}(\|\mathbf{W}-\gamma\mathbf{I}\|_{\mathcal{F}}^2+2{\rm{tr}}[\mathbf{W}\mathbf{T}(\mathbf{z})]).
	\end{align}
	Then, the rank-minimization problem is equivalent to minimizing ${\rm{f}}[\mathbf{W},\mathbf{T}(\mathbf{z}),\gamma]$ under the constraints ${\rm{tr}}[\mathbf{W}\mathbf{T}(\mathbf{z})]\le0$ and $\mathbf{W}\succeq 0$. Hence, both rank functions in \eqref{rankminimization1} are reformulated to the corresponding multi-convex functions. Similar to the DANM, we use the same initialized signal as in \eqref{initialization} for reconstruction. Then, problem \eqref{rankminimization1} is reformulated as
	\begin{align} \label{multiconvex}
	\min \limits_{{\mathbf{z}}_{\rm{p}},{\mathbf{z}}_{\rm{f}},{\mathbf{W}}_{\rm{p}},{\mathbf{W}}_{\rm{f}},\mathbf{X}_{\rm{v}}} \qquad &\text{f}[{\mathbf{W}}_{\rm{p}},{\bf{T}}({\mathbf{z}}_{\rm{p}}),\gamma_{\rm{p}}]+\text{f}[{\mathbf{W}}_{\rm{f}},{\bf{T}}({\mathbf{z}}_{\rm{f}}),\gamma_{\rm{f}}] \nonumber \\
	\text{subject to} \quad &
	\|\mathbf{X}_{\rm{v}}\circ\mathbf{B}-\tilde{\mathbf{X}}_{\rm{v}}\|^2_{\mathcal{F}}\le\eta ,\nonumber\\
	& {\rm{tr}}[{\bf{W_{\rm{p}}T}}({\bf{z_{\rm{p}}}})]\leq 0, {\bf{W}_{\rm{p}}}\succeq 0,   \nonumber \\
	& {\rm{tr}}[{\bf{W_{\rm{f}}T}}({\bf{z}}_{\rm{f}})]\leq 0, {\bf{W}}_{\rm{f}}\succeq 0,   \nonumber \\
	&	
	\begin{bmatrix}
	{\bf{T({\mathbf{z}}_{\rm{p}}}}_{\rm}) & {\mathbf{X}_{\rm{v}}}    \\
	{\mathbf{X}}_{\rm{v}}^{\rm H} & {\bf{T({\bf{z}}_{\rm{f}}}})
	\end{bmatrix}
	\succeq 0.
	\end{align}
	Notice in \eqref{rankformulation} that the minimization of ${\rm{f}}[\mathbf{W},\mathbf{T}(\mathbf{z}),\gamma]$ involves the minimization of ${\rm{tr}}[\mathbf{W}\mathbf{T}(\mathbf{z})]$. The constraints with respect to the latter can be utilized as a stopping condition for the algorithm. In addition, as function $\text{f}[\mathbf{W},\mathbf{T}(\mathbf{z}),\gamma]$ is multi-convex, \eqref{multiconvex} is a multi-convex optimization problem. That is, if we fix $\mathbf{W}_{\rm{p}}$ and $\mathbf{W}_{\rm{f}}$, variables ${\mathbf{z}}_{\rm{p}}$, ${\mathbf{z}}_{\rm{f}}$, and $\mathbf{X}_{\rm{v}}$ can be conveniently optimized. Subsequently,  $\mathbf{W}_{\rm{p}}$ and $\mathbf{W}_{\rm{f}}$ can be optimized alternately by fixing the other.
	
	We further reformulate the first constraint in \eqref{multiconvex} as a regularization term of the optimization function controlled by $\mu$. As such, the proposed alternative optimization problems for CRM to minimize as per \eqref{multiconvex} are respectively given as
	\begin{align}\label{iter1}
	{\bf{z}}_{\rm{p}}^{(i)},{\bf{z}}_{\rm{f}}^{(i)},\mathbf{X}_{\rm{v}}^{(i)}\!&=\!\mathop{\arg\!\min_{{\bf{z}}_{\rm{p}},{\bf{z}}_{\rm{f}},\mathbf{X}_{\rm{v}}}}\!\text{f}[{\bf{W}}_{\rm{p}}^{(i-1)},{\bf{T}}({\bf{z}_{\rm{p}}}),\gamma_{\rm{p}}]\nonumber \\
	&+\text{f}[{\bf{W}}_{\rm{f}}^{(i-1)},{\bf{T}}({\bf{z}_{\rm{f}}}),\gamma_{\rm{f}}]+\mu \|\mathbf{X}_{\rm{v}}\circ\mathbf{B}-\tilde{\mathbf{X}}_{\rm{v}}\|^2_{\mathcal{F}} \nonumber \\
	&\text{subject to} \quad
	\begin{bmatrix}
	{\bf{T({\mathbf{z}}_{\rm{p}}}}_{\rm}) & {\mathbf{X}_{\rm{v}}}    \\
	{\mathbf{X}}_{\rm{v}}^{\rm H} & {\bf{T({\bf{z}}_{\rm{f}}}})
	\end{bmatrix}
	\succeq 0,
	\end{align}
	\begin{subequations}\label{iter2}
		\begin{align}\label{iter2a}
		{\bf{W}}_{\rm{p}}^{(i)}=&\mathop{\arg\min_{{\bf{W}}_{\rm{p}}}} \text{f}[{\bf{W}}_{\rm{p}},{\bf{T}}({{\bf{z}}^{(i)}_{\rm{p}}}),\gamma_{\rm{p}}]
		&\text{subject to} \quad
		\mathbf{W}_{\rm{p}}\succeq 0,
		\end{align}
		\begin{align}\label{iter2b}
		{\bf{W}}_{\rm{f}}^{(i)}=&\mathop{\arg\min_{{\bf{W}}_{\rm{f}}}} \text{f}[{\bf{W}}_{\rm{f}},{\bf{T}}({{\bf{z}}^{(i)}_{\rm{f}}}),\gamma_{\rm{f}}]
		&\text{subject to} \quad
		\mathbf{W}_{\rm{f}}\succeq 0,
		\end{align}
	\end{subequations}
	where $(\cdot)^{(i)}$ denotes the respective value of the variables in the $i$-th iteration.
	
	Regarding the computational complexity, the algorithm terminates when ${\rm{tr}}[\mathbf{W}^{(i)}_{\rm{p}}\mathbf{T}(\mathbf{z}^{(i)}_{\rm{p}})+\mathbf{W}^{\left(i\right)}_{\rm{f}}\mathbf{T}(\mathbf{z}^{(i)}_{\rm{f}})]$ converges or when the maximum number of iterations $N_{\text{max}}^{\text{iter}}$ is reached. Both \eqref{iter1} and \eqref{iter2} are convex SDP problems. In order to further lower the complexity, a closed-form solution\cite{Cai2010Singular} of \eqref{iter2} can be obtained by the eigen-decomposition of Hermitian and Toeplitz matrices $\mathbf{T}(\mathbf{z}_{\rm{p}})$ and $\mathbf{T}(\mathbf{z}_{\rm{f}})$. We first perform the following eigen-decomposition,
	\begin{align}
	\mathbf{T}(\mathbf{z})=\mathbf{U}\mathbf{\Sigma}\mathbf{U}^{\rm{H}},\qquad\mathbf{\Sigma}={\rm{diag}}[\left\{\lambda_{l}[\mathbf{T(\mathbf{z})}] \right\}_{l=1}^{2L+1}],
	\end{align}
	where $\lambda_{1}[\mathbf{T(\mathbf{z})}]\ge\lambda_{2}[\mathbf{T(\mathbf{z})}]\ge\cdots\ge\lambda_{2L+1}[\mathbf{T(\mathbf{z})}]\in\mathbb{R}$ are the eigenvalues of $\mathbf{T}(\mathbf{z})$. We then define an operator as $\Omega_{\gamma,0}[\mathbf{T}(\mathbf{z})]\triangleq\mathbf{U}\mathbf{\Gamma}_{\mathbf{\Sigma}}\mathbf{U}^{\rm{H}}$, where
	\begin{align}
	\mathbf{\Gamma}_{{\mathbf\Sigma}}\!=\!{\rm{diag}}\left[\{{\rm{max}} [\gamma-\lambda_{2L-l}[{\mathbf{T}(\mathbf{z})}],0]\}^{2L+1}_{l=1}\right].
	\end{align}
	As such, the closed-from solution of \eqref{iter2} can be derived as
	\begin{equation}\label{closed-form}
		{\bf{W}}_{\rm{p}}^{(i)}=\Omega_{\gamma,0}\left[\mathbf{T}\left(\mathbf{z}^{\left(i\right)}_{\rm{p}}\right)\right],
\
		{\bf{W}}_{\rm{f}}^{(i)}=\Omega_{\gamma,0}\left[\mathbf{T}\left(\mathbf{z}^{\left(i\right)}_{\rm{f}}\right)\right].
	\end{equation}
	As for \eqref{iter1}, there is no closed-form solution but it can be easily solved using available SDP solvers. Overall, the minimization problem \eqref{multiconvex} can be efficiently solved within a few iterations. After the coarray signal $\mathbf{X}_{\rm{v}}$ is reconstructed, we can jointly estimate the DoA and range of the targets based on SST and 2D MUSIC of $\mathbf{X}_{\rm{v}}$. The proposed method is summarized in {\bf Algorithm \ref{alg2}}.
	
	\begin{algorithm}[t]
		{	
			\caption{DoF-enhanced joint DoA-range estimation method based on cyclic rank-minimization.}
			\label{alg2}
			\textbf{Input:} Received signal $\left\{{\bf x}(t)\right\}^{T}_{t=1}.$\\
			\textbf{Output:}  DoAs $\theta_k$ and ranges $r_k,k=1,\cdots,K$. \\
			\textbf{Initialize:} ${\mathbf{W}}_{\rm{p}}$, ${\mathbf{W}}_{\rm{f}}$ $\leftarrow$ Two random Hermitian matrices, and define $\gamma_{\rm{p}}$, $\gamma_{\rm{f}}$ $\mu$, $\epsilon$, and $N^{\text{iter}}_{\rm{max}}$.
			\begin{algorithmic}[1]
				\item Derive the covariance matrix $\hat{{\bf R}}_{\mathbf{x}}$ using \eqref{covariancematrix};
				\item Obtain the equivalent coarray signal vector $\dot{\mathbf{x}}_{\rm{v}}$ using \eqref{vec};
				\item Initialize the interpolated coarray signal vector $\bar{\mathbf{x}}_{\rm{v}}$ using \eqref{initialization};
				\item Reshape $\tilde{\mathbf{x}}_{\rm{v}}$ as coarray signal matrix  $\tilde{\mathbf{X}}_{\rm{v}}={\rm{matrix}}(\tilde{\mathbf{x}}_{\rm{v}})$;
				\item Use a binary matrix $\mathbf{B}$ to distinguish the elements in the interpolated coarray;
				\For{$i=1$ to $N^{\text{iter}}_{\rm{max}}$}
				\State Optimize ${\bf{T}}({\bf{z}}^{(i)}_{\rm{p}})$ and ${\bf{T}}({\bf{z}}^{(i)}_{\rm{f}})$ using (\ref{iter1});
				\If{	\vspace{-0.5em}
					\begin{align}
					&\left|{\rm{tr}}\left[\mathbf{W}^{\left(i\right)}_{\rm{p}}\mathbf{T}\left(\mathbf{z}^{\left(i\right)}_{\rm{p}}\right)+\mathbf{W}^{\left(i\right)}_{\rm{f}}\mathbf{T}\left(\mathbf{z}^{\left(i\right)}_{\rm{f}}\right)\right] \right.\nonumber \\
					&\left. -{\rm{tr}}\left[\mathbf{W}^{\left(i\right)}_{\rm{p}}\mathbf{T}\left(\mathbf{z}^{\left(i-1\right)}_{\rm{p}}\right)+\mathbf{W}^{\left(i-1\right)}_{\rm{f}}\mathbf{T}\left(\mathbf{z}^{\left(i-1\right)}_{\rm{f}}\right)\right]\right| > \epsilon \nonumber
					\end{align}}\\
				\vspace{-0.5em}
				\qquad\qquad Optimize ${\bf W}^{(i)}_{\rm{p}}$ and ${\bf W}^{(i)}_{\rm{f}}$ using \eqref{closed-form};
				\Else
				\item \qquad \qquad Break;
				\EndIf
				\EndFor
				\item Perform SST and 2D MUSIC on  $\mathbf{X_{\rm{v}}}$ in order to estimate $\theta_{k}$ and $r_k$ according to Fig.\;\ref{SST_figure} and \eqref{2DMUSIC}.
		\end{algorithmic}}
	\end{algorithm}	
	
	$\textit{Remark 5}$: We argue that the CRM performs better than the DANM due to the following reasons:
	
	\begin{itemize}
		\item  First, optimization problem \eqref{multiconvex} introduces two new matrices $\mathbf{W}_{\rm{p}}$ and $\mathbf{W}_{\rm{f}}$. As both $\mathbf{T}\left(\mathbf{z}\right)$ and $\mathbf{W}$ are PSD, ${\rm{tr}}\left[\mathbf{W}\mathbf{T}\left(\mathbf{z}\right)\right]\le 0$ implies that $\mathbf{W}$ must be in the noise subspace of $\mathbf{T}\left(\mathbf{z}\right)$, i.e., ${\rm{tr}}\left[\mathbf{W}\mathbf{T}\left(\mathbf{z}\right)\right] = 0$.
		Thus, each problem in \eqref{iter1} and \eqref{iter2} cyclically estimates the signal and noise subspaces until they become orthogonal.
		
		\item To account for the effect of weight coefficients in the performance, we denote $\mathbf{N}_{\rm{p}}$ and $\mathbf{N}_{\rm{f}}$ as the basis vectors of the noise subspaces for $\mathbf{T}(\mathbf{z}_{\rm{p}})$ and $\mathbf{T}(\mathbf{z}_{\rm{f}})$, respectively. We additionally set $\mathbf{W}_{\rm{p}}=\gamma_{\rm{p}}\mathbf{N}_{\rm{p}}\mathbf{N}_{\rm{p}}^{\rm{H}}$ and $\mathbf{W}_{\rm{f}}=\gamma_{\rm{f}}\mathbf{N}_{\rm{f}}\mathbf{N}_{\rm{f}}^{\rm{H}}$  in problem \eqref{rankformulation}. Then, the optimization terms ${\rm{tr}}[\mathbf{W}_{\rm{p}}\mathbf{T}\left(\mathbf{z}_{\rm{p}}\right)]$ and ${\rm{tr}}[\mathbf{W}_{\rm{f}}\mathbf{T}\left(\mathbf{z}_{\rm{f}}\right)]$ can be rewritten as ${\rm{tr}}[\gamma_{\rm{p}}\mathbf{N}_{\rm{p}}\mathbf{N}_{\rm{p}}^{\rm{H}}\mathbf{T}(\mathbf{z}_{\rm{p}})]$ and ${\rm{tr}}[\gamma_{\rm{f}}\mathbf{N}_{\rm{f}}\mathbf{N}_{\rm{f}}^{\rm{H}}\mathbf{T}(\mathbf{z}_{\rm{f}})]$, respectively. Thus, the impact of the weight coefficients on the optimization terms is revealed, which leads to performance trade-off between the DoA and range estimations. We will show this impact in the numerical simulations.

		\item In addition, the DANM employs convex-relaxation to approximate the original atomic $l_0$-norm. From the viewpoint of low-rank matrix reconstruction, the DANM minimizes the sum of the eigenvalues of $\mathbf{T}\left(\mathbf{z}_{\rm{p}}\right)$ and $\mathbf{T}\left(\mathbf{z}_{\rm{f}}\right)$. In the noisy case, the errors of eigenvalues increase. This in turn leads to perceptible approximation losses of the DANM. Different from the DANM, the CRM instead minimizes the number of nonzero eigenvalues of  $\mathbf{T}\left(\mathbf{z}_{\rm{p}}\right)$ and $\mathbf{T}\left(\mathbf{z}_{\rm{f}}\right)$. On this basis, the equivalent rank-minimization problem is reformulated to a multi-convex form and is solved using cyclic iterations. Therefore, the coarray signal $\mathbf{X}_{\rm{v}}$ can be reconstructed more accurately.
	\end{itemize}
	\vspace{-1.25em}

	\subsection{Low Complexity Solution Using ADMM}
	Compared to the DANM which solves the SDP problem \eqref{DANM1} for only once, the SDP problem \eqref{iter1} involves much higher computational complexity because of the iterations involved in the CRM, even though partial closed-form solutions are derived in \eqref{closed-form} to reduce the computational load. To further improve the efficiency, we derive the following ADMM-based solver:
	\begin{align}\label{ADMM1}
	&{\bf{z}}_{\rm{p}}^{\left(i\right)},{\bf{z}}_{\rm{f}}^{(i)},\mathbf{X}_{\rm{v}}^{\left(i\right)}\!=\nonumber \\
	&\mathop{\arg\!\min_{{\bf{z}}_{\rm{p}},{\bf{z}}_{\rm{f}},\mathbf{X}_{\rm{v}}}}\!\text{f}[{\bf{W}}_{\rm{p}}^{(i-1)},{\bf{T}}({\mathbf{z}_{\rm{p}}}),\gamma_{\rm{p}}]
	+\text{f}[{\bf{W}}_{\rm{f}}^{(i-1)},{\bf{T}}({\bf{z}_{\rm{f}}}),\gamma_{\rm{f}}] \nonumber \\
	&\text{subject to} \quad
	\overline{\mathbf{C}}=\begin{bmatrix}
	{\mu\mathbf{I}} & {\mathbf{X}_{\rm{v}}\circ\mathbf{B}-\tilde{\mathbf{X}}_{\rm{v}}}    \\
	{\mathbf{X}^{\rm{H}}_{\rm{v}}\circ\mathbf{B}-\tilde{\mathbf{X}}^{\rm{H}}_{\rm{v}}} & {\mu\mathbf{I}}
	\end{bmatrix},\nonumber \\
	&\qquad\qquad\quad
	\underline{\mathbf{C}}=\begin{bmatrix}
	{\bf{T({\mathbf{z}}_{\rm{p}}}}_{\rm}) & {\mathbf{X}_{\rm{v}}}    \\
	{\mathbf{X}}_{\rm{v}}^{\rm H} & {\bf{T({\bf{z}}_{\rm{f}}}})
	\end{bmatrix}, \overline{\mathbf{C}}\succeq 0, \underline{\mathbf{C}}\succeq 0.
	\end{align}
	Next, the equality constraints in \eqref{ADMM1} are combined in the augmented Lagrangian as
	\begin{align}\label{Largrangian}
	&\mathcal{L}\left[\mathbf{z}_{\rm{p}},\mathbf{z}_{\rm{f}},\mathbf{X}_{\rm{v}},\overline{\mathbf{C}},\underline{\mathbf{C}},\overline{\mathbf{R}},\underline{\mathbf{R}}\right]\nonumber \\
	&=\text{f}[{\bf{W}}_{\rm{p}}^{(i-1)},{\bf{T}}({\bf{z}_{\rm{p}}}),\gamma_{\rm{p}}]+\text{f}[{\bf{W}}_{\rm{f}}^{(i-1)},{\bf{T}}({\bf{z}_{\rm{f}}}),\gamma_{\rm{f}}]\nonumber \\
	& \!+\!\frac{\rho}{2}\left\|\overline{\mathbf{C}}-\begin{bmatrix}
	{\mu\mathbf{I}} & {\mathbf{X}_{\rm{v}}\circ\mathbf{B}-\tilde{\mathbf{X}}_{\rm{v}}}    \\
	{\mathbf{X}^{\rm{H}}_{\rm{v}}\circ\mathbf{B}-\tilde{\mathbf{X}}^{\rm{H}}_{\rm{v}}} & {\mu\mathbf{I}}
	\end{bmatrix}\!+\!\rho^{-1}\overline{\mathbf{R}}\right\|_{\mathcal{F}}^2\nonumber \\
	&\!+\!\frac{\rho}{2}\left\|\underline{\mathbf{C}}-\begin{bmatrix}
	{\bf{T({\mathbf{z}}_{\rm{p}}}}_{\rm}) & {\mathbf{X}_{\rm{v}}}    \\
	{\mathbf{X}}_{\rm{v}}^{\rm H} & {\bf{T({\bf{z}}_{\rm{f}}}})
	\end{bmatrix}\!+\!\rho^{-1}\underline{\mathbf{R}}\right\|_{\mathcal{F}}^2,
	\end{align}
	where $\rho >0$ is the augmented Lagrangian parameter, whereas $\overline{\mathbf{R}}$ and $\underline{\mathbf{R}}$ are dual variables. As such, the ADMM for \eqref{iter1} consists of the following optimization problems, each with a single variable:
	\begin{align}
	&\mathbf{z}_{\rm{p}}^{\left(j\right)}=\mathop{\arg\min_{{\bf{z}}_{\rm{p}}}}\mathcal{L}\left[\mathbf{z}_{\rm{p}},\left(\mathbf{z}_{\rm{f}},\mathbf{X}_{\rm{v}},\overline{\mathbf{C}},\underline{\mathbf{C}},\overline{\mathbf{R}},\underline{\mathbf{R}}\right)^{\left(j-1\right)}\right],
	\end{align}
	\begin{align}
	\mathbf{z}_{\rm{f}}^{\left(j\right)}=\mathop{\arg\min_{{\bf{z}}_{\rm{f}}}}\mathcal{L}\left[\mathbf{z}_{\rm{p}}^{\left(j\right)},\mathbf{z}_{\rm{f}},\left(\mathbf{X}_{\rm{v}},\overline{\mathbf{C}},\underline{\mathbf{C}},\overline{\mathbf{R}},\underline{\mathbf{R}}\right)^{\left(j-1\right)}\right],
	\end{align}
	\begin{align}
	\mathbf{X}_{\rm{v}}^{\left(j\right)}=\mathop{\arg\min_{{\bf{X}}_{\rm{v}}}}\mathcal{L}\left[\left(\mathbf{z}_{\rm{p}},\mathbf{z}_{\rm{f}}\right)^{\left(j\right)}\!,\!\mathbf{X}_{\rm{v}}\!,\!\left(\overline{\mathbf{C}},\underline{\mathbf{C}},\overline{\mathbf{R}},\underline{\mathbf{R}}\right)^{\left(j-1\right)}\right],
	\end{align}
	\begin{align}
	\overline{\mathbf{C}}^{\left(j\right)}=\mathop{\arg\min_{\overline{\mathbf{C}}\succeq 0}}\mathcal{L}\left[\left(\mathbf{z}_{\rm{p}},\mathbf{z}_{\rm{f}},\mathbf{X}_{\rm{v}}\right)^{\left(j\right)}\!,\!\overline{\mathbf{C}}\!,\!\left(\underline{\mathbf{C}},\overline{\mathbf{R}},\underline{\mathbf{R}}\right)^{\left(j-1\right)}\right],
	\end{align}
	\begin{align}
	\underline{\mathbf{C}}^{\left(j\right)}=\mathop{\arg\min_{\overline{\mathbf{C}}\succeq 0}}\mathcal{L}\left[\left(\mathbf{z}_{\rm{p}},\mathbf{z}_{\rm{f}},\mathbf{X}_{\rm{v}}\overline{\mathbf{C}}\right)^{\left(j\right)}\!,\!\underline{\mathbf{C}},\left(\overline{\mathbf{R}},\underline{\mathbf{R}}\right)^{\left(j-1\right)}\right],
	\end{align}
	\begin{align}
	&\overline{\mathbf{R}}^{\left(j\right)}=\overline{\mathbf{R}}^{\left(j-1\right)}\nonumber \\
	&+\rho\left(\overline{\mathbf{C}}^{\left(j\right)}\!-\!\begin{bmatrix}
	{\mu\mathbf{I}} & {\mathbf{X}^{\left(j\right)}_{\rm{v}}\circ\mathbf{B}\!-\!\tilde{\mathbf{X}}_{\rm{v}}}    \\
	\left({{\mathbf{X}}_{\rm{v}}^{\left(j\right)}}\right)^{\rm H}\circ\mathbf{B}\!-\!\tilde{\mathbf{X}}^{\rm{H}}_{\rm{v}} & {\mu\mathbf{I}}
	\end{bmatrix}\right),
	\end{align}
	\begin{align}
	\underline{\mathbf{R}}^{\left(j\right)}=\underline{\mathbf{R}}^{\left(j-1\right)}+\rho\left(\underline{\mathbf{C}}^{\left(j\right)}-\begin{bmatrix}
	{\mathbf{T}({\mathbf{z}}^{\left(j\right)}_{\rm{p}}}_{\rm}) & {\mathbf{X}^{\left(j\right)}_{\rm{v}}}    \\
	\left({{\mathbf{X}}_{\rm{v}}^{\left(j\right)}}\right)^{\rm H} & {\mathbf{T}({\bf{z}}^{\left(j\right)}_{\rm{f}}})
	\end{bmatrix}\right),
	\end{align}
	where  $(\cdot)^{\left(j\right)}$ denotes the respective value of the variables in the $j$-th iteration of the ADMM. Each optimization problem of the ADMM can be computed in a closed form iteratively until convergence. As a result, the optimization result of the SDP problem \eqref{iter1} can be obtained by iterations of several closed-form solutions. Due to the limit of space, the detailed derivations of the closed-form solutions are given in the Supplement File.

\vspace{-0.25em}	
	\section{Performance analysis of Space-Frequency Coarray}\label{sec:Performance}
In this section, we provide the theoretical analyses in terms of the identifiability, the convergence of the CRM method, and the reconstruction performance, to demonstrate the effectiveness of the proposed joint estimation framework with an increased number of DoFs.
\vspace{-0.25em}	
	\subsection{Identifiability}
		
		As introduced in {\bf Algorithm \ref{alg1}} and {\bf Algorithm \ref{alg2}}, the subspace-based 2D MUSIC algorithm is applied to the smoothed coarray signal $\mathbf{R}_{\rm{ss}}$ to identify the sources. We first assume that the interpolated virtual coarray perfectly fits the ideal fully observed one. In this case, the only difference between the virtual coarray and the physical array is that the coarray sensors are virtually generated using fewer number of physical sensors \cite{Nested2012Piya}. Thus, we explore the identifiability conditions for the space-frequency coarray as same as that of traditional subspace-based methods applied to physical arrays.
		
		Assume that $K$ sources are randomly distributed at $K$ DoA-range pairs $\mathbb{T}=\{(\theta_k,r_k),1\le k \le K\}$  with $K \le L^2+2L$ as described in \textit{Remark} 2 of Section \ref{sec:DANM-analysis}. The identifiability of the space-frequency coarray is provided by the following theorem.
		
		\textit{Theorem 1}: Let the $K$ DoA-range pairs $\mathbb{T}=\{(\theta_k,r_k),1\le k \le K\}$ satisfy both of the following conditions:
		\begin{itemize}
			\item C1: Assume that $\mathbb{T}_1=\{(\theta,r_k),1\le k \le K_1\}$  and $\mathbb{T}_2=\{(\theta_k,r),1\le k \le K_2\}$ are two types of subsets of $\mathbb{T}$
			such that sources in $\mathbb{T}_1$ share the identical DoA value but have different ranges, whereas those in $\mathbb{T}_2$ have the identical range but have distinct DoAs. The size of the subsets satisfies $K_1\le L$ and $K_2\le L$.
			\item C2: Assume that $\mathbb{T}_3=\{(\theta_k,r_k)|{\rm{e}}^{-\jmath\pi\sin \theta_k}={\rm{e}}^{{\rho\jmath4\pi\Delta f r_k}/{\rm{c}}},1\le k \le K_3\}$ is a subset of  $\mathbb{T}$. The size of the subset satisfies $K_3\le\rho(L-1)+L$.
		\end{itemize}
		Then, the 2D MUSIC spectrum will exhibit a peak if and only if the steering vector $\mathbf{a}_{\rm{p,f}}$ corresponds to one of the $K$ sources in $\mathbb{T}=\{(\theta_k,r_k),1\le k \le K\}$ with probability one almost surely (a.s.).
		
		\begin{IEEEproof}
			The following lemma is utilized to provide a statistical bound on the rank of the 2D array manifold.	
			
			\textit{Lemma 1}\cite{Almost2002Liu}: For a pair of Vandermonde matrices $\mathbf{A}\in\mathbb{C}^{K\times F}$ and $\mathbf{B}\in\mathbb{C}^{L\times F}$, with generators on the unit circle,
			\begin{align}
			{\rm{rank}}(\mathbf{A}\odot\mathbf{B})={\rm{min}}(KL, F), \quad  P_{\mathcal{L}}(\mathbb{U}^{2F}) \text{-a.s.},
			\end{align}
			where $\mathbb{U}$ is the unit circle, and $P_{\mathcal{L}}(\mathbb{U}^{2F})$ is the distribution used to draw the $2F$ generators for $\mathbf{A}$ and $\mathbf{B}$, which is assumed continuous with respect to the Lebesgue measure in $\mathbb{U}^{2F}$.
			
			This lemma with respect to the Khatri-Rao product ensures that the array manifold of the space-frequency coarray $\mathbf{A}_{\rm{p,f}}=\mathbf{A}_{\rm{p}}\odot\mathbf{A}_{\rm{f}}\in\mathbb{C}^{{L^2}\times K}$ is full rank with probability one. However, when the distribution of the sources do not follow the condition C1 or C2, the array manifold will be rank-deficient. In this case, false peaks may appear in the spectrum.
			
			\textit{Reason}: For the first case, the $K_1$ sources in $\mathbb{T}_1$ give rise to the $K_1$ columns of the array manifold in the form
			\begin{align}
			&\mathbf{A}_{\rm{p,f}}=\nonumber \\
			&\begin{bmatrix}
			1\\{\rm{e}}^{-\jmath\pi\sin \theta}\\\vdots\\{\rm{e}}^{-\jmath(L-1)\pi\sin \theta}
			\end{bmatrix}\otimes
			\underbrace{\begin{bmatrix}
				1&\cdots&1\\
				{\rm{e}}^{\frac{\jmath 4\pi\Delta f r_1}{\rm{c}}}&\cdots&{\rm{e}}^{\frac{\jmath 4\pi\Delta f r_{K_1}}{\rm{c}}} \\
				\vdots& &\vdots\\
				{\rm{e}}^{\frac{\jmath 4\pi(L-1)\Delta f r_1}{\rm{c}}}&\cdots&{\rm{e}}^{\frac{\jmath 4\pi(L-1)\Delta f r_{K_1}}{\rm{c}}}
				\end{bmatrix}}_{\mathbf{A}_{\rm{f}}\in \mathbb{C}^{L\times K_1}}.
			\end{align}
			
			If $K_1>L$, $\mathbf{A}_{\rm{f}}$ is a Vandermonde matrix of rank $L$, which indicates that these $K_1$ columns are linearly dependent. Thus, the array manifold $\mathbf{A}_{\rm{p,f}}$ is rank-deficient in this case. Likewise, the $K_2$ sources in $\mathbb{T}_2$ leads to a similar result. For the second case, the rank of array manifold $\mathbf{A}_{\rm{p,f}}$ corresponding to $\mathbb{T}_3$ is $\min({\rho(L-1)+L,K_3})$\cite{Nested2012Piya} and has the same rows when $K_3>\rho(L-1)+L$, which leads to the linear-dependence between steering vectors.
			
			On the basis of \textit{Lemma 1}, we assume that the $K$ sources in $\mathbb{T}$ follow both C1 and C2. Then, the array manifold $\mathbf{A}_{\rm{p,f}}$ is full column rank a.s. As such, the null-space $\mathbf{U}_{\rm{N}}$ is spanned by $(L^2-K)$-dimensional orthogonal components of the range space of $\mathbf{A}_{\rm{p,f}}$.  For each source in $\mathbb{T}$, we have $\mathbf{a}^{\rm{H}}_{\rm{p,f}}(\theta_{k},r_k)\mathbf{U}_{\rm{N}}=\boldsymbol{0}^{\rm{T}}$, which leads to a peak in the 2D MUSIC spectrum. Then, we consider a DoA-range pair $(\theta_{K+1},r_{K+1}) \notin \mathbb{T}$ and assume that $\mathbf{a}^{\rm{H}}_{\rm{p,f}}(\theta_{K+1},r_{K+1})\mathbf{U}_{\rm{N}}=\boldsymbol{0}^{\rm{T}}$. The orthogonality between the steering vectors and the null-space can be expressed as
			\begin{align}\label{Orthonogality}
			\begin{bmatrix}
			\mathbf{a}^{\rm{H}}_{\rm{p,f}}(\theta_{1},r_1)\\\mathbf{a}^{\rm{H}}_{\rm{p,f}}(\theta_{2},r_2)\\\vdots\\\mathbf{a}^{\rm{H}}_{\rm{p,f}}(\theta_{K+1},r_{K+1})
			\end{bmatrix}
			\mathbf{U}_{\rm{N}}=
			\begin{bmatrix}
			\boldsymbol{0}^{\rm{T}}\\\boldsymbol{0}^{\rm{T}}\\\vdots\\\boldsymbol{0}^{\rm{T}}
			\end{bmatrix}.
			\end{align}
		We first assume that $\mathbf{a}^{\rm{H}}_{\rm{p,f}}(\theta_{K+1},r_{K+1})$ is linearly independent of the other steering vectors. Combining this assumption with \eqref{Orthonogality} yields to an $(L^2-K-1)$ dimensional null-space, which is contradict with $\mathbf{U}_{\rm{N}}\in\mathbb{C}^{L^2\times (L^2-K)}$. On the other hand, if $\mathbf{a}^{\rm{H}}_{\rm{p,f}}(\theta_{K+1},r_{K+1})$ is linearly dependent on the other steering vectors, neither the condition C1 nor C2 is met. Hence, $\mathbf{a}^{\rm{H}}_{\rm{p,f}}(\theta_{K+1},r_K+1)\mathbf{U}_{\rm{N}}\ne\boldsymbol{0}^{\rm{T}}$ holds for any DoA-range pair falling outside $\mathbb{T}$.
		\end{IEEEproof}
	
		For the interpolation framework, even though the proposed methods uniquely map partial observed signal $\tilde{\mathbf{X}}_{\rm{v}}$ to the hole-free signal $\mathbf{X}_{\rm{v}}$, the interpolated coarray inevitably contains fitting errors. Essentially, the detection is still governed by the  non-negative unique lags. Thus, the number of sources $K$ in \textit{Theorem} 1 should not exceed $D^2+2D$ to guarantee the identifiability for the interpolated coarray. However, the proposed methods have the ability to identify more than $D^2+2D$ sources using the maximum $L^2+2L$ DoFs in some well-reconstructed cases as shown in the simulations. Due to the limit of the space, we provide the simulation results that compare the identifiability between the interpolated coarray, coarray without interpolation, and uniform linear coarray in Section B of the Supplement File. The analysis of the reconstruction performance is provided in the sequel of this section.  This completes the explanation on the identifiability issue in joint DoA-range estimation using a space-frequency virtual difference coarray.

	\subsection{Convergence of CRM}
				
				To prove the convergence of the proposed CRM method, similar to \cite{Daniel2019Low}, we first consider a perturbed version of subproblem \eqref{iter1} and reformulate the regularization term in the constraint of \eqref{multiconvex}. Concretely, we respectively added small perturbation terms $\nu_{\rm{p}}\mathbf{I}$ and $\nu_{\rm{f}}\mathbf{I}$ to matrices $\mathbf{W}^{(i-1)}_{\rm{p}}$ and $\mathbf{W}^{(i-1)}_{\rm{f}}$. Then, the perturbed subproblem is given as
				\begin{align}\label{perturbed}
				{\bf{z}}_{\rm{p}}^{(i)}&,{\bf{z}}_{\rm{f}}^{(i)},\mathbf{X}_{\rm{v}}^{(i)}\!=\!\mathop{\arg\!\min_{{\bf{z}}_{\rm{p}},{\bf{z}}_{\rm{f}},\mathbf{X}_{\rm{v}}}}\!\text{f}[{\bf{W}}_{\rm{p}}^{(i-1)}+\nu_{\rm{p}}\mathbf{I},{\bf{T}}({\bf{z}_{\rm{p}}}),\gamma_{\rm{p}}]\nonumber \\
				&\qquad\qquad+\text{f}[{\bf{W}}_{\rm{f}}^{(i-1)}+\nu_{\rm{f}}\mathbf{I},{\bf{T}}({\bf{z}_{\rm{f}}}),\gamma_{\rm{f}}]\nonumber \\
				&\text{subject to} \quad
				\{\mathbf{X}_{\rm{v}},\mathbf{z}_{\rm{p}},\mathbf{z}_{\rm{f}}\}\in\mathbb{Z},
				\end{align}
				where $\mathbb{Z}\!=\!\{\{\mathbf{X}_{\rm{v}},\mathbf{z}_{\rm{p}},\mathbf{z}_{\rm{f}}\}|\|\mathbf{X}_{\rm{v}}\circ\mathbf{B}-\tilde{\mathbf{X}}_{\rm{v}}\|^2_{\mathcal{F}}\!\le\!\eta,[\mathbf{T}(\mathbf{z}_{\rm{p}})\;\mathbf{X}_{\rm{v}};\mathbf{X}^{\rm{H}}_{\rm{v}}\;\mathbf{T}(\mathbf{z}_{\rm{p}})]\succeq0\}$.
				Then, we derive the following lemma to demonstrate that $\mathbf{T}(\mathbf{z}^{(i)}_{\rm{p}})$ and $\mathbf{T}(\mathbf{z}^{(i)}_{\rm{f}})$ are bounded during the iterations of CRM, which is a condition for the proof of convergence.
				
				$\textit{Lemma 2}$: Let $\{\mathbf{X}^{\rm{D}}_{\rm{v}},\mathbf{z}^{\rm{D}}_{\rm{p}},\mathbf{z}^{\rm{D}}_{\rm{f}}\}$ be the optimal solution of the DANM problem \eqref{DANM1} and $\{\dot{\mathbf{X}}_{\rm{v}},\dot{\mathbf{z}}_{\rm{p}},\dot{\mathbf{z}}_{\rm{f}}\}$ be the optimal solution of the perturbed subproblem \eqref{perturbed}. Then, we have
				\begin{align}\label{lemma2}
				\nu_{\rm{p}}\rm{tr}[\mathbf{T}(\dot{\mathbf{z}}_{\rm{p}})]+\nu_{\rm{f}}\rm{tr}[\mathbf{T}(\dot{\mathbf{z}}_{\rm{f}})]\le(\nu_{\rm{p}}+\gamma_{\rm{p}})\rm{tr}[\mathbf{T}(\mathbf{z}^{\rm{D}}_{\rm{p}})]\\ \nonumber+(\nu_{\rm{f}}+\gamma_{\rm{f}})\rm{tr}[\mathbf{T}(\mathbf{z}^{\rm{D}}_{\rm{f}})].
				\end{align}
				\begin{IEEEproof}
					Notice that, according to \eqref{rankformulation}, the perturbed subproblem \eqref{perturbed} is equivalent to the following problem
					\begin{align}\label{equivalentperturbed}
					&{\bf{z}}_{\rm{p}}^{(i)},{\bf{z}}_{\rm{f}}^{(i)},\mathbf{X}_{\rm{v}}^{(i)}\!=\!\mathop{\arg\!\min_{{\bf{z}}_{\rm{p}},{\bf{z}}_{\rm{f}},\mathbf{X}_{\rm{v}}}}\! {\rm{tr}}[(\mathbf{W}^{(i-1)}_{\rm{p}}+\nu_{\rm{p}}\mathbf{I})\mathbf{T}({\mathbf{z}_{\rm{p}}})]\nonumber \\
					&\qquad\qquad\qquad\qquad\qquad\qquad+{\rm{tr}}[(\mathbf{W}^{(i-1)}_{\rm{f}}+\nu_{\rm{f}}\mathbf{I})\mathbf{T}({\mathbf{z}_{\rm{f}}})] \nonumber \\
					&\qquad\qquad\text{subject to}\quad 	\{\mathbf{X}_{\rm{v}},\mathbf{z}_{\rm{p}},\mathbf{z}_{\rm{f}}\}\in\mathbb{Z}.
					\end{align}
					Let $w$ denote the optimal value of the objective function in \eqref{equivalentperturbed}. Since all matrices in the problem are PSD, we have $w={\rm{tr}}[\mathbf{W}^{(i-1)}_{\rm{p}}\mathbf{T}({\dot{\mathbf{z}}_{\rm{p}}})]+\nu_{\rm{p}}{\rm{tr}}\mathbf{T}({\dot{\mathbf{z}}_{\rm{p}}})+{\rm{tr}}[\mathbf{W}^{(i-1)}_{\rm{f}}\mathbf{T}({\dot{\mathbf{z}}_{\rm{f}}})]+\nu_{\rm{f}}{\rm{tr}}\mathbf{T}({\dot{\mathbf{z}}_{\rm{f}}})\ge\nu_{\rm{p}}{\rm{tr}}\mathbf{T}({\dot{\mathbf{z}}_{\rm{p}}})+\nu_{\rm{f}}{\rm{tr}}\mathbf{T}({\dot{\mathbf{z}}_{\rm{f}}})$. From the closed-form solutions in \eqref{closed-form}, we readily observe that $\gamma_{\rm{p}}\mathbf{I}\succeq\mathbf{W}_{\rm{p}}$ and $\gamma_{\rm{f}}\mathbf{I}\succeq\mathbf{W}_{\rm{f}}$, which means $w\le{\rm{tr}}[(\mathbf{W}^{(i-1)}_{\rm{p}}+\nu_{\rm{p}}\mathbf{I})\mathbf{T}({\mathbf{z}_{\rm{p}}})]+{\rm{tr}}[(\mathbf{W}^{(i-1)}_{\rm{f}}+\nu_{\rm{f}}\mathbf{I})\mathbf{T}({\mathbf{z}_{\rm{f}}})]\le(\nu_{\rm{p}}+\gamma_{\rm{p}})\rm{tr}[\mathbf{T}(\mathbf{z}_{\rm{p}})]+(\nu_{\rm{f}}+\gamma_{\rm{f}})\rm{tr}[\mathbf{T}(\mathbf{z}_{\rm{f}})].$ Letting the matrices equal to the optimal solution of the DANM in \eqref{DANM1}, i.e., $\{\mathbf{X}^{\rm{D}}_{\rm{v}},\mathbf{z}^{\rm{D}}_{\rm{p}},\mathbf{z}^{\rm{D}}_{\rm{f}}\}$, we have $w\le(\nu_{\rm{p}}+\gamma_{\rm{p}})\rm{tr}[\mathbf{T}(\mathbf{z}^{\rm{D}}_{\rm{p}})]+(\nu_{\rm{f}}+\gamma_{\rm{f}})\rm{tr}[\mathbf{T}(\mathbf{z}^{\rm{D}}_{\rm{f}})].$ These results collectively lead to \eqref{lemma2}.
				\end{IEEEproof}
				
				Due to the CRM procedure with the perturbed subproblem (called perturbed CRM) and PSD-ness of the matrices, we have the following inequalities
				\begin{align}
				0&\le\text{f}[{\bf{W}}_{\rm{p}}^{(i)}+\nu_{\rm{p}}\mathbf{I},{\bf{T}}({{\bf{z}}^{(i)}_{\rm{p}}}),\gamma_{\rm{p}}] +\text{f}[{\bf{W}}_{\rm{f}}^{(i)}+\nu_{\rm{f}}\mathbf{I},{\bf{T}}({{\bf{z}}^{(i)}_{\rm{f}}}),\gamma_{\rm{f}}]\nonumber \\
				&=\text{f}[{\bf{W}}_{\rm{p}}^{(i)},{\bf{T}}({{\bf{z}}^{(i)}_{\rm{p}}}),\gamma_{\rm{p}}] +\text{f}[{\bf{W}}_{\rm{f}}^{(i)},{\bf{T}}({{\bf{z}}^{(i)}_{\rm{f}}}),\gamma_{\rm{f}}]\nonumber \\
				&+2\gamma_{\rm{p}}^{-2}\nu_{\rm{p}}{\rm{tr}}[\mathbf{T}({\mathbf{z}}^{(i)}_{\rm{p}})]++2\gamma_{\rm{f}}^{-2}\nu_{\rm{f}}{\rm{tr}}[\mathbf{T}({\mathbf{z}}^{(i)}_{\rm{f}})]\nonumber \\
				&\le\text{f}[{\bf{W}}_{\rm{p}}^{(i-1)},{\bf{T}}({{\bf{z}}^{(i-1)}_{\rm{p}}}),\gamma_{\rm{p}}] +\text{f}[{\bf{W}}_{\rm{f}}^{(i-1)},{\bf{T}}({{\bf{z}}^{(i-1)}_{\rm{f}}}),\gamma_{\rm{f}}]\nonumber \\
				&+2\gamma_{\rm{p}}^{-2}\nu_{\rm{p}}{\rm{tr}}[\mathbf{T}(\mathbf{z}^{(i-1)}_{\rm{p}})]++2\gamma_{\rm{f}}^{-2}\nu_{\rm{f}}{\rm{tr}}[\mathbf{T}(\mathbf{z}^{(i-1)}_{\rm{f}})].
				\end{align}
				Hence, the objective function of the perturbed CRM is nonincreasing and bounded. Therefore, from the monotone convergence theorem, the objective value in the iterations of the perturbed CRM converges to a finite limit.
				
				Next, we provide a proposition which proves that each limit point of the iterations in the perturbed CRM is a stationary point. Toward this end, we first introduce a lemma with respect to the stationary point in an alternative optimization.
				
				\textit{Lemma 3}\cite{Grippo2000On}: Consider the problem
				\begin{align}
				\min_{\mathbf{x},\mathbf{y}} f(\mathbf{x},\mathbf{y}) \quad  \text{subject to} \;\mathbf{x}\in\mathbb{X},\mathbf{y}\in\mathbb{Y},
				\end{align}
				where $f(\mathbf{x},\mathbf{y})$ is a continuously differentiable function, and $\mathbb{X}$, $\mathbb{Y}$ are closed, nonempty and convex sets. Assuming that $\{(\mathbf{x}^{(i)}),(\mathbf{y}^{(i)})\}$ is the sequence generated by the alternative optimization and has limit points. Then, every limit point of $\{(\mathbf{x}^{(i)}),(\mathbf{y}^{(i)})\}$ is a stationary point.
				
				\textit{Proposition 2}: Let $\{(\mathbf{z}^{(i)}_{\rm{p}},\mathbf{z}^{(i)}_{\rm{f}},\mathbf{X}^{(i)}_{\rm{v}}),(\mathbf{W}^{(i)}_{\rm{p}},\mathbf{W}^{(i)}_{\rm{f}})\}$ be the sequence generated in the iterations of the perturbed CRM and $\{(\mathbf{z}^{(i,s)}_{\rm{p}},\mathbf{z}^{(i,s)}_{\rm{f}},\mathbf{X}^{(i,s)}_{\rm{v}}),(\mathbf{W}^{(i,s)}_{\rm{p}},\mathbf{W}^{(i,s)}_{\rm{f}})\}$ be a subsequence converging to a limit point $((\bar{\mathbf{z}}_{\rm{p}},\bar{\mathbf{z}}_{\rm{f}},\bar{\mathbf{X}}_{\rm{v}}),(\bar{\mathbf{W}}_{\rm{p}},\bar{\mathbf{W}}_{\rm{f}}))$. Then, $((\bar{\mathbf{z}}_{\rm{p}},\bar{\mathbf{z}}_{\rm{f}},\bar{\mathbf{X}}_{\rm{v}}),(\bar{\mathbf{W}}_{\rm{p}},\bar{\mathbf{W}}_{\rm{f}}))$ satisfies the Karush–Kuhn–Tucker (KKT) condition.
				
				\begin{IEEEproof}
					As we will demonstrate in the ADMM solution (see next subsection), the objective function of the CRM is continuously differentiable. Besides,  the feasible subset is closed, nonempty, and convex, thus satisfying the conditions of \textit{Lemma 3}. We have already observed that the matrices $\{(\mathbf{z}^{(i)}_{\rm{p}},\mathbf{z}^{(i)}_{\rm{f}},\mathbf{X}^{(i)}_{\rm{v}}),(\mathbf{W}^{(i)}_{\rm{p}},\mathbf{W}^{(i)}_{\rm{f}})\}$ are bounded because of the constraint $\|\mathbf{X}_{\rm{v}}\circ\mathbf{B}-\tilde{\mathbf{X}}_{\rm{v}}\|^2_{\mathcal{F}}\le\eta$ and the result $\nu_{\rm{p}}\rm{tr}[\mathbf{T}(\dot{\mathbf{z}}_{\rm{p}})]+\nu_{\rm{f}}\rm{tr}[\mathbf{T}(\dot{\mathbf{z}}_{\rm{f}})]\le(\nu_{\rm{p}}+\gamma_{\rm{p}})\rm{tr}[\mathbf{T}(\mathbf{z}^{\rm{D}}_{\rm{p}})]+(\nu_{\rm{f}}+\gamma_{\rm{f}})\rm{tr}[\mathbf{T}(\mathbf{z}^{\rm{D}}_{\rm{f}})]$ in \textit{Lemma 2}. As such, we can conclude that $((\bar{\mathbf{z}}_{\rm{p}},\bar{\mathbf{z}}_{\rm{f}},\bar{\mathbf{X}}_{\rm{v}}),(\bar{\mathbf{W}}_{\rm{p}},\bar{\mathbf{W}}_{\rm{f}}))$ is a stationary point of the perturbed CRM and satisfies the KKT conditions. This implies the convergence of the perturbed CRM.
				\end{IEEEproof}
				
				Finally, we prove the equivalence between the perturbed CRM and the CRM in \eqref{multiconvex}. To this end, we introduce the following proposition.
				
				\textit{Proposition 3}: Let $\{\mathring{\mathbf{X}}_{\rm{v}},\mathring{\mathbf{z}}_{\rm{p}},\mathring{\mathbf{z}}_{\rm{f}}\}$ be the minimum trace solution of non-perturbed subproblem \eqref{perturbed} ($\nu_{\rm{p}}$=$\nu_{\rm{f}}=0$), i.e.,
				\begin{align}\label{mintrace}
				&\mathring{{\bf{z}}}_{\rm{p}},\mathring{{\bf{z}}}_{\rm{f}},\mathring{\mathbf{X}}_{\rm{v}}\!=\!\mathop{\arg\!\min_{{\bf{z}}_{\rm{p}},{\bf{z}}_{\rm{f}},\mathbf{X}_{\rm{v}}}}\! {\rm{tr}}[\mathbf{T}({\mathbf{z}_{\rm{p}}})]
				+{\rm{tr}}[\mathbf{T}({\mathbf{z}_{\rm{f}}})]\quad  \text{subject to} \nonumber \\
				&	(\mathbf{X}_{\rm{v}},\mathbf{z}_{\rm{p}},\mathbf{z}_{\rm{f}}) \;\text{is a solution of non-perturbed  \eqref{perturbed}}.
				\end{align}
				When $\nu_{\rm{p}}$ and $\nu_{\rm{f}}$ satisfy $	2\gamma_{\rm{p}}^{-2}\nu_{\rm{p}}{\rm{tr}}[\mathbf{T}(\mathring{\mathbf{z}}_{\rm{p}})]+2\gamma_{\rm{f}}^{-2}\nu_{\rm{f}}{\rm{tr}}[\mathbf{T}(\mathring{\mathbf{z}}_{\rm{f}})]<1$, the perturbed CRM and the CRM are equivalent.
				
				\begin{IEEEproof}
					Due to the limit of space, the detailed proof is given in Section C of the Supplement File.
				\end{IEEEproof}
				In this case, $((\bar{\mathbf{z}}_{\rm{p}},\bar{\mathbf{z}}_{\rm{f}},\bar{\mathbf{X}}_{\rm{v}}),(\bar{\mathbf{W}}_{\rm{p}},\bar{\mathbf{W}}_{\rm{f}}))$ is also a stationary point of the CRM \eqref{multiconvex} and follows the KKT condition. This concludes the convergence of the proposed CRM method.

			\subsection{Reconstruction Performance of Interpolation}\label{sec:DANMperformance}
			We now analyze the reconstruction performance of the interpolation methods by providing the theoretical  reconstruction  error of the virtual signal matrix. Denote the theoretical virtual signal matrix (without noise) as $\dot{\mathbf{X}}_{\rm{v}}$ and the optimal solution to the DANM problem \eqref{DANM1} as  $\hat{\mathbf{X}}_{\rm{v}}$. Generally, the reconstruction error between the theoretical and the optimal solutions is evaluated in terms of $\|\hat{\mathbf{X}}_{\rm{v}}\circ\mathbf{B}-\dot{\mathbf{X}}_{\rm{v}}\circ\mathbf{B}\|^{2}_{\mathcal{F}}$. Then, the following proposition is derived to theoretically demonstrate the reconstruction performance of the DANM method.
			
			\textit{Proposition 4}: There exists a constant $C>0$ such that the regularization parameter $\mu\ge {L^2}/{\sqrt{T}}(\sum\nolimits^{K}_{k=1}p_k+\sigma^2_{\rm{n}})$ is sufficient to guarantee the reconstruction performance of \eqref{DANM1} as
			\begin{align}\label{Proposition4}
			\|\hat{\mathbf{X}}_{\rm{v}}\circ\mathbf{B}-\dot{\mathbf{X}}_{\rm{v}}\circ\mathbf{B}\|^{2}_{\mathcal{F}}\le\mu+\sqrt{\mu^2+\frac{L}{2\mu}\left(\sum\nolimits^{K}_{k=1}p_k+\sigma^2_{\rm{n}}\right)}
			\end{align}
			with probability at least $1-2{\rm{e}}^{-2C\sqrt{T}}$.
			\begin{IEEEproof}
				Due to the limit of space, the detailed proof is given in Section D of the Supplement File.
			\end{IEEEproof}
			According to \textit{Proposition 4}, we conclude that the reconstruction performance of the DANM method is related to the number of snapshots and the power of the receive signal, i.e., the trace of the covariance matrix $\mathbf{R}_{\rm{v}}$. Notice that, the objective function of the DANM method \eqref{DANM1} is also the trace function. This observation reveals again that the essence of sparse reconstruction using DANM is power minimization.
			
			Similar to the analysis of DANM, we provide the theoretical reconstruction error of the CRM method, given in the following proposition.
			
			\textit{Proposition 5}: There exists a positive constant $C$ such that the regularization parameter $\mu\ge {L^2}/{\sqrt{T}}(\sum\nolimits^{K}_{k=1}p_k+\sigma^2_{\rm{n}})$ is sufficient to guarantee the reconstruction of \eqref{multiconvex} as
			\begin{align}\label{Proposition5-CRM}
			\|\hat{\mathbf{X}}_{\rm{v}}\circ\mathbf{B}-\dot{\mathbf{X}}_{\rm{v}}\circ\mathbf{B}\|^{2}_{\mathcal{F}}\le\mu+\sqrt{\mu^2+\frac{2K}{\mu}}
			\end{align}
			with probability at least $1-2{\rm{e}}^{-2C\sqrt{T}}$.
			\begin{IEEEproof}
				Due to the limit of space, the detailed proof is given in Section D the Supplement File.
			\end{IEEEproof}
			
			We observe in \eqref{Proposition5-CRM} that the reconstruction performance of the CRM method is mostly related to the number of snapshots and the rank of the covariance matrix $\mathbf{R}_{\mathbf{x}}$. This observation further reveals that the essence of the CRM method is $l_0$-norm minimization.
	
		\begin{figure*}[t]
		\centering
		\subfloat[]{
			\includegraphics[width=0.235\linewidth]{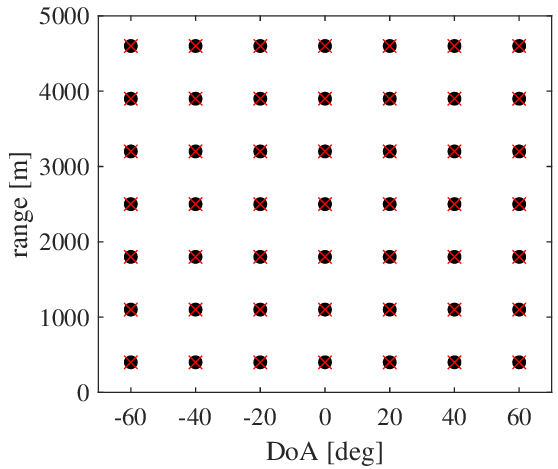}}
		\subfloat[]{
			\includegraphics[width=0.235\linewidth]{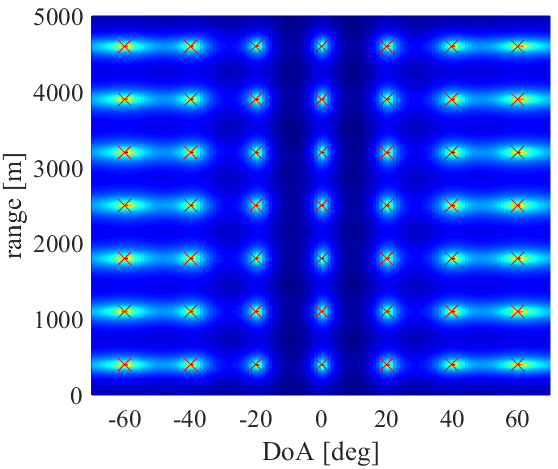}}
		\subfloat[]{
			\includegraphics[width=0.235\linewidth]{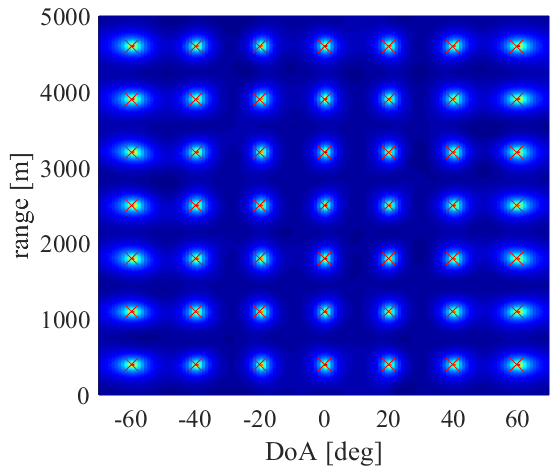}}
		\subfloat[]{
			\includegraphics[width=0.235\linewidth]{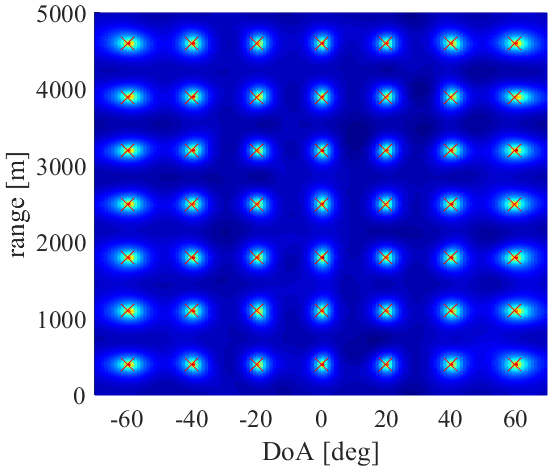}}
		\vspace{-1em}
		\hfill
		\subfloat[]{
			\includegraphics[width=0.235\linewidth]{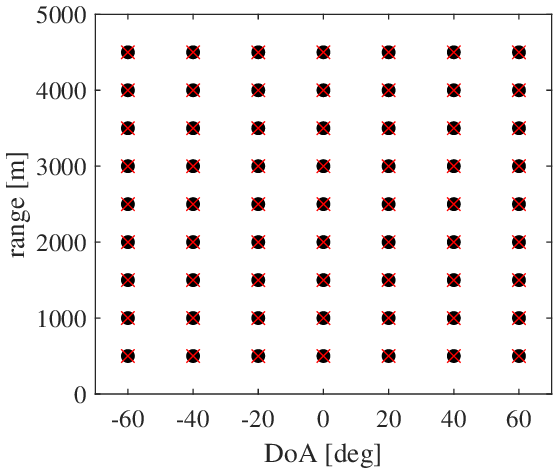}}
		\subfloat[]{
			\includegraphics[width=0.235\linewidth]{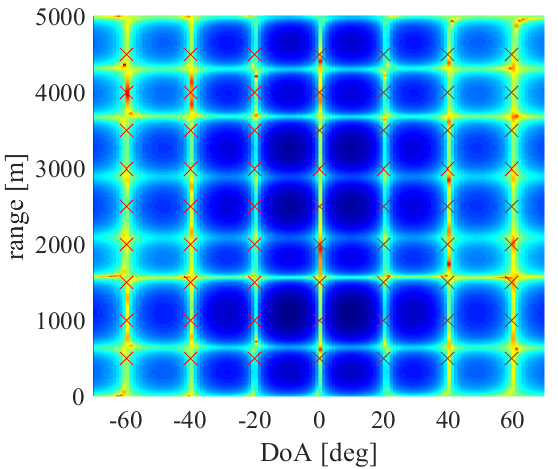}}
		\subfloat[]{
			\includegraphics[width=0.235\linewidth]{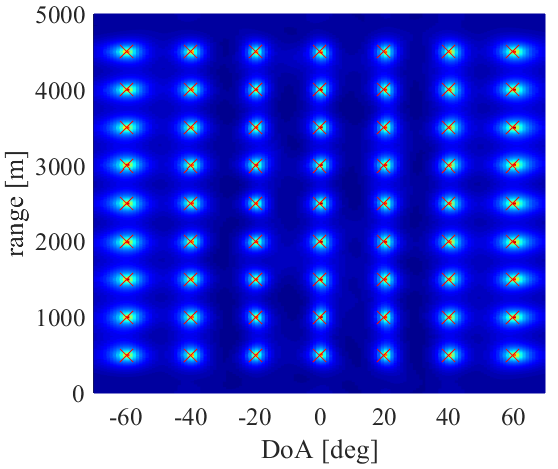}}
		\subfloat[]{
			\includegraphics[width=0.235\linewidth]{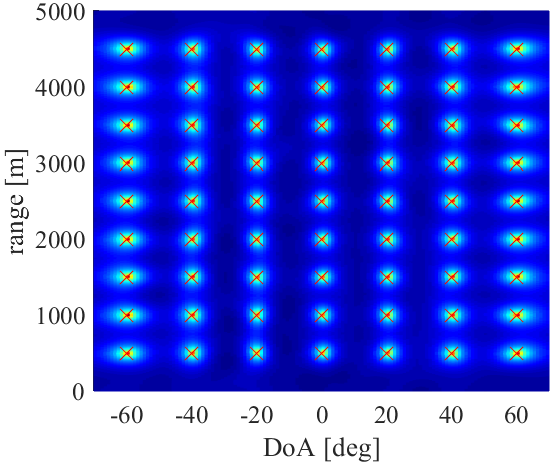}}
		\vspace{-1em}
		\hfill
		\subfloat[]{
			\includegraphics[width=0.237\linewidth]{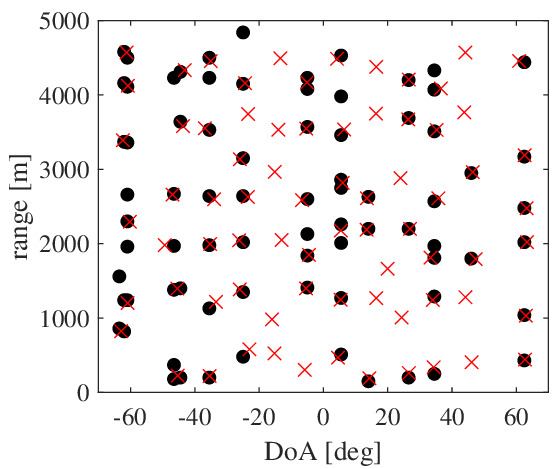}}
		\subfloat[]{
			\includegraphics[width=0.235\linewidth]{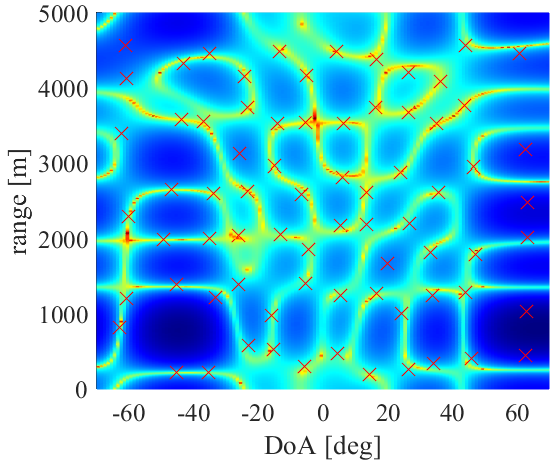}}
		\subfloat[]{
			\includegraphics[width=0.235\linewidth]{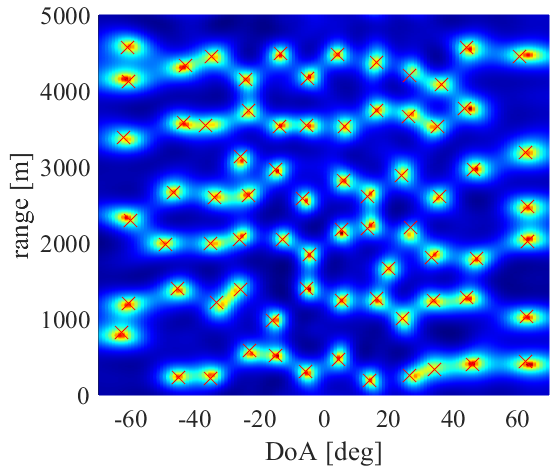}}
		\subfloat[]{
			\includegraphics[width=0.235\linewidth]{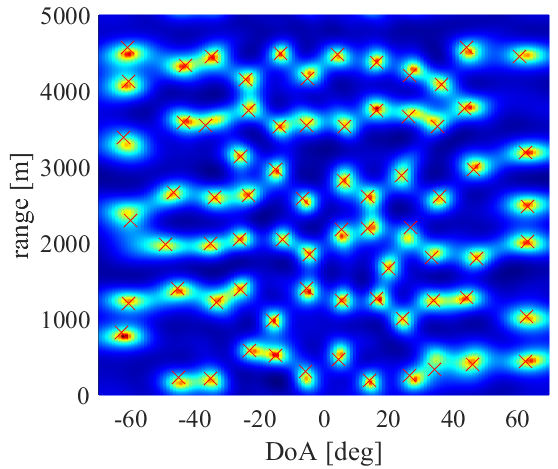}}
		\vspace{-1em}
		\hfill
		\subfloat[]{
			\includegraphics[width=0.235\linewidth]{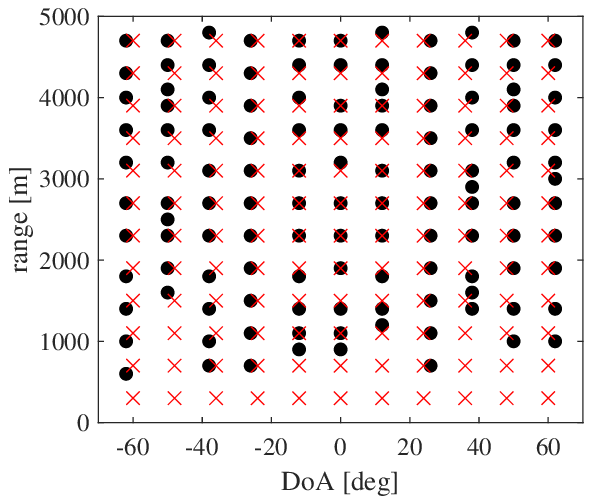}}
		\subfloat[]{
			\includegraphics[width=0.235\linewidth]{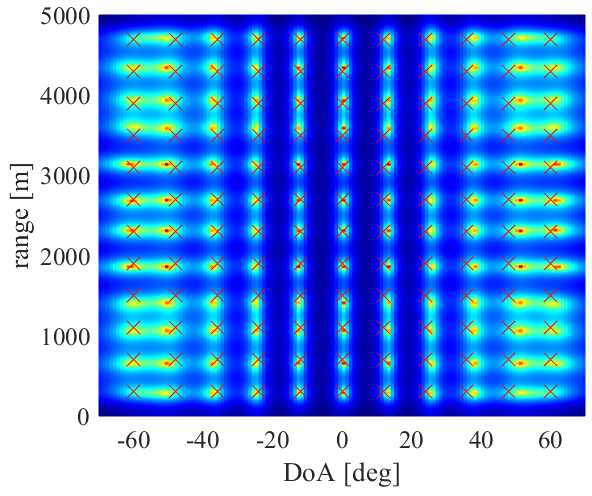}}
		\subfloat[]{
			\includegraphics[width=0.235\linewidth]{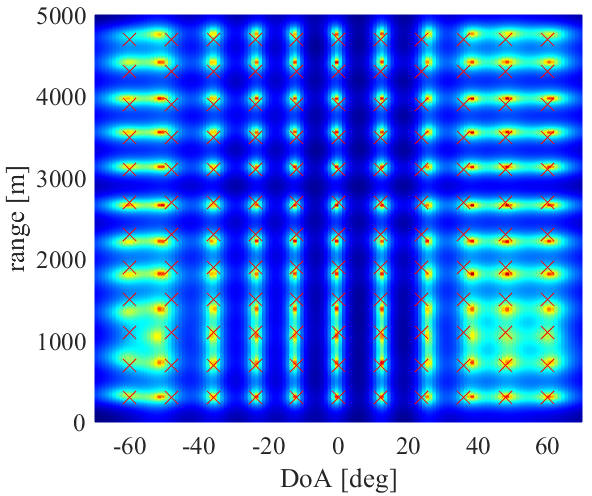}}
		\caption{2D estimation results of the space-frequency difference coarray. Red `x' marks denote the true target locations. (a) CMT-BCS, 49 targets; (b) SST, 49 targets; (c) DANM, 49 targets; (d) CRM, 49 targets; (e) CMT-BCS, 63 targets; (f) SST, 63 targets; (g) DANM, 63 targets; (h) CRM, 63 targets; (i) CMT-BCS, 74 targets; (j) SST, 74 targets; (k) DANM, 74 targets; (l) CRM, 74 targets; (m) CMT-BCS, 132 targets; (n) DANM, 132 targets; (o) CRM, 132 targets.}
		\label{DoF}
	\end{figure*}
	\section{Simulation Results and Analyses}\label{sec:simu}
	
	\begin{figure*}[!htpb]
		\centering
		\vspace{-1em}
		\subfloat[]{
			\includegraphics[width=0.4\linewidth]{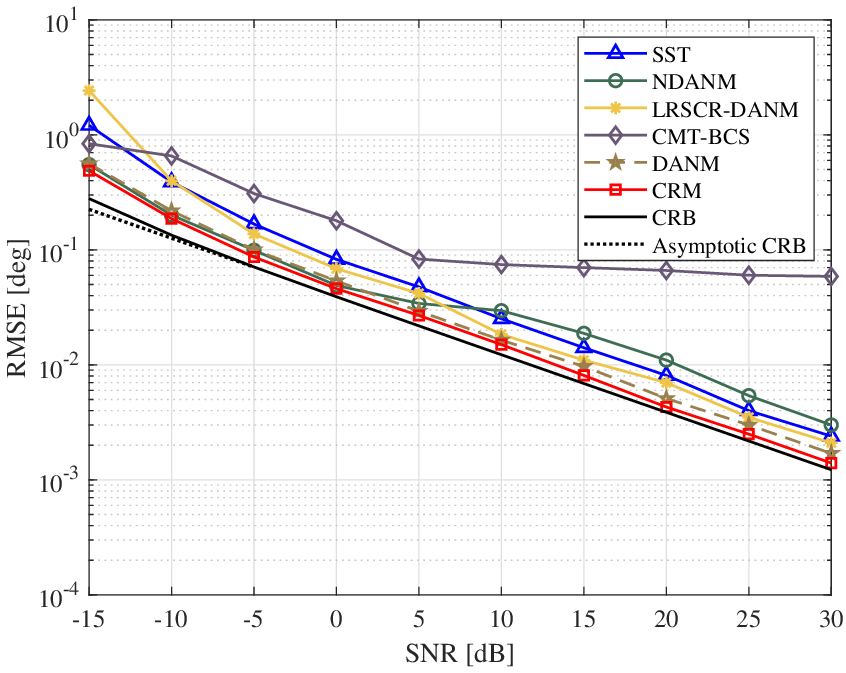}}
		\vspace{-1em}
		\hfill
		\subfloat[]{
			\includegraphics[width=0.4\linewidth]{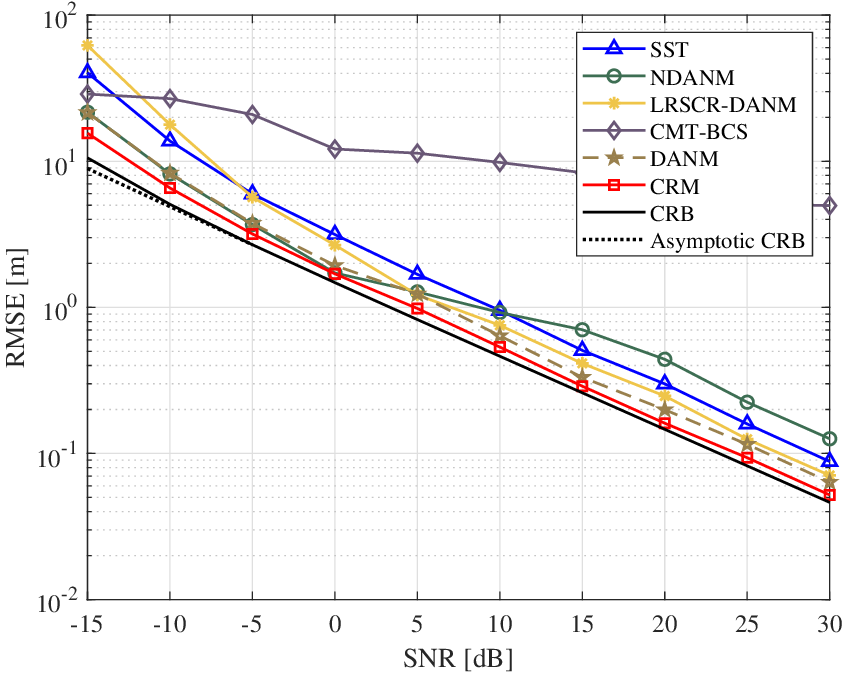}}
		\hfill
		\\
		\subfloat[]{
			\includegraphics[width=0.4\linewidth]{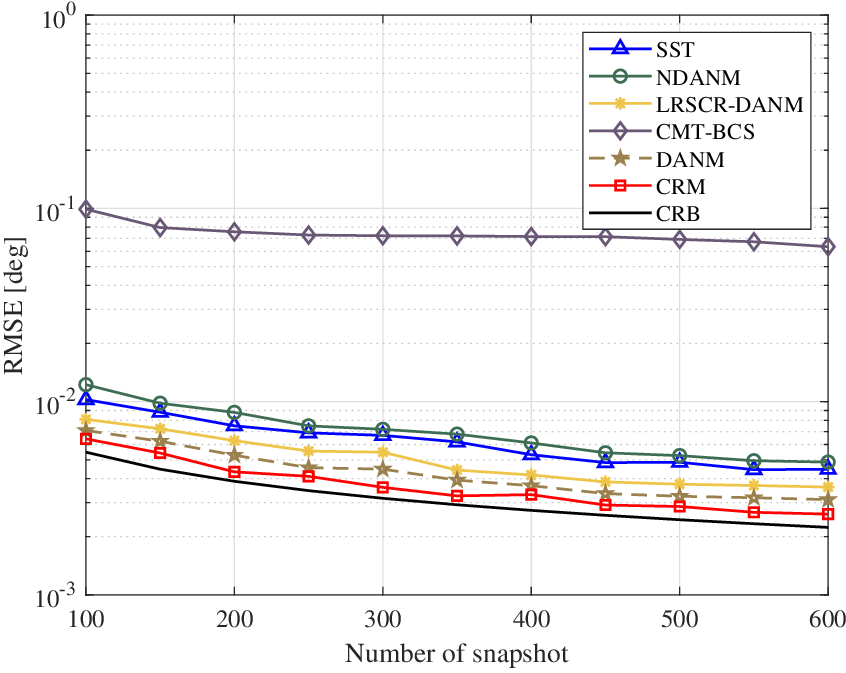}}
		\hfill
		\subfloat[]{
			\includegraphics[width=0.4\linewidth]{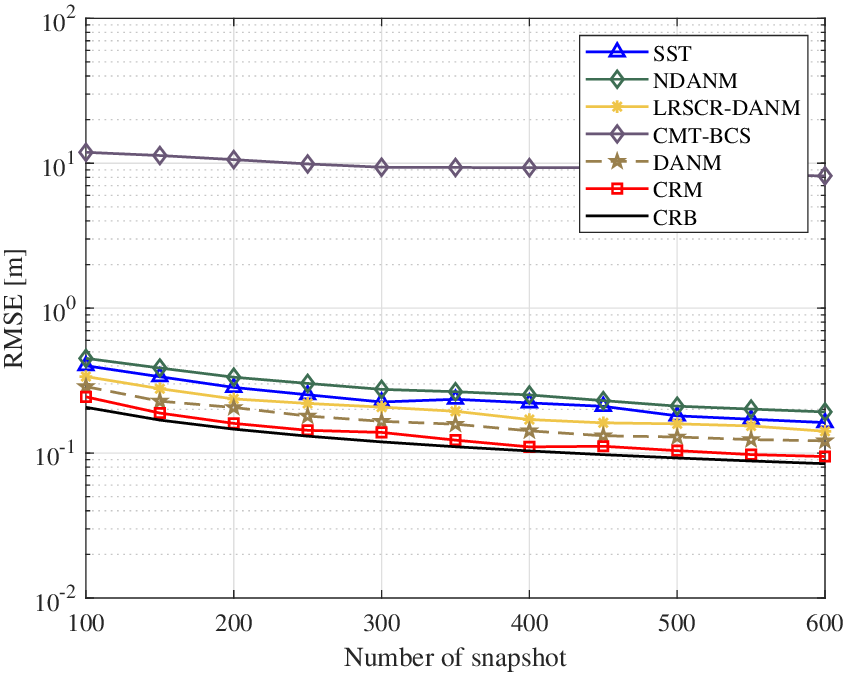}}
		\caption{Performance comparison between different methods. (a) RMSE of DoA versus SNR; (b) RMSE of range versus SNR; (c) RMSE of DoA versus number of snapshots; (d) RMSE of range versus number of snapshots.}\label{fig:RMSE_single}
		\vspace{-1.0em}
	\end{figure*}
	
	\begin{table}[b]
		\centering
		\caption{Number of Sensors and Frequency Offsets for Different Array Models}
		\label{table1}
		\begin{tabular}{c|c|c|c|c}
			\hline
			& & & &\\[-6pt]
			Type&Method&Sensors&Frequencies&DoF\\
			\hline
			& & & &\\[-6pt]
			FDCA&2D MUSIC&7&7&48\\
			\hline
			& & & &\\[-6pt]
			Difference Coarray\!&\!CMT-BCS&21&21&121\\
			\hline
			& & & &\\[-6pt]
			Consecutive Coarray&SST&15&15&63\\
			\hline
			& & & &\\[-6pt]
			Interpolated Coarray&DANM \& CRM&25&25&168\\
			\hline
		\end{tabular}
	\end{table}
	
	In this section, we demonstrate the effectiveness of joint DoA-range estimation using space-frequency difference coarray of FDCA. We consider coprime numbers with $M=3$ and $N=5$. Thus, the FDCA under simulation has a total of $M+N-1=7$ physical sensors located at $\left\{0d,3d,5d,6d,9d,10d,12d\right\}$, and the frequency offsets are set as $\left\{0\Delta f,3\Delta f,5\Delta f,6\Delta f,9\Delta f,10\Delta f,12\Delta f\right\}$. We assume that the FDCA operates at the X-band with a carrier frequency $f_0=10\;\text{GHz}$ and the unit frequency offset is $\Delta f=30\;\text{KHz}$. As such, the observed space-frequency difference coarray has a total of 21 sensors located from $\left[-12d,12d\right]$ with `holes' at $\left\{-11d,-8d,8d,11d\right\}$. Likewise, the frequency offset of the coarray are set as $\left[-12\Delta f,12\Delta f\right]$ with `holes' at $\left\{-11\Delta f,-8\Delta f,8\Delta f,11\Delta f\right\}$. The numbers of sensors, frequency offsets, and theoretic DoFs are listed in Table \ref{table1} for the four types of signal models and the corresponding methods under investigation, namely, the physical FDCA (2D MUSIC), the difference coarray (CMT-BCS), the consecutive coarray (SST), and the interpolated coarray (DANM and CRM).
	
	In the following simulations, the regularization coefficient $\mu$ for DANM and CRM is empirically set to $50$ (except for the simulations in Figs.\;\ref{fig:RMSE_gamma}(c) and \ref{fig:RMSE_gamma}(e), where $\mu$ varies), and the weight coefficients $\gamma_{\rm{p}}$ and $\gamma_{\rm{f}}$ are respectively set to $0.6$ and $0.4$ normalized by $\|\tilde{\mathbf{X}}_{\rm{v}}\|_{\mathcal{F}}$ (except for the simulations in Figs.\;\ref{fig:RMSE_gamma}(a), \ref{fig:RMSE_gamma}(b), \ref{fig:RMSE_gamma}(d) and \ref{fig:RMSE_gamma}(e), where $\gamma_{\rm{p}}$ and $\gamma_{\rm{f}}$ vary). The SDP problems in DANM and CRM are solved by CVX \cite{cvx} (except for ADMM in Fig.\;\ref{computationtime}).

	\subsection{Number of DoFs}
	
	In the first set of simulations, we confirm the increased number of DoFs achieved by the space-difference coarray of the FDCA. The input signal-to-noise ratio (SNR) and the number of snapshots are respectively fixed to $15\;\rm{dB}$ and $400$.  We first consider $49$ uncorrelated targets that are uniformly distributed in seven azimuth angles between $\left[-60^{\circ},60^{\circ}\right]$ and seven ranges between $\left[400\;\text{m},4600\;\text{m}\right]$. Four different methods are adopted to localize the targets, which are CMT-BCS, SST, DANM and CRM based on the space-frequency difference coarray. The dictionary matrices of CMT-BCS are assumed to contain all possible grid entries within $[-70^{\circ},-70^{\circ}]$ and $[0\;\text{m},5000\;\text{m}]$ with uniform intervals $\Delta\theta=1^{\circ}$ and $\Delta r=100\;\text{m}$, respectively. The estimation results are given in Figs.~\ref{DoF}(a)--\ref{DoF}(d), which clearly showcase the effectiveness of the space-frequency difference coarray in increasing the number of DoFs. For the on-grid CMT-BCS, the algorithm resolves all targets when the locations are included in the dictionary matrices. For the gridless methods, the 2D spectra of interpolation-based methods (DANM and CRM) are more focused than that of the SST because of the utilization of non-consecutive part.
	
	Next, we consider the case that the number of uncorrelated targets reaches the maximum number of DoFs of the coarray without interpolation, which is $63$. For clear illustration, we assume that $63$ targets are uniformly distributed in seven azimuths between $\left[-60^{\circ},60^{\circ}\right]$ and nine ranges between $\left[500\;\text{m},4500\;\text{m}\right]$. The result shown in Fig.~\ref{DoF}(f) confirms that the SST fail to function in this case because of the information loss as the non-consecutive part is discarded. In comparison, CMT-BCS and the interpolation-based methods still accurately estimate all $63$ targets.

	\begin{figure*}[!htpb]
		\centering
		\subfloat[]{
			\includegraphics[width=0.36\linewidth]{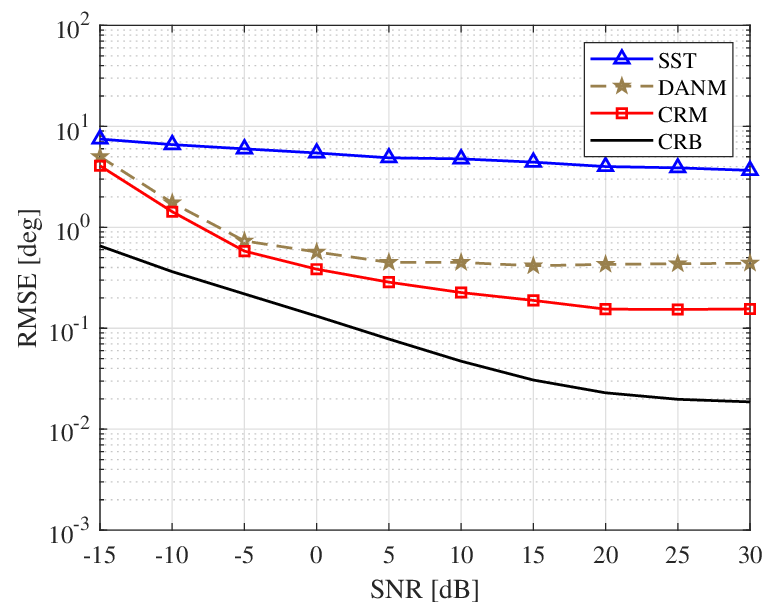}}
		\hfil
		\subfloat[]{
		\includegraphics[width=0.36\linewidth]{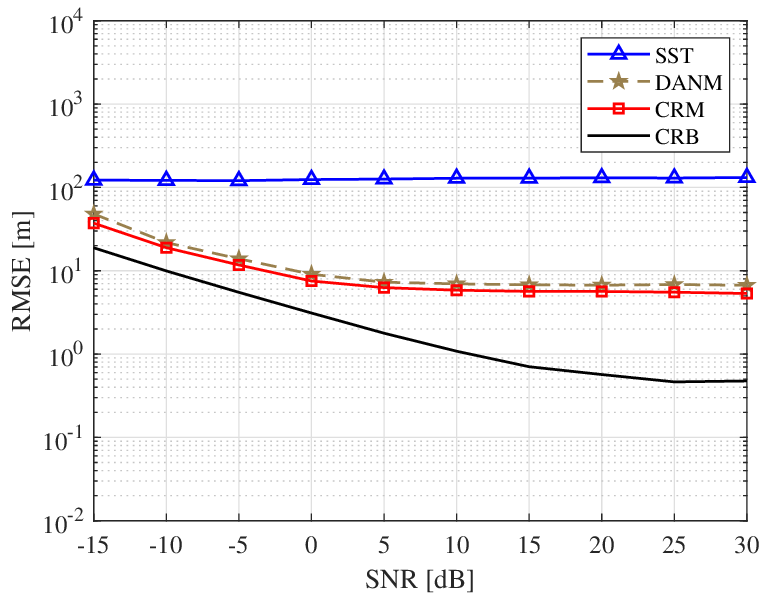}}
		\hfill
		\caption{ Performance comparison of different methods, $63$ targets. (a) RMSE of DoA versus SNR; (b) RMSE of range versus SNR.}\label{fig:RMSE_sixtythree}
		\vspace{-1em}
	\end{figure*}
	In the following, we assume that $74$ targets are randomly and non-uniformly distributed in the DoA-range plane, which is an off-grid case for CMT-BCS. As shown in Fig.~\ref{DoF}(i), CMT-BCS fails to resolve all targets due to grid mismatch, and some false and miss alarms can be observed. By contrast, the proposed interpolation-based methods can still accurately estimate all the peaks of $74$ targets in the 2D spatial spectra.

	Finally, we confirm the feasibility of the proposed interpolation-based methods, i.e., DANM and CRM, when the number of targets exceeds the maximum number of DoFs of the difference coarray.  We provide the DoF comparison for CMT-BCS and the proposed interpolation methods (DANM and CRM) in a $132$-targets case, which is already beyond the capacity of the CMT-BCS. We assume that 132 targets are uniformly distributed in 11 azimuths between $[-60^{\circ},60^{\circ}]$ and 12 ranges between $[300{\rm m}, 4700{\rm m}]$. The results shown in  Figs.~\ref{DoF}(m)--\ref{DoF}(o) confirm that the CMT-BCS performs incorrectly as the maximum DoF of the difference coarray is limited by the number of non-negative unique lags. However, the proposed interpolation methods can  still roughly resolve all the targets. These results further demonstrate the advantage of the interpolation-based methods in terms of DoF compared to CMT-BCS.

	\subsection{Coarray CRB and RMSE}
	To theoretically evaluate and analyze the performance of the proposed space-frequency coarray scheme for joint DoA-range estimation as well as the proposed reconstruction algorithms, we derive the CRB as a statistical benchmark. For the FDCA receive signal \eqref{subproof1}, the unknown parameters can be written in a vector form
		\begin{align} {\boldsymbol{\zeta}}&=\left[\theta_1,\cdots,\theta_K,r_1,\cdots,r_K,p_1,\cdots,p_K,\sigma^2_{\rm{n}}\right]^{\rm{T}}\nonumber\\ &=\left[\boldsymbol{\theta}^{\rm{T}},\mathbf{r}^{\rm{T}},\mathbf{p}^{T},\sigma^2_{\rm{n}}\right]^{\rm{T}}\in \mathbb{R}^{3K+1}.
		\end{align}
		As indicated in \eqref{vec}, the covariance matrix $\mathbf{R}_{\mathbf{x}}$ generates the virtual signal of the space-frequency coarray. Thus, the Fisher information is obtained from $\mathbf{R}_{\mathbf{x}}$. The Fisher information with respect to the $\tilde{i}$-th and $\tilde{j}$-th parameters $\zeta_{\tilde{i}}$ and $\zeta_{\tilde{j}}$ in $\boldsymbol{\zeta}$ can be represented as
		\begin{align}\label{Fisher1}	J_{\tilde{i},\tilde{j}}=T\cdot{\rm{tr}}\left[\mathbf{R}_{\mathbf{x}}^{-1}\frac{\partial\mathbf{R}_{\mathbf{x}}}{\partial\zeta_{\tilde{i}}}\mathbf{R}_{\mathbf{x}}^{-1}\frac{\partial\mathbf{R}_{\mathbf{x}}}{\partial\zeta_{\tilde{j}}}\right].
		\end{align}
	When the number of targets is higher than the number of DoFs provided by the physical FDCA, the Fisher information matrix (FIM) $\mathbf{J}$ in \eqref{Fisher1} is singular \cite{Liu2016Cramer}. As a result, a CRB cannot be derived from \eqref{Fisher1}.
	To the best of our knowledge, the CRB has not been derived for space-frequency coarrays in such an underdetermined case. To provide a benchmark for performance evaluation, therefore, we derive a coarray CRB analysis in the Supplement File.
	
	In the second set of simulations, statistical results in terms of the RMSE are used to compare the estimation accuracy of CMT-BCS, SST, DANM and CRM. The covariance-based DANM methods, normal DANM (NDANM)\cite{Lu2020Efficient}, and low rank structured covariance reconstruction DANM (LRSCR-DANM)\cite{Wang2020Efficient} are also included for comparison. The RMSE of a certain parameter $\xi$ is defined as
	\begin{align}
	{{\rm{RMSE}}}\left(\xi\right)=\sqrt{\frac{1}{KP}\sum\nolimits_{k=1}^K\sum\nolimits_{p=1}^P\left(\hat{\xi}_k\left(p\right)-\xi_k\right)^2},
	\end{align}
	where $\hat{\xi}_k\left(p\right)$ denotes the estimation result of the $k$-th target in the $p$-th Monte Carlo trial, and $P$ is the total number of Monte Carlo trials. 
	
	We first examine the RMSE performance in a single-target case. The DoA and range are randomly generated for 1000 Monte Carlo trials from their respective Gaussian distributions $\theta\sim\mathcal{N}({30^{\circ},(1^{\circ})^2})$ and $r\sim\mathcal{N}(2500 \ \rm{m},(10 \ \rm{m})^2)$. The dictionary matrices of CMT-BCS contain steering vectors over all possible values in $[25^{\circ},35^{\circ}]$ and $[2450\;\text{m},2550\;\text{m}]$ with uniform intervals $\Delta\theta=0.2^{\circ}$ and $\Delta r=1\;\text{m}$, respectively. We first fix the number of snapshots to 200 and let the input SNR vary between $-15\;{\rm{dB}}$ and $30\;{\rm{dB}}$, and the RMSE results are shown in Fig.\;\ref{fig:RMSE_single}. As indicated in Fig.\;\ref{fig:RMSE_single}(a), as the input SNR increase, the coarray CRB is asymptotically linear and approaches the asymptotic CRB which is precisely linear \cite{Wang2017Coarrays}. This is one of the typical behaviors of coarray CRB, which explains the similar trends of RMSEs for both SST and interpolation-based methods. In particular, the RMSE with respect to DoA asymptotically approaches to the coarray CRB when the input SNR is between $-15\;{\rm{dB}}$ and $0\;{\rm{dB}}$. The gaps between the RMSE and the CRB almost remain unchanged when the input SNR is between $0\;{\rm{dB}}$ and $30\;{\rm{dB}}$.
	
		\begin{figure*}[!htpb]
		\centering
		\vspace{-1em}
		\subfloat[MAPE of DoA, single target.]{
			\includegraphics[width=0.32\linewidth]{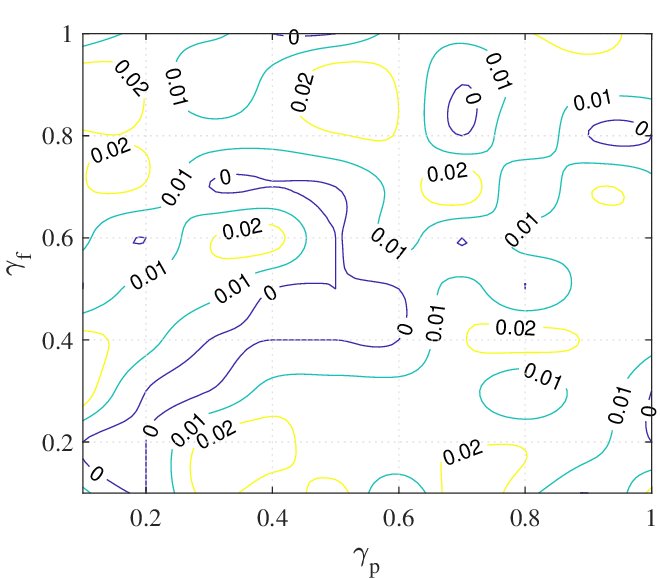}}
		\hfill
		\vspace{-1em}
		\subfloat[MAPE of DoA, 63 targets.]{
			\includegraphics[width=0.32\linewidth]{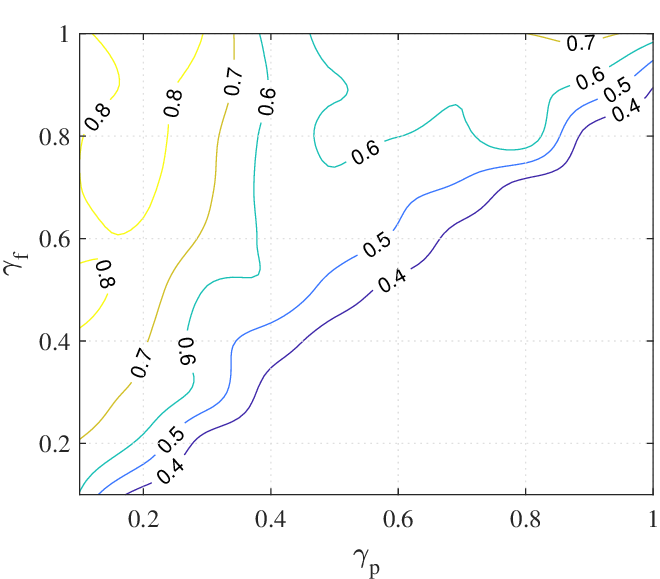}}
		\hfill
		\subfloat[RMSEs of DoA versus $\mu$.]{				        	
			\includegraphics[width=0.32\linewidth]{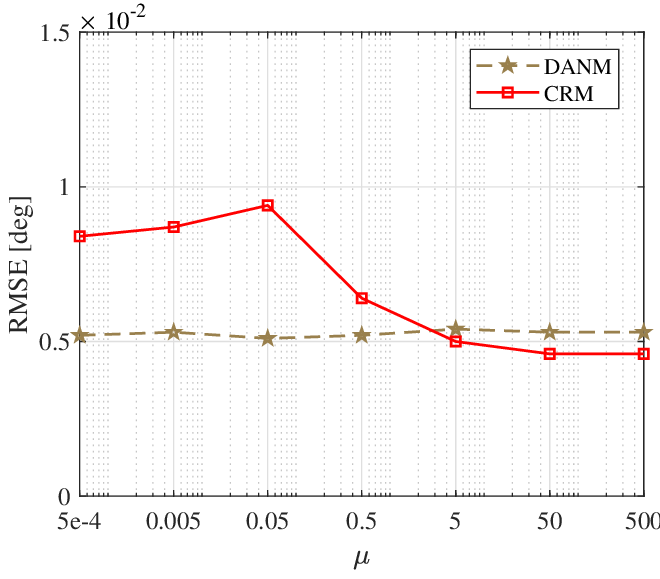}}
		\hfill
		\subfloat[MAPE of range, single target.]{
			\includegraphics[width=0.32\linewidth]{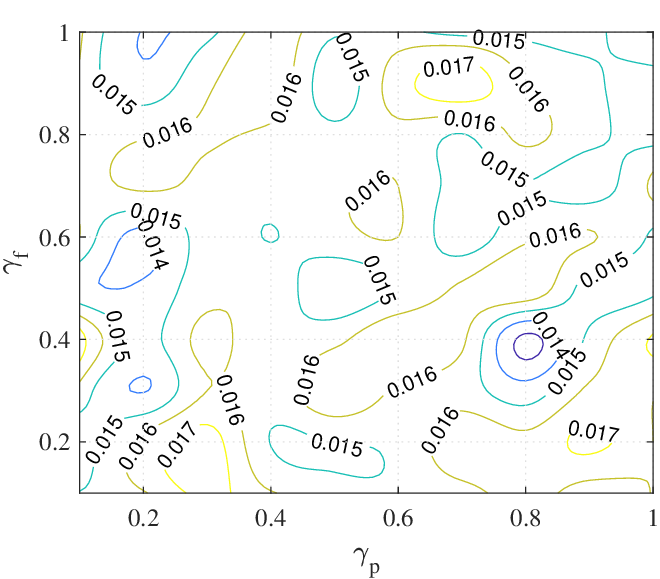}
		}
		\subfloat[MAPE of range, 63 targets.]{
			\includegraphics[width=0.32\linewidth]{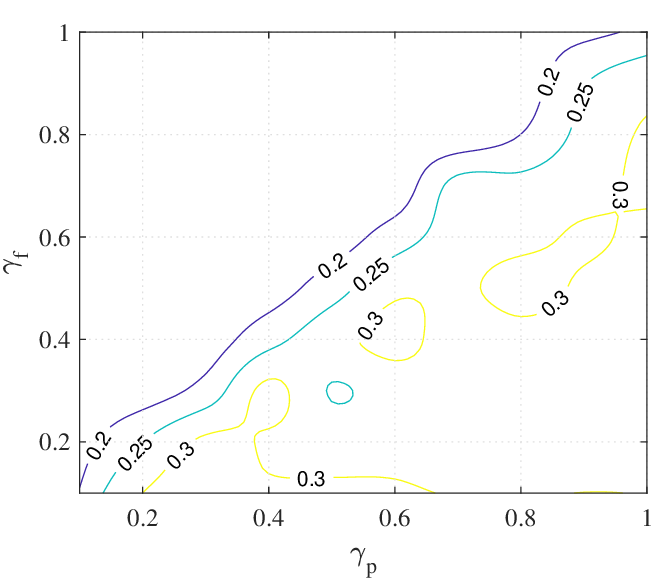}}
		\hfill
		\subfloat[RMSEs of range versus $\mu$.]{
			\includegraphics[width=0.325\linewidth]{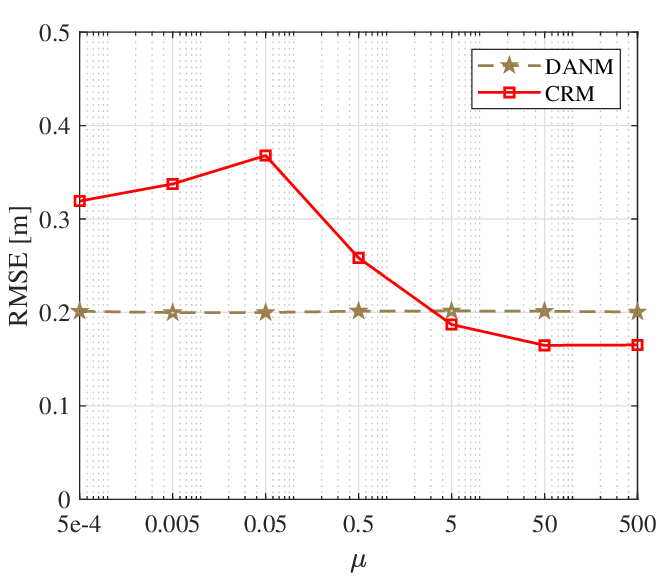}}
		\hfill
		\caption{Impact of different selections of optimization coefficients $\gamma_{\rm{p}}$, $\gamma_{\rm{f}}$ and $\mu$. The results with respect to $\gamma_{\rm{p}}$ and $\gamma_{\rm{f}}$ are presented via the contour plots of MAPE. The results with respect to $\mu$ are compared via RMSE.}\label{fig:RMSE_gamma}	\vspace{-1.0em}
	\end{figure*}

	Next, we compare the RMSE performance for different gridless algorithms. Compared to the other methods, the gap between the RMSE and the coarray CRB in Fig.\;\ref{fig:RMSE_single}(a) is approximately twice as large in SST, except for LRSCR-DANM in the low-SNR region and NDANM in the high-SNR region. This is because the fitting threshold of NDANM is fixed, while the proposed DANM can adaptively optimize the fitting term owing to the regularization. As for LRSCR-DANM, the measurement is not observed directly but rather over a linear compressive kernel, which may suffer from information loss in the low-SNR region\cite{Pakrooh2015Analysis}. For the proposed interpolation-based algorithms, the RMSE of CRM is consistently lower than that of DANM; and in the DANM case, the gap between coarray CRB is approximately twice that of CRM for input SNR between $-5\;{\rm{dB}}$ and $30\;{\rm{dB}}$ as well. The improved DoA estimation performance is obtained because the problem is more accurately reformulated and solved without approximation in CRM. On the other hand, the floor of the RMSE performance for CMT-BCS is because this method is grid-based and, thus, suffers from performance loss due to basis mismatch.
	
	When we fix the input SNR to $20\;\rm{dB}$ and vary the number of snapshots, the results in Fig.\;\ref{fig:RMSE_single}(c) confirm again that CMT-BCS renders higher error due to the basis mismatch. Whereas, the off-grid methods achieve better performance and, in particular, CRM outperforms other methods. As for the behavior of coarray CRB, (87) in the Supplement File indicates that the Fisher information is proportional to $T$, which is consistent with traditional CRB. We now observe the RMSE curves with respect to range in joint estimation, which are given in Figs.\;\ref{fig:RMSE_single}(b) and \ref{fig:RMSE_single}(d). Compared with Figs.\;\ref{fig:RMSE_single}(a) and \ref{fig:RMSE_single}(c), very similar trends can be observed for gridless methods based on the specific decoupled design in \eqref{signalmatrix}, thereby demonstrating the superiority of joint estimation. However, the RMSE of CMT-BCS performs worse due to error propogation in muti-task BCS, i.e., errors in the DoA estimation stage yield additional perturbations in the range estimation.
	
	In the following, we consider the scenario where the number of targets is greater than the number of DoFs of the physical FDCA, i.e., $48$. The parameters $\left(\boldsymbol{\theta},\mathbf{r}\right)$ of the targets are similar to those $(\breve{\boldsymbol{\theta}},\breve{\mathbf{r}})$ used in Figs.\ \ref{DoF}(c)--\ref{DoF}(e), with two additional two targets generated from Gaussian distributions  $[\boldsymbol{\theta}\sim\mathcal{N}({\breve{\boldsymbol{\theta}},(1^{\circ})^2}),\mathbf{r}\sim\mathcal{N}(\breve{\mathbf{r}},(5 \ \rm{m})^2)]$. The results are given in Figs.\;\ref{fig:RMSE_sixtythree}(a) and \ref{fig:RMSE_sixtythree}(b). In this scenario, as shown in Fig.\;\ref{DoF}(d), the SST does not resolve all targets and renders a high RMSE with respect to both DoA and range in all SNR region.
	Fig.\;\ref{fig:RMSE_sixtythree} also includes the coarray CRB which gradually converges as the input SNR increases, which shows the \emph{saturation} behaviors of coarray CRB due to the positive definiteness of the FIM in the underdetermined problem with $K$ exceeding the number of DoFs of physical FDCA\cite{Wang2017Coarrays}.  In addition, the RMSEs of DANM and CRM exhibit similar trends, and CRM consistently outperforms DANM in both DoA and range estimation. In particular, the advantage of CRM with respect to DoA is more evident owing to the selection of the weight coefficients $\gamma_{\rm{p}}$ and $\gamma_{\rm{f}}$.
	
	\subsection{Impact of Coefficient Settings}
	
	In the third set of simulations, we observe and analyze the impact of coefficient settings in DANM and CRM. Furthermore, we provide both theoretical and empirical guidelines to reduce the cost of cross-validation in practice. The impact of the weight coefficients on the performance of the CRM is evaluated in terms of the mean absolute percentage error (MAPE), which is defined as
	\begin{equation}
	{\rm{MAPE}}\left(\xi\right)=\frac{\sum\nolimits_{k=1}^K\sum\nolimits_{p=1}^P\vert\hat{\xi}_{k}\left(p\right)-\xi_k\vert}{P\sum\nolimits_{k=1}^K\vert\xi_{k}\vert}\times 100\%.
	\end{equation}
	
	The input SNR and the number of snapshots are respectively fixed to $20\;\rm{dB}$ and $200$. We first analyze the single target case where the target location remains the same as that used in Fig.\;\ref{fig:RMSE_single}. Note that, the parameters $\gamma_{\rm{p}}$ and $\gamma_{\rm{f}}$ are normalized by $\|\tilde{\mathbf{X}}_{\rm{v}}\|_{\mathcal{F}}$. The results in Figs.\;\ref{fig:RMSE_gamma} (a) and \ref{fig:RMSE_gamma}(c) show that CRM is insensitive to $\gamma_{\rm{p}}$ and $\gamma_{\rm{f}}$ in the single target case as the MAPE of both DoA and range remains low (${\rm{MAPE}}\left(\theta\right)<0.03\%$ and ${\rm{MAPE}}\left(r\right)<0.018\%$), and the contour plots are irregular. Nevertheless, this is not the case in the other observations when the number of targets is increased to $63$. We similarly assume that the locations of targets are identical to those in Fig.\;\ref{fig:RMSE_sixtythree}.  The contour plots in Figs.\;\ref{fig:RMSE_gamma}(b) and \ref{fig:RMSE_gamma}(e) display regular boundaries for both ${\rm{MAPE}}\left(\theta\right)$ and ${\rm{MAPE}}\left(r\right)$ in the sense that line $\gamma_{\rm{p}}=\gamma_{\rm{f}}$ divides the contour plots into two regions. Specifically, ${\rm{MAPE}\left(\theta\right)}$ decreases as $\gamma_{\rm{p}}/\gamma_{\rm{f}}$ increases, while ${\rm{MAPE}}\left(r\right)$ shows a reverse result. As mentioned in \emph{Remark 5} of Section \ref{sec:rankreformulation}, the optimization terms ${\rm{tr}}[\mathbf{W_{\rm{p}}}\mathbf{T}(\mathbf{z}_{\rm{p}})]$ and ${\rm{tr}}[\mathbf{W_{\rm{f}}}\mathbf{T}(\mathbf{z}_{\rm{f}})]$ are respectively weighted by $\gamma_{\rm{p}}$ and $\gamma_{\rm{f}}$. Thus, ratio $\gamma_{\rm{p}}/\gamma_{\rm{f}}$ represents the trade-off preference between the DoA and range estimation performance. By comparing Figs.\;\ref{fig:RMSE_gamma}(a) and \ref{fig:RMSE_gamma}(d) to \ref{fig:RMSE_gamma}(b) and \ref{fig:RMSE_gamma}(e), we observe that the CRM-based method exhibits the coupling effect between DoA and range estimation exists in multiple-target cases. To summarize, in the single-target case, fine tuning of $\gamma_{\rm p}$ and $\gamma_{\rm f}$ is not required. In multi-target scenarios, the values of $\gamma_{\rm p}$ and $\gamma_{\rm f}$ should be adjusted to trade off between the DoA and range estimation performances. Concretely, a higher $\gamma_{\rm{p}}/\gamma_{\rm{f}}$ ratio provides a better DoA estimation, whereas a larger $\gamma_{\rm{f}}/\gamma_{\rm{p}}$ ratio yields a higher range estimation accuracy.

	In the following, we evaluate the RMSE performance with respect to the regularization coefficient $\mu$ in a single target case, where the input SNR is $20\;\rm{dB}$ and the number of snapshots is $T=200$. The other parameters remain unchanged from those used in Fig.\;\ref{fig:RMSE_single}. DANM and CRM are examined. As shown in  Figs.\;\ref{fig:RMSE_gamma}(c) and \ref{fig:RMSE_gamma}(f), the RMSE of the DANM remains unchanged for most of the values of $\mu$. Different from DANM, the RMSE of CRM fluctuates and becomes lower when $\mu>5$. This suggests higher dependency of the performance for CRM on the regularization term $\mu$ than the DANM. Empirically, $\mu\ge 50$ is recommended for CRM to achieve a satisfactory performance.
	\vspace{-1em}
	\subsection{Computational Complexity}
	
\begin{figure}[t]
		\centering
		{\includegraphics[width=0.82\linewidth]{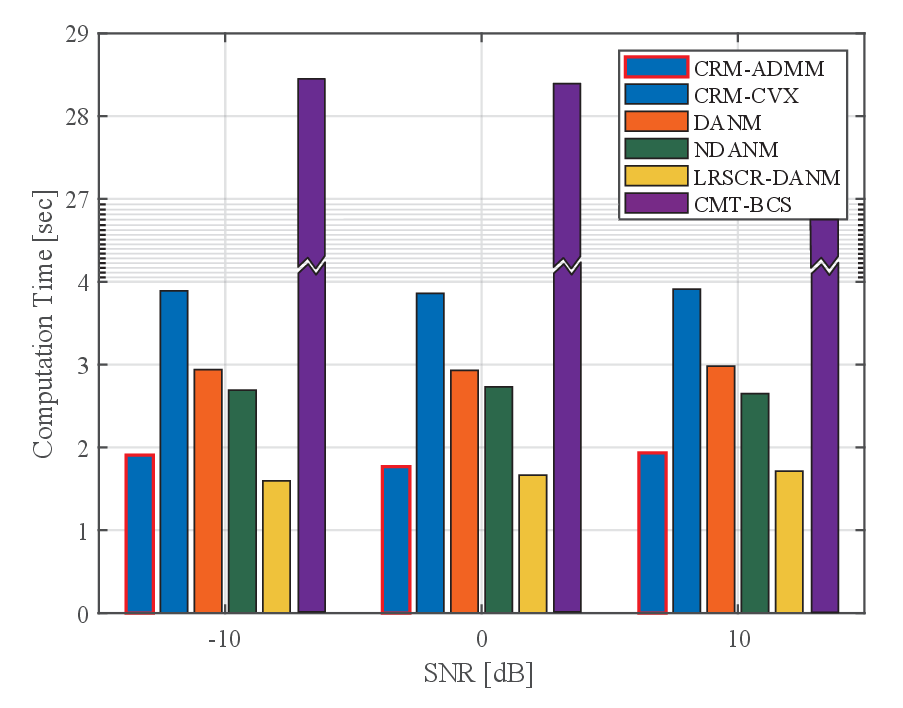}}
\vspace{-0.5em}
		\caption{{Comparison of computation time between different algorithms.}}\label{computationtime}
\vspace{-1em}
	\end{figure}

	The last simulation compares the computational efficiency of DANM, NDANM, LRSCR-DANM, CRM-CVX, CRM-ADMM, CMT-BCS through 100 Monte Carlo trials using an Intel(R) Core(TM) i7-8700 CPU in three different SNR cases. The dictionary matrices of the CMT-BCS are consistent with the setting in Fig.\;\ref{fig:RMSE_single}. The result is shown in Fig.\;\ref{computationtime}. Compared to  DANM, NDANM, and LRSCR-DANM which only optimize the signal subspace, the computational complexity of CRM is slightly higher as it needs to optimize both the signal and noise subspaces. Particularly, the runtime of  LRSCR-DANM  is the lowest by utilizing compressed sensing. On the other hand, the computational efficiency is significantly improved by using ADMM-based closed-form solutions, rendering a comparable computational time as the LRSCR-DANM method. Compared to the off-grid algorithms, the complexity of CMT-BCS is much higher due to the multi-task structure and the grid density.

	\section{Conclusion}\label{sec:conclusion}
	In this paper, we presented joint DoA-range estimation algorithms using space-frequency virtual difference coarray of the FDCA. Compared to existing 2D SST method which only utilizes consecutive coarray lags and frequency shifts, we propose interpolating the virtual signal to fully utilize the underlying received information in the presence of missing elements in the coarray and achieve a higher number of DoFs, and the joint DoA-range estimation is solved using three novel algorithms. The first one, termed DANM, simplifies the computationally prohibitive doubly-Toeplitz reconstruction problem as a decoupled atomic norm minimization problem which is then solved utilizing convex relaxation. To avoid the approximation loss, the second approach recasts the problem as a dual-variable rank minimization problem  which is solved using an iterative CRM-based algorithm. The third approach provides ADMM-based closed-form solutions for the SDP problem with significant reduction in the computational complexity. The improved joint DoA-range estimation performance of the proposed techniques is clearly demonstrated by theoretical analyses using coarray CRB and is verified using extensive numerical results.

	\bibliographystyle{IEEEbib}

\begin{thebibliography}{99}
	
		\balance
		
		\bibitem{Cao2021Doubly}
		R. Cao, S. Liu, Z. Mao and Y. Huang, ``Doubly-Toeplitz-based interpolation for joint DOA-range estimation using coprime FDA,'' in \emph{Proc. 2021 IEEE Radar Conf.}, Atlanta, GA, USA, May\ 2021, pp. 1--6.
		
		\bibitem{Patole2017Automotive}
		S.~M.\ Patole, M.~Torlak, D.~Wang and M.~Ali, ``Automotive radars: A review of signal processing techniques,'' \emph{IEEE Signal Process. Mag.}, vol.\ 34, no.\ 2, pp.\ 22--35, Mar.\ 2017.
		
		\bibitem{Garcia2017Direct}
		N.~Garcia, H.~Wymeersch, E.~G.\ Larsson, A.~M.\ Haimovich and M.~Coulon, ``Direct localization for massive MIMO," \emph{IEEE Trans. Signal Process.}, vol.\ 65, no.\ 10, pp.\ 2475--2487, May\ 2017.
		
		\bibitem{Wan2018Context}
		J.~Wan, S.~Tang, Q.~Hua, D.~Li, C.~Liu and J.~Lloret, ``Context-aware cloud robotics for material handling in cognitive industrial Internet of Things," \emph{IEEE Internet Things J.}, vol.\ 5, no.\ 4, pp.\ 2272--2281, Aug.\ 2018.
		
		\bibitem{Wang2013Phased}
		W.-Q.\ Wang, ``Phased-MIMO radar with frequency diversity for range-dependent beamforming," \emph{IEEE Sens. J.}, vol.\ 13, no.\ 4, pp.\ 1320--1328, Apr.\ 2013.
		
		\bibitem{Liao2019Frequency}
		Y. ~Liao, W.-Q.\ Wang and Z.~Zheng, ``Frequency diverse array beampattern synthesis using symmetrical logarithmic frequency offsets for target indication," \emph{IEEE Trans. Antennas Propaga.}, vol.\ 67, no.\ 5, pp.\ 3505--509, Feb.\ 2019.
		
		\bibitem{Antonik2006Range}
		P. Antonik, M. C. Wicks, H. D. Griffiths and C. J. Baker, ``Range dependent beamforming using element level waveform diversity,"  \emph{in Proc. IEEE Int. Waveform Diversity Design Conf.}, Las Vegas, NV, USA, Jan.\ 2006, pp.\ 140--144.
		
		\bibitem{Antonik2007Frequency}
		P. Antonik, M. C. Wicks, H. D. Griffiths and C. J. Baker, ``Frequency diverse array radars," in \emph{Proc. IEEE Radar Conf.}, Verona, NY, USA, Apr.\ 2006, pp.\ 215--217.
		
		\bibitem{Wang2014Transmit}
		W.-Q.\ Wang and H.~C.\ So, ``Transmit subaperturing for range and angle estimation in frequency diverse array radar,'' \emph{IEEE Trans. Signal Process.}, vol.\ 62, no.\ 8, pp.\ 2000--2011, Feb.\ 2014.
		
		\bibitem{Xu2015Joint}
		J.~Xu, G.~Liao, S.~Zhu, L.~Huang and H.~C.\ So, ``Joint range and angle estimation using {MIMO} radar with frequency diverse array,'' \emph{IEEE Trans. Signal Process.}, vol.\ 63, no.\ 13, pp.\ 3396--3410, Apr.\ 2015.
		
		\bibitem{Lan2020Trans}
		L.~Lan, G.~Liao, J.~Xu, Y.~Zhang and B.~Liao, ``Transceive beamforming with accurate nulling in FDA-MIMO radar for imaging," \emph{IEEE Trans. Geosci. Remote Sens.}, vol.\ 58, no.\ 6, pp. 4145--4159, Jun.\ 2020.
		
		\bibitem{Xiong17}
		J.~Xiong,  W.-Q.\ Wang and K.~Gao, ``FDA-MIMO radar range-angle estimation: CRLB, MSE, and  resolution analysis," \emph{IEEE Trans. Aerosp. Electron. Syst.}, vol.\ 54, no.\ 1, pp. 284--294, Feb.\ 2017.
		
		\bibitem{Kotaru2015SpotFi}
		M.~Kotaru, K.~Joshi, D.~Bharadia and S.~Katti, ``SpotFi: Decimeter level localization using WiFi,'' in \emph{Proc. SIGCOMM'15}, London, UK, Aug. 2015, pp.\ 269--282.
		
		\bibitem{Zheng2020Padded}
		W.~Zheng, X.~Zhang, Y.~Wang, J.~Shen and B.~Champagne, ``Padded coprime arrays for improved DOA estimation: Exploiting hole representation and filling strategies," \emph{IEEE Trans. Signal Process.}, vol.\ 68, pp.\ 4597--4611, Jul.\ 2020.
		
		\bibitem{Zhang2013Sparsity}
		Y.~D.\ Zhang, M.~G.\ Amin and B.~Himed, ``Sparsity-based DOA estimation using co-prime arrays,'' in \emph{Proc. IEEE Int. Conf. Acoust. Speech Signal Process. (ICASSP)}, Vancouver, Canada, May 2013, pp.\ 3967--3971.
		
		\bibitem{Qin2015Genralized}
		S.~Qin, Y.~D.\ Zhang and M.~G.\ Amin, ``Generalized coprime array configurations for direction-of-arrival estimation," \emph{IEEE Trans. Signal Process.}, vol.\ 63, no.\ 6, pp.\ 1377--1390, Mar.\ 2015.
		
		\bibitem{Liu2016Super}
		C.~Liu and P.~P.\ Vaidyanathan, ``Super nested arrays: Linear sparse arrays with reduced mutual coupling—Part II: High-order extensions," \emph{IEEE Trans. Signal Process.}, vol.\ 64, no.\ 16, pp.\ 4203--4217, Aug.\ 2016
		
		\bibitem{Guo2018DOA}
		M.~Guo, Y.~D.\ Zhang and T.~Chen, ``DOA estimation using compressed sparse array," \emph{IEEE Trans. Signal Process.}, vol.\ 66, no.\ 15, pp.\ 4133--4146, Aug.\ 2018
		
		\bibitem{Zheng2018DoA}
		W. Zheng, X. Zhang, P. Gong and H. Zhai, ``DOA estimation for coprime linear arrays: An ambiguity-free method involving full DOFs,” \emph{IEEE Commun. Lett.}, vol. 12, no. 3, pp. 562--565, Dec.\ 2018.
		
		\bibitem{Qin2017Frequency}
		S.~Qin, Y.~D.\ Zhang, M.~G.\ Amin and F.~Gini, ``Frequency diverse coprime arrays with coprime frequency offsets for multitarget localization,'' \emph{IEEE J. Sel. Top. Signal Process.}, vol.\ 11, no.\ 2, pp.\ 321--335, Mar.\ 2017.
		
		\bibitem{Ni2021Range}
		T. Ni, S. Liu, Z. Mao and Y. Huang, ``Range-dependent beamforming using space-frequency virtual difference coarray," in \emph{Proc. IEEE Radar	Conf.}, Atlanta, GA, USA, May 2021, pp. 1--6.
		
		\bibitem{Vaidyanathan2011Theory}
		P.~P.\ Vaidyanathan and P.\ Pal, ``Theory of sparse coprime sensing in multiple dimensions,'' \emph{IEEE Trans. Signal Process.}, vol.\ 59, no.\ 8, pp.\ 3592--3608, Apr.\ 2011.
		
		\bibitem{BouDaher2015Multi}
		E.~BouDaher, Y.~Jia, F.~Ahmad and M.~G.\ Amin, ``Multi-frequency coprime arrays for high-resolution direction-of-arrival estimation," \emph{IEEE Trans. Signal Process.}, vol.\ 63, no.\ 14, pp.\ 3797--3808, Jul.\ 2015.
		
		\bibitem{misc}
		Z.~Zheng, W.-Q.~Wang, Y.~Kong, and Y.~D.~Zhang, ``MISC Array: A new sparse array design achieving increased degrees of freedom and reduced mutual coupling effect," \emph{IEEE Trans. Signal Process.}, vol.\ 67, no.\ 7, pp.\ 1728--1741, April 2019.
		
		\bibitem{Cohen2020Sparse}
		R.~Cohen and Y.~C.\ Eldar, ``Sparse array design via fractal geometries,'' \emph{IEEE Trans.\ Signal Process.}, vol.\ 68, pp.\ 4797--4812, Aug.\ 2020.
		
		\bibitem{Liu2016Coprime}
		C.-L.\ Liu, P.~P.\ Vaidyanathan and P.~Pal, ``Coprime coarray interpolation for DOA estimation via nuclear norm minimization,'' in \emph{Proc. IEEE Int. Symp. Circuits Syst. (ISCAS)}, Montr\'eal, Canada, May.\ 2016, pp.\ 2639--2642.
		
		\bibitem{Qiao2017Unified}
		H.~Qiao and P.~Pal, ``Unified analysis of co-array interpolation for direction-of-arrival estimation,'' in \emph{Proc. IEEE Int. Conf. Acoust. Speech Signal Process. (ICASSP)}, New Orleans, LA, Mar.\ 2017, pp.\ 3056--3060.
		
		\bibitem{Yang2016Enhancing}
		Z.~Yang and L.~Xie, ``Enhancing sparsity and resolution via reweighted atomic norm minimization,'' \emph{IEEE Trans. Signal Process.}, vol.\ 64, no.\ 4, pp.\ 995--1006, Feb.\ 2016.
		
		\bibitem{Zhou2018Direction}
		C.~Zhou, Y.~Gu, X.~Fan, Z.~Shi, G.~Mao and Y.~D.\ Zhang, ``Direction-of-arrival estimation for coprime array via virtual array interpolation,'' \emph{IEEE Trans. Signal Process.}, vol.\ 66, no.\ 22, pp.\ 5956--5971, Sept.\ 2018.
		
		\bibitem{Zheng2020DoA}
		Z.~Zheng, Y.~Huang, W.-Q.\ Wang and H.~C.\ So, ``Direction-of-arrival estimation of coherent signals via coprime array interpolation,'' \emph{IEEE Signal Process. Lett.}, vol.\ 27, pp.\ 585--589, Mar.\ 2020.
		
		
		\bibitem{Schmidt1986Multiple}
		R.~Schmidt, ``Multiple emitter location and signal parameter estimation," \emph{IEEE Trans. Antennas Propaga.,} vol.\ 34, no.\ 3, pp.\ 276--280, Mar.\ 1986.
		
		\bibitem{Liu2021Rank}
		S.~Liu, Z.~Mao, Y.~D.\ Zhang and Y.~Huang, ``Rank minimization-based Toeplitz reconstruction for DoA estimation using coprime array,'' \emph{IEEE Commun. Lett.}, vol.\ 25, no.\ 7, pp.\ 2265--2269, July 2021.
		
		\bibitem{Daniel2019Low}
		D.~Castanheira and A.~Gameiro, ``Low complexity and high-resolution line spectral estimation using cyclic minimization,'' \emph{IEEE Trans. Signal Process.}, vol.\ 67, no.\ 24, pp.\ 6285--6300, Dec.\ 2019.
		
		\bibitem{Tian2017Low}
		Z.~Tian, Z.~Zhang and Y.~Wang, ``Low-complexity optimization for two-dimensional direction-of-arrival estimation via decoupled atomic norm minimization,'' in  \emph{Proc. IEEE Int. Conf. Acoust., Speech Signal Process. (ICASSP)}, New Orleans, LA, USA, Mar.\ 2017, pp.\ 3071--3075.
		
		\bibitem{Yang2016Vandermonde}
		Z.~Yang, L.~Xie and P.~Stoica, ``Vandermonde decomposition of multilevel Toeplitz matrices with application to multidimensional super-Resolution,''  \emph{IEEE Trans. Inf. Theory}, vol.\ 62, no.\ 6, pp.\ 3685--3701, Jun.\ 2016.	
		
		\bibitem{Lu2020Efficient}
		A.~Lu, Y.~Guo, N.~Li and S.~Yang. "Efficient gridless 2-D Direction-of-Arrival estimation for coprime array based on decoupled atomic norm minimization." \emph{IEEE Access}, vol.\ 8, pp.\ 57786--57795, Mar.\ 2020.
		
		
	    \bibitem{Wang2020Efficient}
		Y.~Wang, Y.~Zhang, Z.~Tian, G.~Leus and G.~Zhang, ``Efficient super-resolution two-dimensional harmonic retrieval via enhanced low-rank structured covariance reconstruction,'' in  \emph{Proc. IEEE Int. Conf. Acoust., Speech Signal Process. (ICASSP)}, Barcelona, ES, May.\ 2020, pp.\ 5720--5724.
	
		
		
		\bibitem{Cai2010Singular}
		J.~Cai, E.~Cand\'es and Z.~Shen, ``A singular value thresholding algorithm
		for matrix completion," \emph{SIAM J. Optim.}, vol.\ 20, no.\ 4, pp.\ 1956–1982,
		Mar.\ 2010.
		
	
		\bibitem{Nested2012Piya}
		P.~Pal and  P.~P.\ Vaidyanathan, ``Nested arrays in two dimensions, part II: Application in two dimensional array processing,'' \emph{IEEE Trans. Signal Process.}, vol.\ 60, no.\ 9, pp.\ 4706--4718, Sept.\ 2012.
	
		\bibitem{Almost2002Liu}
		X.~Liu and  N.~D.\ Sidiropoulos, ``Almost sure identifiability of constant modulus multidimensional harmonic retrieval,'' \emph{IEEE Trans. Signal Process.}, vol.\ 50, no.\ 9, pp.\ 4706--4718, Sept.\ 2002.
	
		\bibitem{Grippo2000On}
		L.~Grippo and M.~Sciandrone, ``On the convergence of the block nonlinear Gauss–-Seidel method under convex constraints,'' \emph{Oper. Res. Lett.}, vol.\ 26, no.\ 3, pp.\ 127--136, Apr.\ 2000.
		
		\bibitem{cvx}
		M.~Grant and S.~Boyd, \emph{CVX: MATLAB Software for Disciplined Convex Programming, Version 2.1}. Available at
		{http://cvxr.com/cvx}.
		
		
		
		\bibitem{Liu2016Cramer}
		C.-L.\ Liu and P.~P.\ Vaidyanathan, ``{Cram\'er-Rao} bounds for coprime and other sparse arrays, which find more sources than sensors,'' \emph{Digital Signal Process.}, vol.\ 61, pp.\ 43--61, Feb.\ 2016.
		
		\bibitem{Pakrooh2015Analysis}
		P.~Pakrooh, A.~Pezeshki, L.~L.\ Scharf, D.~ Cochran and S.~D.\ Howard, ``Analysis of Fisher information and the {Cram\'er-–Rao} bound for nonlinear parameter estimation afte random compression,'' \emph{IEEE Trans Signal Process.}, vol.\ 63, no.\ 23, pp.\ 6423--6428, Dec.\ 2015.
		
		\bibitem{Wang2017Coarrays}
		M.~Wang and A.~Nehorai, ``Coarrays, MUSIC, and the {Cram\'er-Rao} bound,'' \emph{IEEE Trans Signal Process.}, vol.\ 65, no.\ 4, pp.\ 933--946, Feb.\ 2017.
		
		
		\bibitem{Boyd2010Distributed}
		S.~Boyd, N.~Parikh, E.~Chu, B.~Peleato and J.~Eckstein, ``Distributed optimization and statistical learning via the alternating direction method of multipliers," \emph{Found. Trends Mach. Learn.}, vol.\ 3, no.\ 1, pp. 1--122, 2010.
		
		\bibitem{Wei2020Gridless}
		Z.~Wei, W.~Wang, F. ~Dong and Q.~ Liu, ``Gridless one-bit direction-of-arrival estimation via atomic norm denoising," \emph{IEEE Commun. Lett.}, vol.\ 24, no.\ 10, pp. 2177-2181, Oct.\ 2020.
		
		\bibitem{Qiao2017Gridless}
		H.~Qiao and P.~Pal, ``Gridless line spectrum estimation and low-rank Toeplitz matrix compression using structured samplers: A regularization-free approach,'' \emph{IEEE Trans Signal Process.}, vol.\ 65, no.\ 9, pp.\ 2221--2236, May\ 2017.
	
		
	\end{thebibliography}

\end{document}


\title{Joint DoA-Range Estimation Using Space-Frequency Virtual Difference Coarray: \\ Supplement File
	}
	
	\author{Zihuan~Mao,~\IEEEmembership{Student Member,~IEEE},
	Shengheng~Liu,~\IEEEmembership{Senior Member,~IEEE},\\
	Yimin~D.\ Zhang,~\IEEEmembership{Fellow,~IEEE}, 
	Leixin~Han,~\IEEEmembership{Student Member,~IEEE},
	Yongming~Huang,~\IEEEmembership{Senior Member,~IEEE}}

	\markboth{IEEE Transactions on Signal Processing,~Vol.~XX, No.~XX, XXX~2021}%
	{MAO \MakeLowercase{\textit{et al.}}: Joint DoA-Range Estimation Using Space-Frequency Virtual Difference Coarray}
	
\maketitle
In this supplement file, we provide the closed-form solutions of ADMM for the SDP problem (33), the simulation result with respect to identifiability, the proof of \textit{Proposition 3}, the derivation of reconstruction performance and the derivation of the coarray CRB.



\subsection{Closed-Form Solutions of ADMM Approach}
\enlargethispage{-5.5cm}
\noindent\begin{picture}(0,0)
\put(0,-380){\begin{minipage}{\textwidth}
	\centering
	\hrulefill
	\newcounter{MYtempeqncnt}
	\setcounter{MYtempeqncnt}{\value{equation}}
	\setcounter{equation}{57}
	\begin{align}\label{alpha}
	&\boldsymbol{\alpha}\left(\mathbf{W}_{\rm{p}}\right)=\nonumber \\
	&\begin{bmatrix}		{{\rm{tr}}\left(\mathbf{W}_{\rm{p}}\right)}&{{\rm{tr}}\left(\left[\mathbf{W}^{\rm{T}}_{\rm{p}}\right]_{2L+1,1}\right)}&{{\rm{tr}}\left(\left[\left[\mathbf{W}^{\rm{T}}_{\rm{p}}\right]_{1,2L+1}\right]_{2L,2}\right)}&{\cdots}&\left(\mathbf{W}_{\rm{p}}\right)_{2L+1,1}\\
	{{\rm{tr}}\left(\left[\mathbf{W}^{\rm{T}}_{\rm{p}}\right]_{1,2L+1}\right)}&{{\rm{tr}}\left(\mathbf{W}_{\rm{p}}\right)}&{\ddots}&{\ddots}&\vdots\\
	{{\rm{tr}}\left(\left[\left[\mathbf{W}^{\rm{T}}_{\rm{p}}\right]_{1,2L+1}\right]_{2,2L}\right)}&\ddots&\ddots&\ddots&{{\rm{tr}}\left(\left[\left[\mathbf{W}^{\rm{T}}_{\rm{p}}\right]_{1,2L+1}\right]_{2L,2}\right)}\\
	\vdots&\ddots&\ddots&\ddots&{{\rm{tr}}\left(\left[\mathbf{W}^{\rm{T}}_{\rm{p}}\right]_{2L+1,1}\right)}\\
	\left(\mathbf{W}_{\rm{p}}\right)_{1,2L+1}&\cdots&{{\rm{tr}}\left(\left[\left[\mathbf{W}^{\rm{T}}_{\rm{p}}\right]_{1,2L+1}\right]_{2,2L}\right)}&{{\rm{tr}}\left(\left[\mathbf{W}^{\rm{T}}_{\rm{p}}\right]_{1,2L+1}\right)}&{{\rm{tr}}\left(\mathbf{W}_{\rm{p}}\right)}
	\end{bmatrix}.
	\end{align}
	\setcounter{equation}{\value{MYtempeqncnt}}
	\end{minipage}}
\end{picture}

%

The closed-form solutions can be obtained by the derivatives with respect to each optimization variable\cite{Boyd2010Distributed}. Concretely, the closed-form solution of one particular variable is computed by fixing the others. Hence, we first derive the solution of $\mathbf{T}\left(\mathbf{z}_{\rm{p}}\right)$. Denote $\left\|\underline{\mathbf{C}}-\begin{bmatrix}
{\bf{T({\mathbf{z}}_{\rm{p}}}}_{\rm}) & {\mathbf{X}_{\rm{v}}}    \\
{\mathbf{X}}_{\rm{v}}^{\rm H} & {\bf{T({\bf{z}}_{\rm{f}}}})
\end{bmatrix}\!+\!\rho^{-1}\underline{\mathbf{R}}\right\|_{\mathcal{F}}^2$ as ${\rm{g}}(\widetilde{\mathbf{M}})$. According to the chain rule and the derivative of a Toeplitz matrix, we have
\setcounter{equation}{56}
\begin{align}
\frac{\partial\mathcal{L}\left(\mathbf{T}\left({\mathbf{z}_{\rm{p}}}\right)\right)}{\partial\mathbf{T}\left({\mathbf{z}_{\rm{p}}}\right)}&=\frac{2\partial{\rm{tr}}\left[\mathbf{W}_{\rm{p}}\mathbf{T}\left(\mathbf{z}_{\rm{p}}\right)\right]}{\partial \mathbf{T}\left(\mathbf{z}_{\rm{p}}\right)}+\frac{\rho}{2}\cdot\frac{\partial\rm{g}(\widetilde{\mathbf{M}})}{\partial \mathbf{T}\left(\mathbf{z}_{\rm{p}}\right)}\nonumber \\
&=2\boldsymbol{\alpha}\left(\mathbf{W}_{\rm{p}}\right)+\rho\cdot{\rm{tr}}\left[\left(\widetilde{\mathbf{M}}\right)^{\rm{T}}\cdot\frac{\partial\widetilde{\mathbf{M}}}{\partial\mathbf{T}\left(\mathbf{z}_{\rm{p}}\right)}\right],
\end{align}
where $\boldsymbol{\alpha}\left(\mathbf{W}_{\rm{p}}\right)$ is given in \eqref{alpha}, shown on the bottom of this page. The second term of (57) can be reformulated as
\setcounter{equation}{58}
\begin{align}
&\rho\cdot{\rm{tr}}\left[\left(\widetilde{\mathbf{M}}\right)^{\rm{T}}\cdot\frac{\partial\widetilde{\mathbf{M}}}{\partial\left(\mathbf{T}\left(\mathbf{z}_{\rm{p}}\right)\right)_{i,j}}\right]\nonumber \\
&=\rho\cdot{\rm{tr}}\left[\left(\underline{\mathbf{C}}^{\rm{T}}\!-\!\begin{bmatrix}
{\mathbf{T}^{\rm{T}}\left(\mathbf{z}_{\rm{p}}\right)}&{\mathbf{X}^{*}_{\rm{v}}}\\
{\mathbf{X}^{\rm{T}}_{\rm{v}}}&{\mathbf{T}^{\rm{T}}\left(\mathbf{z}_{\rm{f}}\right)}
\end{bmatrix}\!+\!\rho^{-1}\underline{\mathbf{R}}^{\rm{T}}\right)\cdot\mathring{\mathbf{1}}\right],
\end{align}

where $\mathring{\mathbf{1}}=\begin{bmatrix}
{\dot{\mathbf{1}}_{i,j}}&{\mathbf{0}}\\
{\mathbf{0}}&{\mathbf{0}}
\end{bmatrix}$, and $\dot{\mathbf{1}}_{i,j}$ represents a matrix in which the entries on the $\left(i-j\right)$-th diagonal are 1 whereas the others are 0. 
We further rewrite $\underline{\mathbf{C}}$ and $\underline{\mathbf{R}}$ as
$\underline{\mathbf{C}} = \begin{bmatrix}
{\underline{\mathbf{C}}_1}&{\underline{\mathbf{C}}_2}\\
{\underline{\mathbf{C}}_3}&{\underline{\mathbf{C}}_4}
\end{bmatrix}$
and
$\underline{\mathbf{R}}=\begin{bmatrix}
{\underline{\mathbf{R}}_1}&{\underline{\mathbf{R}}_2}\\
{\underline{\mathbf{R}}_3}&{\underline{\mathbf{R}}_4}
\end{bmatrix}$. Then, (54) can be reduced to

\begin{align}
&\rho\cdot{\rm{tr}}\left[\left(\underline{\mathbf{C}}^{\rm{T}}-\begin{bmatrix}
{\mathbf{T}^{\rm{T}}\left(\mathbf{z}_{\rm{p}}\right)}&{\mathbf{X}^{*}_{\rm{v}}}\\
{\mathbf{X}^{\rm{T}}_{\rm{v}}}&{\mathbf{T}^{\rm{T}}\left(\mathbf{z}_{\rm{f}}\right)}
\end{bmatrix}+\rho^{-1}\underline{\mathbf{R}}^{\rm{T}}\right)\cdot\mathring{\mathbf{1}}\right]\nonumber\\
&=\rho\cdot\left(\underline{\mathbf{C}}_1^{\rm{T}}-\mathbf{T}^{\rm{T}}\left(\mathbf{z}_{\rm{p}}\right)+\rho^{-1}\underline{\mathbf{R}}_1^{\rm{T}}\right)_{\Sigma\left(i-j\right)},
\end{align}
where the operation $\left(\cdot\right)_{\Sigma\left(i-j\right)}$ sums up the $\left(i-j\right)$-th diagonal. Finally, when (57) equals to zero, $\mathbf{T}\left(\mathbf{z}_{\rm{p}}\right)$ can be obtained by its entries as
\begin{align}
&\left(\mathbf{T}\left(\mathbf{z}_{\rm{p}}\right)\right)_{i,j}=\nonumber \\
&\frac{1}{2L+1-|i-j|}\left[\left(\underline{\mathbf{C}}_1+\rho^{-1}\underline{\mathbf{R}}_1\right)_{\Sigma\left(i-j\right)}-\frac{2\boldsymbol{\alpha}^{\rm{T}}\left(\mathbf{W}_{\rm{p}}\right)}{\rho}\right].
\end{align}

Similarly, a closed-form solution of $\mathbf{T}\left(\mathbf{z}_{\rm{f}}\right)$ is obtained as
\begin{align}
&\left(\mathbf{T}\left(\mathbf{z}_{\rm{f}}\right)\right)_{i,j}=\nonumber \\
&\frac{1}{2L+1-|i-j|}\left[\left(\underline{\mathbf{C}}_4+\rho^{-1}\underline{\mathbf{R}}_4\right)_{\Sigma\left(i-j\right)}-\frac{2\boldsymbol{\alpha}^{\rm{T}}\left(\mathbf{W}_{\rm{f}}\right)}{\rho}\right].
\end{align}

Next, we derive a closed-form solution for $\mathbf{X}_{\rm{v}}$. Likewise, we denote $\left\|\overline{\mathbf{C}}-\begin{bmatrix}
{\mu\mathbf{I}} & {\mathbf{X}_{\rm{v}}\circ\mathbf{B}-\tilde{\mathbf{X}}_{\rm{v}}}    \\
{\mathbf{X}^{\rm{H}}_{\rm{v}}\circ\mathbf{B}-\tilde{\mathbf{X}}^{\rm{H}}_{\rm{v}}} & {\mu\mathbf{I}}
\end{bmatrix}\!+\!\rho^{-1}\overline{\mathbf{R}}\right\|_{\mathcal{F}}^2$ as ${\rm{g}}(\widetilde{\mathbf{N}})$. Then, the derivative of $\mathbf{X}_{\rm{v}}$ can be written as 
\vfill\eject
\begin{align}	&\frac{\partial\mathcal{L}\left(\mathbf{X}_{\rm{v}}\right)}{\partial\mathbf{X}_{\rm{v}}}=\frac{\rho}{2}\cdot\frac{\partial\rm{g}(\widetilde{\mathbf{M}})}{\partial \mathbf{X}_{\rm{v}}}+\frac{\rho}{2}\cdot\frac{\partial\rm{g}(\widetilde{\mathbf{N}})}{\partial \mathbf{X}_{\rm{v}}}\nonumber \\
&=\rho\cdot\left\{{\rm{tr}}\left[\left(\widetilde{\mathbf{N}}\right)^{\rm{T}}\cdot\frac{\partial\widetilde{\mathbf{N}}}{\partial \mathbf{X}_{\rm{v}}}\right]+{\rm{tr}}\left[\left(\widetilde{\mathbf{M}}\right)^{\rm{T}}\cdot\frac{\partial\widetilde{\mathbf{M}}}{\partial \mathbf{X}_{\rm{v}}}\right]\right\}.
\end{align}
Again, we rewrite $\overline{\mathbf{C}}$ and $\overline{\mathbf{R}}$ respectively  as
$\overline{\mathbf{C}}=\begin{bmatrix}
{\overline{\mathbf{C}}_1}&{\overline{\mathbf{C}}_2}\\
{\overline{\mathbf{C}}_3}&{\overline{\mathbf{C}}_4}
\end{bmatrix}$
and
$\overline{\mathbf{R}}=\begin{bmatrix}
{\overline{\mathbf{R}}_1}&{\overline{\mathbf{R}}_2}\\
{\overline{\mathbf{R}}_3}&{\overline{\mathbf{R}}_4}
\end{bmatrix}$. Then, (63) is further expressed as
\begin{align}
&{\rm{tr}}\left(\begin{bmatrix}
{\mathbf{0}}&{\mathbf{1}_{i,j}}\\
{\mathbf{1}_{j,i}}&{\mathbf{0}}
\end{bmatrix}\cdot\right.\nonumber \\
&\left.\begin{bmatrix}
{\underline{\mathbf{C}}_1-\mathbf{T\left(\mathbf{z}_{\rm{p}}\right)}+\rho^{-1}\underline{\mathbf{R}}_1}&{\underline{\mathbf{C}}_2-\mathbf{X}_{\rm{v}}+\rho^{-1}\underline{\mathbf{R}}_2}\\
{\underline{\mathbf{C}}_3-\mathbf{X}^{\rm{H}}_{\rm{v}}+\rho^{-1}\underline{\mathbf{R}}_3}&{\underline{\mathbf{C}}_4-\mathbf{T\left(\mathbf{z}_{\rm{p}}\right)}+\rho^{-1}\underline{\mathbf{R}}_4}
\end{bmatrix}\right)\nonumber \\
&={\rm{tr}}\left(\begin{bmatrix}
{\mathbf{0}}&{-\mathbf{B}_{i,j}}\\
{-\mathbf{B}_{j,i}}&{\mathbf{0}}
\end{bmatrix}\right.\cdot\nonumber \\
&\left.\begin{bmatrix}
{\overline{\mathbf{C}}_1\!-\!\mu \mathbf{I}\!+\!\rho^{-1}\overline{\mathbf{R}}_1}\!&\!{\overline{\mathbf{C}}_2\!-\!\mathbf{X}_{\rm{v}}\circ\mathbf{B}\!-\!\tilde{\mathbf{X}}_{\rm{v}}\!+\!\rho^{-1}\overline{\mathbf{R}}_2}\\
{\overline{\mathbf{C}}_3\!-\!\mathbf{X}^{\rm{H}}_{\rm{v}}\circ\mathbf{B}\!-\!\tilde{\mathbf{X}}^{\rm{H}}_{\rm{v}}\!+\!\rho^{-1}\overline{\mathbf{R}}_3}\!&\!{\overline{\mathbf{C}}_4\!-\!\mu \mathbf{I}\!+\!\rho^{-1}\overline{\mathbf{R}}_4}
\end{bmatrix}\right),
\end{align}
where $\mathbf{1}_{i,j}$ and $\mathbf{B}_{i,j}$ represent two selective matrices with the $(i,j)$-th entries respectively equal to 1 and $B_{i,j}$ while the others are zero. Notice that, ${\rm{tr}}\left(\mathbf{1}_{i,j}\cdot\mathbf{A}\right)=A_{j,i}$ and, thus, we have
\begin{align}
&\left(\underline{\mathbf{C}}_{2}-\mathbf{X}_{\rm{v}}+\rho^{-1}\underline{\mathbf{R}}_2+\underline{\mathbf{C}}_{3}-\mathbf{X}^{\rm{H}}_{\rm{v}}+\rho^{-1}\underline{\mathbf{R}}_3\right)_{i,j}\nonumber \\
&=-\left(\overline{\mathbf{C}}_2-\mathbf{X}_{\rm{v}}\circ\mathbf{B}-\tilde{\mathbf{X}}_{\rm{v}}+\rho^{-1}\overline{\mathbf{R}}_2\right.\nonumber \\
&\qquad\qquad\qquad\left.+\overline{\mathbf{C}}_3-\mathbf{X}^{\rm{H}}_{\rm{v}}\circ\mathbf{B}-\tilde{\mathbf{X}}^{\rm{H}}_{\rm{v}}+\rho^{-1}\overline{\mathbf{R}}_3\right)_{i,j}.
\end{align}

According to the augmented Lagrangian (33), we readily observe that $\overline{\mathbf{C}}_2=\overline{\mathbf{C}}_3^{\rm{H}}$, $\underline{\mathbf{C}}_2=\underline{\mathbf{C}}_3^{\rm{H}}$,$\overline{\mathbf{R}}_2=\overline{\mathbf{R}}_3^{\rm{H}}$ and $\underline{\mathbf{R}}_2=\underline{\mathbf{R}}_3^{\rm{H}}$. As such, the closed-form solution of $\mathbf{X}_{\rm{v}}$ is given by
\begin{align}
&\mathbf{X}_{\rm{v}}=\left[\left(\underline{\mathbf{C}}_2+\rho^{-1}\underline{\mathbf{R}}_2+\overline{\mathbf{C}}_2+\rho^{-1}\overline{\mathbf{R}}_2-\tilde{\mathbf{X}}_{\rm{v}}\right)\right.\nonumber \\
&+\left.\left(\underline{\mathbf{C}}_3+\rho^{-1}\underline{\mathbf{R}}_3+\overline{\mathbf{C}}_3+\rho^{-1}\overline{\mathbf{R}}_3-\tilde{\mathbf{X}}^{\rm{H}}_{\rm{v}}\right)^{\rm{H}}\right]\circ\left(\mathbf{1}+\mathbf{B}\right)^{-},
\end{align}
where $(\left(\mathbf{1}+\mathbf{B}\right)^{-})_{i,j}=\frac{1}{\left(\mathbf{1}+\mathbf{B}\right)_{i,j}}$.
Finally, the derivatives with respect to $\overline{\mathbf{C}}$ and $\underline{\mathbf{C}}$ are respectively derived as
\begin{align}
&\left(\frac{\partial\mathcal{L}\left(\overline{\mathbf{C}}\right)}{\partial\overline{\mathbf{C}}}\right)_{i,j}=\nonumber \\
&\rho\left(\left[\overline{\mathbf{C}}-\begin{bmatrix}
{\mu\mathbf{I}} & {\mathbf{X}_{\rm{v}}\circ\mathbf{B}\!-\!\tilde{\mathbf{X}}_{\rm{v}}}    \\
{\mathbf{X}^{\rm{H}}_{\rm{v}}\circ\mathbf{B}\!-\!\tilde{\mathbf{X}}^{\rm{H}}_{\rm{v}}} & {\mu\mathbf{I}}
\end{bmatrix}\!+\!\rho^{-1}\overline{\mathbf{R}}\right]\right)_{j,i},
\end{align}
\begin{align}
\left(\frac{\partial\mathcal{L}\left(\underline{\mathbf{C}}\right)}{\partial\underline{\mathbf{C}}}\right)_{i,j}\!=\!\rho\left(\left[\underline{\mathbf{C}}\!-\!\begin{bmatrix}
{\bf{T({\mathbf{z}}_{\rm{p}}}}_{\rm}) & {\mathbf{X}_{\rm{v}}}    \\
{\mathbf{X}}_{\rm{v}}^{\rm H} & {\bf{T({\bf{z}}_{\rm{f}}}})
\end{bmatrix}\!+\!\rho^{-1}\underline{\mathbf{R}}\right]\right)_{j,i}.
\end{align}
Additionally, taking into account the similarities among the solutions of such problems \cite{Wei2020Gridless}, the solutions of $\overline{\mathbf{C}}$ and $\underline{\mathbf{C}}$ are respectively derived as
\begin{align}
\overline{\mathbf{C}}=\left(\begin{bmatrix}
{\mu\mathbf{I}} & {\mathbf{X}_{\rm{v}}\circ\mathbf{B}-\tilde{\mathbf{X}}_{\rm{v}}}    \\
{\mathbf{X}^{\rm{H}}_{\rm{v}}\circ\mathbf{B}-\tilde{\mathbf{X}}^{\rm{H}}_{\rm{v}}} & {\mu\mathbf{I}}
\end{bmatrix}\!-\!\rho^{-1}\overline{\mathbf{R}}\right)_{+},
\end{align}
\begin{align}
\underline{\mathbf{C}}=\left(\begin{bmatrix}
{\bf{T({\mathbf{z}}_{\rm{p}}}}_{\rm}) & {\mathbf{X}_{\rm{v}}}    \\
{\mathbf{X}}_{\rm{v}}^{\rm H} & {\bf{T({\bf{z}}_{\rm{f}}}})
\end{bmatrix}-\rho^{-1}\underline{\mathbf{R}}\right)_{+},
\end{align}
where the operator $\left(\cdot\right)_{+}$ computes the eigen-decomposition of the matrix and let the negative eigenvalues be zero to guarantee PSD. $\hfill \blacksquare$

	\vspace{-0.5em}
	\subsection{Comparison of Identifiability}
	
	To further validate the approximate identifiability of the proposed  interpolation-based framework and compare it to the uniform linear coarray case, we investigate a non-uniform physical array which is associated with a hole-free uniform linear coarray. In particular, we place receive sensors at position 11$d$ to form a different non-uniform configuration compared to the FDCA depicted in Fig.\;1. Correspondingly, the holes at $\pm 8d$ and $\pm 11d$ in the coarray are respectively padded by the correlation of ($3d,11d$) and ($0d,11d$). This uniform linear coarray serves as a reference in the comparison of the identifiability. Similar with the parameter setting used in Figs.\;3(m)--3(o), we first assume that 132 targets are distributed in 11 azimuth angles $\boldsymbol{\theta} \in [-60^{\circ},60^{\circ}]$ and 12 ranges $\mathbf{r} \in [300\;{\rm m},4700\;{\rm m}]$ with respective uniform intervals $\Delta\theta=12^{\circ}$ and $\Delta r=400\;\text{m}$. The DoAs and ranges of the 132 targets are randomly generated for 1,000 Monte Carlo trials from their respective Gaussian distributions $\tilde{\boldsymbol{\theta}}\sim\mathcal{N}({\boldsymbol{\theta}},(1^{\circ})^2)$ and $\tilde{\mathbf{r}}\sim\mathcal{N}(\mathbf{r}, (10 \ \rm{m})^2)$. Then, we compare the identifiability of the interpolated coarray, the coarray without interpolation, and the uniform linear coarray in the 132-target case. The cumulative distribution function (CDF) curves of the number of unidentified targets are compared in Fig.\;\ref{addFig1}.
	
	\begin{figure}[!hpt]
		\newcounter{TempEqCnt}                         
		\setcounter{TempEqCnt}{\value{figure}} 
		\setcounter{figure}{8}                           
		\centering	
		{\includegraphics[scale=0.5]{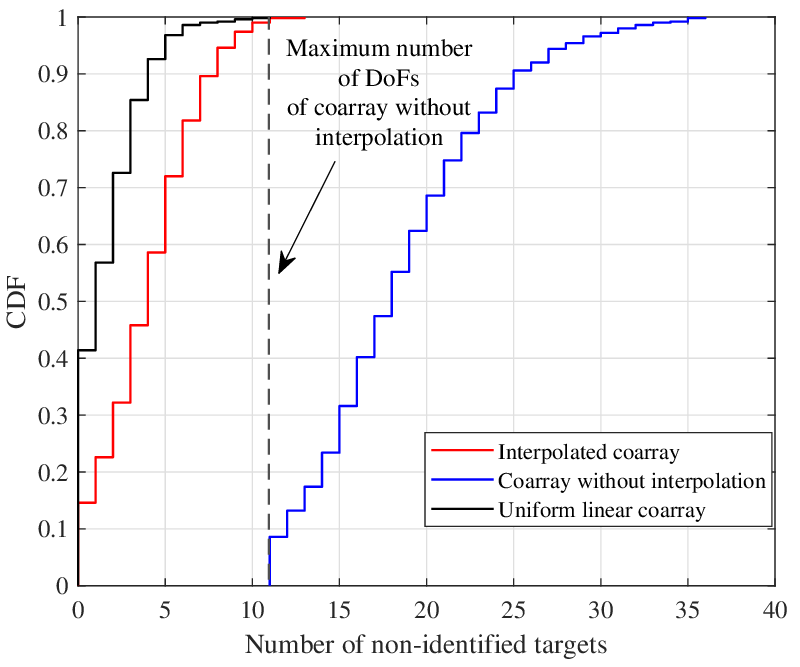}}
		\caption{Comparison of identifiability for different types of coarray.}
		\label{addFig1}
	\end{figure}	
	The following facts can be readily observed from Fig.\;\ref{addFig1}.
	\begin{itemize}
	
	\item As expected, the uniform linear coarray has the best performance with a 41.4\% probability to identify all 132 targets. For the worst case, the CDF curve of the uniform linear coarray terminates at 11, which indicates that at least 121 targets are identified in all the Monte Carlo trials. 
	
	\item For the interpolation framework, the probability of identifying all the targets is 14.6\%, which implies that the extra DoFs of the interpolated coarray sensors can be potentially obtained. Besides, the CDF curve of the interpolated coarray terminates at 13, which is slightly worse than that of the uniform linear coarray. 
	
	\item As for the coarray without interpolation, the starting point of the corresponding CDF curve is 11. This phenomenon agrees with our common sense that the maximum number of DoFs of the coarray without interpolation cannot exceed the number of the non-negative unique lags. Besides, due to the grid mismatch of the CMT-BCS, the maximal number of unidentified targets reaches 36. The missing lags and the grid mismatch conspire the worst identifiability of the coarray without interpolation. 

	\end{itemize}
	
	These results validate the extra number of DoFs available to improve the identifiability using the interpolation compared to the coarray without performing interpolation.
	
	\subsection{Proof of Proposition 3}
	From the definitions of the two group solutions  $\{\dot{\mathbf{X}}_{\rm{v}},\dot{\mathbf{z}}_{\rm{p}},\dot{\mathbf{z}}_{\rm{f}}\}$ and $\{\mathring{\mathbf{X}}_{\rm{v}},\mathring{\mathbf{z}}_{\rm{p}},\mathring{\mathbf{z}}_{\rm{f}}\}$, we respectively have 
	\setcounter{equation}{70}
	\begin{align}\label{ApB1}
	\text{f}[{\mathring{\bf{W}}}_{\rm{p}},{\bf{T}}({\mathring{{\bf{z}}}_{\rm{p}}}),\gamma_{\rm{p}}] +\text{f}[\mathring{{\bf{W}}}_{\rm{f}},{\bf{T}}(\mathring{{{\bf{z}}}_{\rm{f}}}),\gamma_{\rm{f}}]\le\nonumber \\
	\text{f}[{\dot{\bf{W}}}_{\rm{p}},{\bf{T}}({\dot{{\bf{z}}}_{\rm{p}}}),\gamma_{\rm{p}}] +\text{f}[\dot{{\bf{W}}}_{\rm{f}},{\bf{T}}(\dot{{{\bf{z}}}_{\rm{f}}}),\gamma_{\rm{f}}],
	\end{align}
	and 
	\begin{align}\label{ApB2}
	\text{f}[{\mathring{\bf{W}}}_{\rm{p}}&,{\bf{T}}({\mathring{{\bf{z}}}_{\rm{p}}}),\gamma_{\rm{p}}] +\text{f}[\mathring{{\bf{W}}}_{\rm{f}},{\bf{T}}(\mathring{{{\bf{z}}}_{\rm{f}}}),\gamma_{\rm{f}}]\nonumber \\
	&+2\gamma_{\rm{p}}^{-2}\nu_{\rm{p}}{\rm{tr}}[\mathbf{T}(\mathring{\mathbf{z}}_{\rm{p}})]+2\gamma_{\rm{f}}^{-2}\nu_{\rm{f}}{\rm{tr}}[\mathbf{T}(\mathring{\mathbf{z}}_{\rm{f}})]\ge\nonumber\\
	\text{f}[{\dot{\bf{W}}}_{\rm{p}}&,{\bf{T}}({\dot{{\bf{z}}}_{\rm{p}}}),\gamma_{\rm{p}}] +\text{f}[\dot{{\bf{W}}}_{\rm{f}},{\bf{T}}(\dot{{{\bf{z}}}_{\rm{f}}}),\gamma_{\rm{f}}]\nonumber \\
	&+2\gamma_{\rm{p}}^{-2}\nu_{\rm{p}}{\rm{tr}}[\mathbf{T}(\dot{\mathbf{z}}_{\rm{p}})]+2\gamma_{\rm{f}}^{-2}\nu_{\rm{f}}{\rm{tr}}[\mathbf{T}(\dot{\mathbf{z}}_{\rm{f}})].
	\end{align}
	The inequalities \eqref{ApB1} and \eqref{ApB2} lead to the following expression
	\begin{align}
	0\le \text{f}[{\mathring{\bf{W}}}_{\rm{p}},{\bf{T}}({\mathring{{\bf{z}}}_{\rm{p}}}),\gamma_{\rm{p}}] +\text{f}[\mathring{{\bf{W}}}_{\rm{f}},{\bf{T}}(\mathring{{{\bf{z}}}_{\rm{f}}}),\gamma_{\rm{f}}]-\nonumber \\
	(\text{f}[{\dot{\bf{W}}}_{\rm{p}},{\bf{T}}({\dot{{\bf{z}}}_{\rm{p}}}),\gamma_{\rm{p}}] +\text{f}[\dot{{\bf{W}}}_{\rm{f}},{\bf{T}}(\dot{{{\bf{z}}}_{\rm{f}}}),\gamma_{\rm{f}}])\nonumber \\
	\le2\gamma_{\rm{p}}^{-2}\nu_{\rm{p}}{\rm{tr}}[\mathbf{T}(\mathring{\mathbf{z}}_{\rm{p}})]+2\gamma_{\rm{f}}^{-2}\nu_{\rm{f}}{\rm{tr}}[\mathbf{T}(\mathring{\mathbf{z}}_{\rm{f}})].
	\end{align}
	When $2\gamma_{\rm{p}}^{-2}\nu_{\rm{p}}{\rm{tr}}[\mathbf{T}(\mathring{\mathbf{z}}_{\rm{p}})]+2\gamma_{\rm{f}}^{-2}\nu_{\rm{f}}{\rm{tr}}[\mathbf{T}(\mathring{\mathbf{z}}_{\rm{f}})]<1$, we have ${\rm{rank}}\left[\mathbf{T}(\mathring{\mathbf{z}}_{\rm{p}})\right]+{\rm{rank}}\left[\mathbf{T}(\mathring{\mathbf{z}}_{\rm{f}})\right]={\rm{rank}}\left[\mathbf{T}(\dot{\mathbf{z}}_{\rm{p}})\right]+{\rm{rank}}\left[\mathbf{T}(\dot{\mathbf{z}}_{\rm{f}})\right]$, which implies that the objective functions of CRM and perturbed CRM are equivalent.$\hfill \blacksquare$

	\subsection{Proof of Propositions 4 and 5}
	To derive the reconstruction error $	\|\hat{\mathbf{X}}_{\rm{v}}\circ\mathbf{B}-\dot{\mathbf{X}}_{\rm{v}}\circ\mathbf{B}\|^{2}_{\mathcal{F}}$, we first introduce the following lemma.
	
	\textit{Lemma 4}\cite{Qiao2017Gridless}: Let $\mathbf{x}(t),t=1,\cdots,T$, be zero mean i.i.d.\ Gaussian random vectors distributed as $\mathbf{x}(t)\sim\mathcal{N}(\mathbf{0},\mathbf{R}_{\mathbf{x}})$. Then,
	\begin{align}\label{ApA1}
	{\rm{P}}\left\{\|\mathbf{R}_{\mathbf{x}}-\hat{\mathbf{R}}_{\mathbf{x}}\|^2_{\mathcal{F}}\ge\frac{{\rm{tr}}(\mathbf{R}_{\mathbf{x}})}{\sqrt{T}}\right\}\le2{\rm{e}}^{-2C\sqrt{T}},
	\end{align}
	where $\rm{P}\{\cdot\}$ denotes the probability. 
	
	In Section II, the virtual coarray signal is initialized by selecting the corresponding cross-correlation in the estimated covariance matrix $\hat{\mathbf{R}}_{\mathbf{x}}$. Thus, by considering the element-wise property of the Frobenius norm, the relationship can be generalized from the physical domain to the virtual domain as
	\begin{align}\label{ApA2}
	{\rm{P}}\left\{\|\dot{\mathbf{X}}_{\rm{v}}\circ\mathbf{B}-\tilde{\mathbf{X}}_{\rm{v}}\|^2_{\mathcal{F}}\le\frac{{\rm{tr}}(\mathbf{R}_{\mathbf{x}})}{\sqrt{T}}\right\}\ge1-2{\rm{e}}^{-2C\sqrt{T}}.
	\end{align}
	Then, the regularization term can be written as
	\begin{align}\label{ApA3}
	\|\hat{\mathbf{X}}_{\rm{v}}\circ\mathbf{B}-\tilde{\mathbf{X}}_{\rm{v}}\|^2_{\mathcal{F}}=\|\hat{\mathbf{X}}_{\rm{v}}\circ\mathbf{B}+\dot{\mathbf{X}}_{\rm{v}}\circ\mathbf{B}-\dot{\mathbf{X}}_{\rm{v}}\circ\mathbf{B}-\tilde{\mathbf{X}}_{\rm{v}}\|^2_{\mathcal{F}}\nonumber \\
	=
	\|\hat{\mathbf{X}}_{\rm{v}}\circ\mathbf{B}-\dot{\mathbf{X}}_{\rm{v}}\circ\mathbf{B}\|^2_{\mathcal{F}}+\|\dot{\mathbf{X}}_{\rm{v}}\circ\mathbf{B}-\tilde{\mathbf{X}}_{\rm{v}}\|^2_{\mathcal{F}}\nonumber \\
	+2 \langle \hat{\mathbf{X}}_{\rm{v}}\circ\mathbf{B}-\dot{\mathbf{X}}_{\rm{v}}\circ\mathbf{B},\dot{\mathbf{X}}_{\rm{v}}\circ\mathbf{B}-\tilde{\mathbf{X}}_{\rm{v}} \rangle_{\mathcal{F}}.
	\end{align}
	From the definition of the optimal solution of (23), we have
	\begin{align}\label{ApA4}
	\frac{1}{2L}({\rm{tr}}[\mathbf{T}({\hat{\mathbf{z}}_{\rm{p}}})]+{\rm{tr}}[\mathbf{T}(\hat{{\mathbf{z}}_{\rm{f}}})])+\mu\|\hat{\mathbf{X}}_{\rm{v}}\circ\mathbf{B}-\tilde{\mathbf{X}}_{\rm{v}}\|^2_{\mathcal{F}}\nonumber \\
	\le \frac{1}{2L}({\rm{tr}}[\mathbf{T}({\dot{\mathbf{z}}_{\rm{p}}})]+{\rm{tr}}[\mathbf{T}(\dot{{\mathbf{z}}_{\rm{f}}})])+\mu\|\dot{\mathbf{X}}_{\rm{v}}\circ\mathbf{B}-\tilde{\mathbf{X}}_{\rm{v}}\|^2_{\mathcal{F}},
	\end{align}
	which implies
	\begin{align}\label{ApA5}
	&\|\hat{\mathbf{X}}_{\rm{v}}\circ\mathbf{B}-\tilde{\mathbf{X}}_{\rm{v}}\|^2_{\mathcal{F}}-\|\dot{\mathbf{X}}_{\rm{v}}\circ\mathbf{B}-\tilde{\mathbf{X}}_{\rm{v}}\|^2_{\mathcal{F}}\nonumber \\
	&\le \frac{1}{2\mu L}({\rm{tr}}[\mathbf{T}({\dot{\mathbf{z}}_{\rm{p}}})]-{\rm{tr}}[\mathbf{T}({\hat{\mathbf{z}}_{\rm{p}}})]+{\rm{tr}}[\mathbf{T}(\dot{{\mathbf{z}}_{\rm{f}}})]-{\rm{tr}}[\mathbf{T}({\hat{\mathbf{z}}_{\rm{f}}})]).
	\end{align}
	Furthermore, combining the Cauchy-Schwarz inequality on the Frobenius inner product term, we have
	\begin{align}\label{ApA6}
	| \langle \hat{\mathbf{X}}_{\rm{v}}&\circ\mathbf{B}-\dot{\mathbf{X}}_{\rm{v}}\circ\mathbf{B},\dot{\mathbf{X}}_{\rm{v}}\circ\mathbf{B}-\tilde{\mathbf{X}}_{\rm{v}}\rangle_{\mathcal{F}}|\nonumber \\
	&\le\|\hat{\mathbf{X}}_{\rm{v}}\circ\mathbf{B}-\dot{\mathbf{X}}_{\rm{v}}\circ\mathbf{B}\|_{\mathcal{F}}\|\dot{\mathbf{X}}_{\rm{v}}\circ\mathbf{B}-\tilde{\mathbf{X}}_{\rm{v}}\|_{\mathcal{F}}.
	\end{align}
	Without loss of generality, we consider the regularization parameter $\mu$ to satisfy
	\begin{align}\label{ApA7}
	\mu\ge\frac{\mathbf{R}_{\mathbf{x}}}{\sqrt{T}}=\frac{L^2}{\sqrt{T}}\left(\sum\nolimits_{k=1}^K p_k+\sigma^2_{\rm{n}}\right).
	\end{align}
	Let $\xi$ denote the reconstruction term $\|\hat{\mathbf{X}}_{\rm{v}}\circ\mathbf{B}-\dot{\mathbf{X}}_{\rm{v}}\circ\mathbf{B}\|_{\mathcal{F}}$. By considering the PSD-ness of the Toeplitz matrices $\mathbf{T}(\hat{\mathbf{z}}_{\rm{p}})$ and $\mathbf{T}(\hat{\mathbf{z}}_{\rm{f}})$ and combining \eqref{ApA3} with \eqref{ApA5}, \eqref{ApA6}, \eqref{ApA7}, we have
	\begin{align}
	&\left(\xi-\mu-\sqrt{\mu^2+\frac{1}{2\mu L}\left({\rm{tr}}[\mathbf{T}({\dot{\mathbf{z}}_{\rm{p}}})]+{\rm{tr}}[\mathbf{T}(\dot{{\mathbf{z}}_{\rm{f}}})]\right)}\right)\nonumber \\
	&\cdot \left(\xi-\mu+\sqrt{\mu^2+\frac{1}{2\mu L}\left({\rm{tr}}[\mathbf{T}({\dot{\mathbf{z}}_{\rm{p}}})]+{\rm{tr}}[\mathbf{T}(\dot{{\mathbf{z}}_{\rm{f}}})]\right)}\right)\le 0.
	\end{align}	
	Thus, we have 
	\begin{align}
	&\|\hat{\mathbf{X}}_{\rm{v}}\circ\mathbf{B}-\dot{\mathbf{X}}_{\rm{v}}\circ\mathbf{B}\|^{2}_{\mathcal{F}}\nonumber \\
	&\le\mu+\sqrt{\mu^2+\frac{1}{2\mu L}\left({\rm{tr}}[\mathbf{T}({\dot{\mathbf{z}}_{\rm{p}}})]+{\rm{tr}}[\mathbf{T}(\dot{{\mathbf{z}}_{\rm{f}}})]\right)}\nonumber \\
	&=\mu+\sqrt{\mu^2+\frac{L}{2\mu}\left(\sum\nolimits^{K}_{k=1}p_k+\sigma^2_{\rm{n}}\right)}.
	\end{align}
	These derivations establish the relationship in \textit{Proposition 4}.

	Likewise, the reconstruction performance of the CRM method can be derived by replacing the objective function and using the same steps. Specifically, \eqref{ApA5} in the CRM method is expressed as
	\begin{align}
	\|\hat{\mathbf{X}}_{\rm{v}}\circ\mathbf{B}&-\tilde{\mathbf{X}}_{\rm{v}}\|^2_{\mathcal{F}}-\|\dot{\mathbf{X}}_{\rm{v}}\circ\mathbf{B}-\tilde{\mathbf{X}}_{\rm{v}}\|^2_{\mathcal{F}}\le\nonumber \\
	&\frac{1}{\mu}\left(\text{f}[{\dot{\bf{W}}}_{\rm{p}},{\bf{T}}({\dot{{\bf{z}}}_{\rm{p}}}),\gamma_{\rm{p}}]+	\text{f}[{\dot{\bf{W}}}_{\rm{f}},{\bf{T}}({\dot{{\bf{z}}}_{\rm{f}}}),\gamma_{\rm{f}}]\right.\nonumber \\
	&\left. -\text{f}[{\hat{\bf{W}}}_{\rm{p}},{\bf{T}}({\hat{{\bf{z}}}_{\rm{p}}}),\gamma_{\rm{p}}]-	\text{f}[{\hat{\bf{W}}}_{\rm{f}},{\bf{T}}({\hat{{\bf{z}}}_{\rm{f}}}),\gamma_{\rm{f}}]\right).
	\end{align}
	As we have mentioned in \textit{Remark 5} of Section IV, the optimal solution of (27) satisfies that $\hat{\mathbf{W}}$ in the null-space of $\mathbf{T}(\hat{\mathbf{z}})$. Further considering the inequality ${\gamma}^{-2}\|\mathbf{W}-\gamma\mathbf{I}\|_{\mathcal{F}}^2\ge 0$ yields
	\begin{align}
	&\left(\xi\!-\!\mu\!-\!\sqrt{\mu^2+\frac{1}{\mu }\left(\text{f}[{\dot{\bf{W}}}_{\rm{p}},{\bf{T}}({\dot{{\bf{z}}}_{\rm{p}}}),\gamma_{\rm{p}}]+	\text{f}[{\dot{\bf{W}}}_{\rm{f}},{\bf{T}}({\dot{{\bf{z}}}_{\rm{f}}}),\gamma_{\rm{f}}]\right)}\right)\nonumber \\
	&\cdot\!\left(\xi\!-\!\mu\!+\!\sqrt{\mu^2+\frac{1}{\mu }\left(\text{f}[{\dot{\bf{W}}}_{\rm{p}},{\bf{T}}({\dot{{\bf{z}}}_{\rm{p}}}),\gamma_{\rm{p}}]+	\text{f}[{\dot{\bf{W}}}_{\rm{f}},{\bf{T}}({\dot{{\bf{z}}}_{\rm{f}}}),\gamma_{\rm{f}}]\right)}\right)\nonumber \\
	&\le 0.
	\end{align}
	Thus, we have 
	\begin{align}
	&\|\hat{\mathbf{X}}_{\rm{v}}\circ\mathbf{B}-\dot{\mathbf{X}}_{\rm{v}}\circ\mathbf{B}\|^{2}_{\mathcal{F}}\nonumber \\
	&\le\mu+\sqrt{\mu^2+\frac{1}{\mu }\left(\text{f}[{\dot{\bf{W}}}_{\rm{p}},{\bf{T}}({\dot{{\bf{z}}}_{\rm{p}}}),\gamma_{\rm{p}}]+	\text{f}[{\dot{\bf{W}}}_{\rm{f}},{\bf{T}}({\dot{{\bf{z}}}_{\rm{f}}}),\gamma_{\rm{f}}]\right)}\nonumber \\
	&=\mu+\sqrt{\mu^2+\frac{2K}{\mu}}.
	\end{align}
	These derivations yield the relationship in \textit{Proposition 5}. $\hfill \blacksquare$

\subsection{Derivation of Coarray CRB}
In this appendix, we derive the CRB for joint DoA-range estimation as a statistical benchmark. For the DoF-enhanced algorithms that utilize the space-frequency difference coarray, we use the matrix identity ${\rm{tr}}[\tilde{\mathbf{A}}\tilde{\mathbf{B}}\tilde{\mathbf{C}}\tilde{\mathbf{D}}]=[{\rm{vec}}(\tilde{\mathbf{B}})]^{\rm{H}}(\tilde{\mathbf{A}}^{\rm{T}}\otimes\tilde{\mathbf{C}})\left[{\rm{vec}\left(\tilde{\mathbf{D}}\right)}\right]$, and reformulate the FIM as
\begin{align}
\mathbf{J}=T\cdot\left[{\rm{vec}}\left(\frac{\partial\mathbf{R}_{\mathbf{x}}}{\partial\boldsymbol{\zeta}}\right)\right]^{\rm{H}}\left(\mathbf{R}_{\mathbf{x}}^{\rm{T}}\otimes\mathbf{R}_{\mathbf{x}}\right)^{-1}\left[{\rm{vec}}\left(\frac{\partial\mathbf{R}_{\mathbf{x}}}{\partial\boldsymbol{\zeta}}\right)\right].
\end{align}
The vectorization of $\mathbf{R}_{\mathbf{x}}$ increases the dimension of $\mathbf{J}$ and leads to valid CRB derivation. Specifically, for the space-frequency difference coarray of FDCA, the FIM can be written as
\begin{align}\label{FIMcoarray}
\mathbf{J}=T\cdot\left[\frac{\partial\dot{\mathbf{x}}_{\rm{v}}}{\partial\boldsymbol{\zeta}}\right]^{\rm{H}}\left(\mathbf{R}_{\mathbf{x}}^{\rm{T}}\otimes\mathbf{R}_{\mathbf{x}}\right)^{-1}\left[\frac{\partial\dot{\mathbf{x}}_{\rm{v}}}{\partial\boldsymbol{\zeta}}\right],
\end{align}
where
\begin{align}	\frac{\partial\dot{\mathbf{x}}_{\rm{v}}}{\partial\boldsymbol{\zeta}}\!=\!\left[\frac{\partial\dot{\mathbf{x}}_{\rm{v}}}{\partial\theta_1},\!\cdots\!,\frac{\partial\dot{\mathbf{x}}_{\rm{v}}}{\partial\theta_K},\frac{\partial\dot{\mathbf{x}}_{\rm{v}}}{\partial r_1},\!\cdots\!,\frac{\partial\dot{\mathbf{x}}_{\rm{v}}}{\partial r_K},\frac{\partial\dot{\mathbf{x}}_{\rm{v}}}{\partial p_1},\!\cdots\!,\frac{\partial\dot{\mathbf{x}}_{\rm{v}}}{\partial p_K},\frac{\partial\dot{\mathbf{x}}_{\rm{v}}}{\partial\sigma_n^2}\right]
\end{align}
with
\begin{align}
\frac{\partial\dot{\mathbf{x}}_{\rm{v}}}{\partial p_k}=\mathbf{h}^{*}\left(\theta_k,r_k\right)\otimes\mathbf{h}\left(\theta_k,r_k\right),
\end{align}
\begin{align}
\frac{\partial\dot{\mathbf{x}}_{\rm{v}}}{\partial \sigma_{\rm{n}}^2}=\mathbf{i}=\underbrace{\left[\underbrace{1\;0\; \cdots\; 0}_{\left(M+N-1\right)^2}\;\underbrace{0\;1\; \cdots\; 0}_{\left(M+N-1\right)^2}\;\cdots\;\underbrace{0\;0\; \cdots\; 1}_{\left(M+N-1\right)^2}\right]}_{\left(M+N-1\right)^4},
\end{align}
\vspace{-1em}	
\begin{align}
\frac{\partial\dot{\mathbf{x}}_{\rm{v}}}{\partial \theta_k}=p_k&\left[\frac{\partial\mathbf{h}_{\rm{p}}^{*}\left(\theta_k\right)}{\partial\theta_k}\otimes\mathbf{h}^{*}_{\rm{f}}\left(r_k\right)\otimes\mathbf{h}\left(\theta_k,r_k\right)\right.\nonumber \\
&\left.+\mathbf{h}^{*}\left(\theta_k,r_k\right)\otimes\frac{\partial\mathbf{h}_{\rm{p}}\left(\theta_k\right)}{\partial\theta_k}\otimes\mathbf{h}_{\rm{f}}\left(r_k\right)\right],
\end{align}
\begin{align}
\frac{\partial\dot{\mathbf{x}}_{\rm{v}}}{\partial r_k}=p_k&\left[\mathbf{h}^{*}_{\rm{p}}\left(\theta_{k}\right)\otimes\frac{\partial\mathbf{h}^{*}_{\rm{f}}\left(r_k\right)}{\partial r_k}\otimes\mathbf{h}\left(\theta_k,r_k\right)\right.\nonumber \\
&\left.+\mathbf{h}^{*}\left(\theta_k,r_k\right)\otimes\mathbf{h}_{\rm{p}}\left(\theta_k\right)\otimes\frac{\partial\mathbf{h}_{\rm{f}}\left(r_k\right)}{\partial r_k}\right].
\end{align}
Hence, the CRB of joint DoA-range estimation for the $k$-th target can be obtained as
\begin{equation}
{\rm{CRB}}\left(\theta_k\right)=\left[\mathbf{J}^{-1}\right]_{k,k},
\ 		{\rm{CRB}}\left(r_k\right)=\left[\mathbf{J}^{-1}\right]_{K+k,K+k},
\end{equation}
for $k=1,\cdots,K$. Note that we use the Slepian-Bangs formula in deriving the CRB to allow the readers to evaluate the RMSE performance of other array configurations, since the proposed framework is general and is extensible to arbitrary holed coarray. \hfill$\blacksquare$

	\bibliographystyle{IEEEbib}